\documentclass[aps,prd,superscriptaddress,amsmath,showpacs,floatfix,nofootinbib,preprintnumbers,notitlepage]{revtex4-1}

\usepackage{pstricks}
\usepackage{color}
\usepackage{graphicx}
\usepackage{amsmath} 
\usepackage[colorlinks=true,linkcolor=linkcolor,citecolor=darkred,urlcolor=darkred, pdfborder={0 0 0}]{hyperref}
\usepackage{amssymb}
\usepackage{multirow}
\usepackage{pifont}

\newcommand{\yes}{\ding{51}}
\newcommand{\no}{\ding{55}}
\def\lfv{lepton flavor violation }

\definecolor{linkcolor}{rgb}{0,0,0.3}
\definecolor{darkred}{rgb}{0.6,0,0}

\graphicspath{{figs/}}

\def\gsim{\raise0.3ex\hbox{$\;>$\kern-0.75em\raise-1.1ex\hbox{$\sim\;$}}}
\def\lsim{\raise0.3ex\hbox{$\;<$\kern-0.75em\raise-1.1ex\hbox{$\sim\;$}}}

\def\beqn#1{\begin{equation}\label{#1}}
\def\eeqn{\end{equation}}

\def\beqa#1{\begin{eqnarray}\label{#1}}
\def\eeqa{\end{eqnarray}}

\def\g{\gamma}
\def\g5{\gamma_5}
\def\21{SU(2) $\otimes$ U(1) }
\def\lr{(2)_L \otimes SU(2)_R \otimes U(1)$}

\def\Z2{$\mathcal{Z_2}$}

\def\vev#1{\left\langle #1\right\rangle}

\def\oneR{\ensuremath{\mathbf{1}}}
\def\twoR{\ensuremath{\mathbf{2}}}

\def\SM{$\mathrm{SU(3)_c \otimes SU(2)_L \otimes U(1)_Y}$ }
\def\lr{$\mathrm{SU(3) \otimes SU(2)_L \otimes SU(2)_R \otimes
    U(1)_{B-L}}$ }
\newcommand{\sm}{{standard model }}

\newcommand{\AHEP}{AHEP Group at IFIC, Instituto de F\'{\i}sica Corpuscular --
 Parque Cient\'{\i}fico, Universitat de Val{\`e}ncia \\
 C/Catedr\'atico Jos\'e Beltr\'an, 2 
 E-46980 Paterna, Val\`encia, Spain.}

\begin{document}

\title{The low-scale approach to neutrino masses}

\author{Sofiane M. Boucenna} \email{msboucenna@gmail.com}
\affiliation{\AHEP}
  \author{Stefano Morisi} \email{stefano.morisi@gmail.com}
\affiliation{DESY, Platanenallee 6, D-15735 Zeuthen, Germany.}
\author{Jos\'e W.F. Valle} \email{valle@ific.uv.es}
 \affiliation{\AHEP}
\date{\today}

\begin{abstract}
  \noindent In this short review we revisit the broad landscape of
  low-scale \SM models of neutrino mass generation, with view on their
  phenomenological potential. This includes signatures associated to
  direct neutrino mass messenger production at the LHC, as well as
  messenger-induced \lfv processes. We also briefly comment on
  the presence of WIMP cold dark matter candidates. 
\end{abstract}

\maketitle
\tableofcontents
\newpage


\section{Introduction}

The flavor problem, namely why we have three families of fermions with
the same standard model quantum numbers, but with very hierarchical
masses and a puzzling pattern of mixing parameters, constitutes one of
the most challenging open problems in particle physics.  In this
regard neutrinos are probably the most mysterious particles.  Indeed,
while the discovery of the Higgs boson by the ATLAS and CMS
experiments at the Large Hadron Collider (LHC) at
CERN~\cite{ATLASandCMSCollaborations:2013pga,Aad:2012tfa,Chatrchyan:2012ufa}
has clarified to some extent the nature of electroweak symmetry
breaking, the origin of neutrino masses remains elusive.
With standard model fields one can induce Majorana neutrino masses
through the non-renormalizable dimension-5 operator
\begin{equation}\label{eq:d5}
{\cal O}_{\rm dim=5}=\frac{\lambda}{\Lambda} LLHH
\end{equation}
or higher order ones, e.~g.  $LLHH(H^\dagger
H)^m$~\cite{Weinberg:1979sa,Bonnet:2012kh,Bonnet:2012kz,Gouvea:2007xp,Bonnet:2009ej,Krauss:2013gy}, where $\lambda$ is a dimensionless
coupling and $\Lambda$ denotes some unknown effective scale.
However, strictly speaking, we still do not know whether neutrinos are
Dirac or Majorana fermions, and many issues remain open regarding the
nature of the associated mass-giving operator, for example,
\begin{itemize}
\item its underlying symmetries, such as total lepton number,
\item its flavor structure which should account for the observed
  oscillation pattern,
\item its dimensionality,
\item its characteristic scale,
\item its underlying mechanism.
\end{itemize}
This leads to considerable theoretical freedom which makes model
building an especially hard task, a difficulty which to a large extent
persists despite the tremendous experimental progress of the last
fifteen years~\cite{Nunokawa:2007qh,Tortola:2012te}.

Indeed the origin of neutrino mass remains so far a mystery. From
oscillation studies we can not know the absolute neutrino mass
scale. Still we know for certain that neutrinos are the lightest known
fermions. Their mass must be below the few eV scale from tritium beta
decay studies at the Katrin experiment~\cite{Osipowicz:2001sq}, with
somewhat stronger, though more model dependent limits coming from
cosmology~\cite{pastor-book} and from negative neutrinoless double
beta decay searches~\cite{Barabash:2011fn}. Unfortunately this vast
body of information is far from sufficient to underpin the nature of
the neutrino mass generation mechanism.

Mechanisms inducing neutrino mass may be broadly divided on the basis
of whether the associated messengers lie at the high energy scale,
related say, to some unification scheme or, in contrast, they involve
new physics at the TeV scale, potentially accessible at the LHC.

For simplicity here we tacitly assume neutrino masses to come from
Weinberg's operator in Eq.~(\ref{eq:d5}).
This operator can arise in a rich variety of different
pathways~\cite{Ma:1998dn}. For instance in the case of the standard
type-I seesaw
mechanism~\cite{Minkowski:1977sc,Yanagida:1979as,GellMann:1980vs,Schechter:1980gr,Mohapatra:1979ia,Schechter:1981cv}
the right-handed neutrino messengers have a Majorana mass at some
large scale, fitting naturally in Grand Unified Theories (GUTs).
There are, however, many alternative realizations of the dimension-5
operator, such as the
type-II~\cite{Schechter:1980gr,Cheng:1980qt,Magg:1980ut,Wetterich:1981bx,Mohapatra:1980yp}
and type-III seesaw~\cite{Foot:1988aq} constructions, in which the
messengers have non-trivial gauge quantum numbers. 
Such schemes are \textit{bona fide} high-scale seesaw in the sense
that, to account for the observed neutrino masses with reasonable
strength for the relevant neutrino Yukawa couplings, one needs very
large scales for the messenger mass, hence inaccessible to collider
experiments.
Of course within such scenarios one may artificially take TeV scales
for the messenger mass by assuming tiny Yukawas, so as to account for
the smallness of neutrino mass~\footnote{One can avoid this in schemes
  where \textit{ad hoc} cancellations~\cite{Kersten:2007vk} or
  symmetries~\cite{Gu:2008yj,Dev:2013wba} prevent
  seesaw-produced masses. We do not consider such a special case in
  this review.  Similarly we will not assume any family symmetry
  restricting the flavor structure of models.}. However by doing so
one erases a number of potential phenomenological implications.  Hence
we call such standard seesaw varieties as {\it high scale seesaw}.
\begin{figure}[t]
\centering
\includegraphics[width=0.35\textwidth,keepaspectratio=true,clip=true]{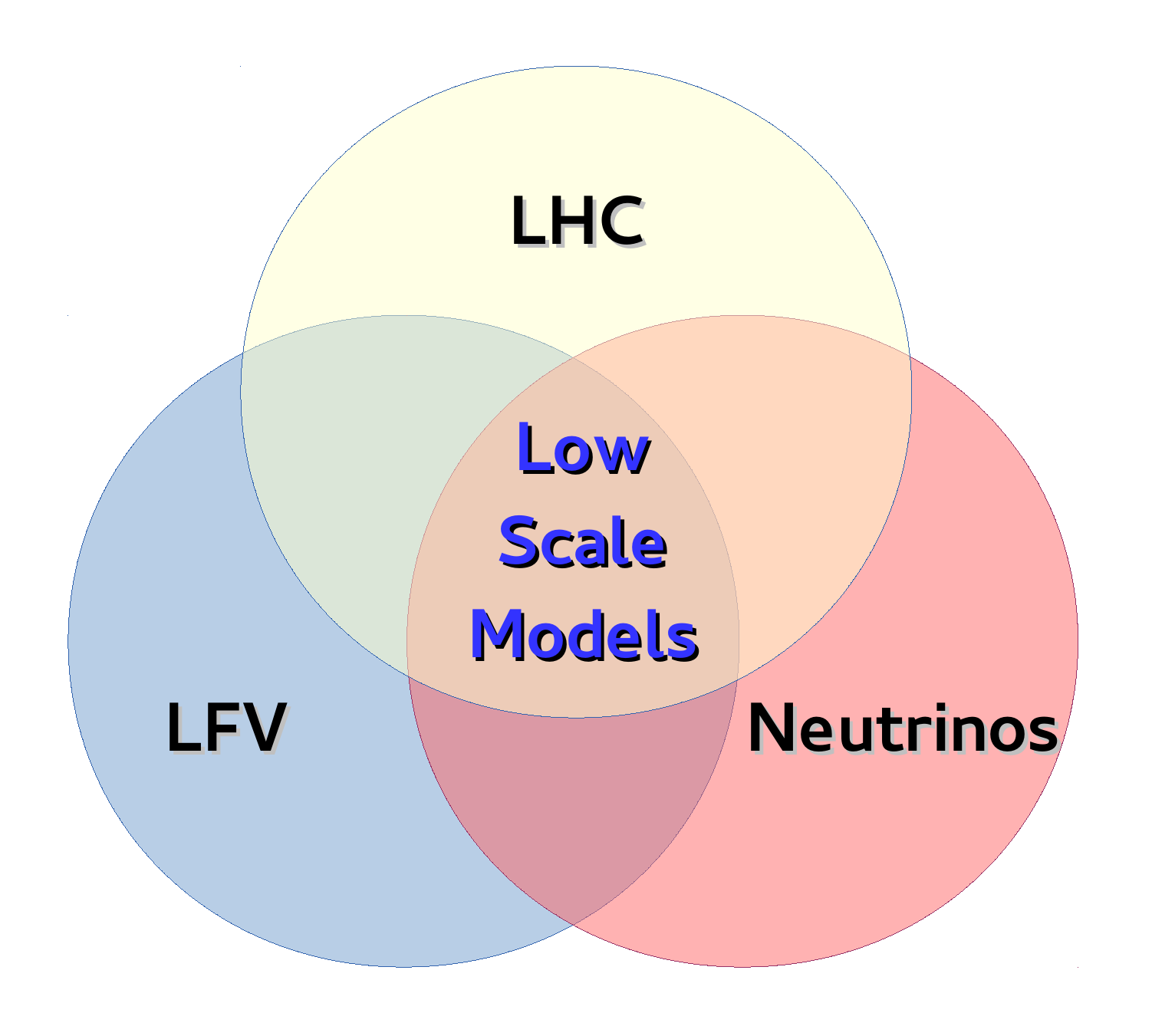}
\caption{Low scale neutrino mass models at the crossroad of high and low energy experiments.}
\label{fig:propaganda}
\end{figure}
It has long ago been realized~\cite{Schechter:1980gr} that, carrying
no anomalies, singlets can be added in an arbitrary number to any
gauge theory. Within the framework of the \sm \SM gauge structure, the
models can be labeled by an integer, $m$, the number of singlets. For
example, to account for current neutrino oscillation data, a type-I
seesaw model with two right-handed neutrinos is sufficient
($m=2$). Likewise for models with $m=1$ in which another mechanism such as
radiative corrections (see below) generates the remaining scale.
Especially interesting are models with $m>3$, where one can exploit the
extra freedom to realize symmetries, such as lepton number L, so as to
avoid seesaw-induced neutrino masses, naturally allowing for TeV-scale
messengers. This is the idea behind the
inverse~\cite{Mohapatra:1986bd} and linear seesaw
schemes~\cite{Akhmedov:1995ip,Akhmedov:1995vm,Malinsky:2005bi}
described in the next section.
We call such schemes as genuine  {\it low scale seesaw} constructions. 
A phenomenologically attractive alternative to low-scale seesaw are
models where neutrino masses arise radiatively~\cite{Babu:1988ki}.

In principle one can assume the presence of supersymmetry in any such
scheme, though in most cases it does not play an essential role for
neutrino mass generation, \textit{per se}. However we give an example
where it could, namely, when the origin of neutrino mass is strictly
supersymmetric because R-parity breaks.
Indeed, neither gauge invariance nor supersymmetry require R-parity
conservation. There are viable models where R-parity is an exact
symmetry of the Lagrangian but breaks spontaneously through the Higgs
mechanism~\cite{masiero:1990uj,Romao:1992vu} by an $L=1$
vacuum expectation value.
As we will explain in the next section this scheme is hybrid in the
sense that it combines seesaw and radiative contributions.
In all of the above one can assume that the neutrino mass messengers
lie at the TeV mass scale, hence have potentially detectable
consequences.

In this review we consider the low-scale approach to neutrino masses.
We choose to map out the possible schemes taking their potential
phenomenological implications as guiding criteria, focusing on
possible signatures at the LHC and lepton flavor violation (LFV)
processes (Fig.\,\ref{fig:seesawI} ).
The paper is organized as follows: in section 2 we review low energy
seesaw schemes, in section 3 we discuss one, two and three-loop
radiative models.  In section 4 we discuss the supersymmetric
mechanism and we sum up in section 5.



\vskip3.mm
\section{Seesaw mechanism}

\subsection{High scale seesaw }

\begin{figure}[b]
\centering
\includegraphics[width=0.3\textwidth,keepaspectratio=true,clip=true]{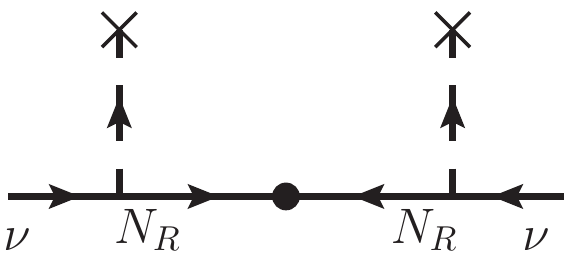}
\hskip5.mm
\includegraphics[width=0.25\textwidth,keepaspectratio=true,clip=true]{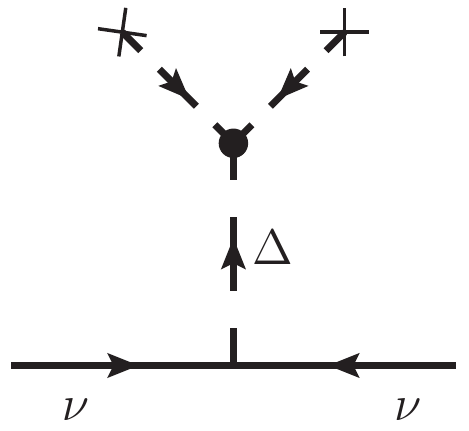}
\caption{Neutrino mass generation in the type-I seesaw (left) and type-II seesaw (right). The black disks show where lepton number violation takes place.}
\label{fig:seesawI}
\end{figure}

Within minimal unified models such as SO(10), without gauge singlets,
one automatically encounters the presence of new scalar or fermion
states that can act as neutrino mass mediators inducing Weinberg's
operator in Eq.~(\ref{eq:d5}).
This leads to different variants of the so-called seesaw
mechanism. One possibility is to employ the right-handed neutrinos
present in the {\bf 16} of SO(10), and are broadly called type-I
seesaw
schemes~\cite{Minkowski:1977sc,Yanagida:1979as,GellMann:1980vs,Schechter:1980gr,Mohapatra:1979ia,Schechter:1981cv}
(see Fig.\,\ref{fig:seesawI}).
Similar unified constructions can also be made substituting the
right-handed neutrino exchange by that of an exotic
hypercharge-neutral isotriplet lepton~\cite{Foot:1988aq},
$$\Sigma=(\Sigma^+,\Sigma^0,\Sigma^-),$$ which is
called type-III seesaw~\cite{Foot:1988aq}.
An alternative mediator is provided by a hypercharge-carrying
isotriplet coming from the {\bf 126} of SO(10), and goes by the name
type-II seesaw
mechanism~\cite{Schechter:1980gr,Cheng:1980qt,Magg:1980ut,Mohapatra:1980yp}
(see Fig.~(\ref{fig:seesawI})).

The three options all involve new physics at high scale, typically
close to the unification scale. While model-dependent, the expected
magnitude of the mass of such messengers is typically expected to be
high, say, associated to the breaking of extra gauge symmetries, such
as the $B-L$ generator.

Within standard type-I or type-III seesaw mechanism with three
right-handed neutrinos the isodoublet neutrinos get mixed with the new
messenger fermions by a 6 $\times$ 6 seesaw block diagonalization
matrix that can be determined perturbatively using the general method
in~\cite{Schechter:1981cv}.
For example in the conventional type-I seesaw case the $6\times 6$
matrix $U$ that diagonalizes the neutrino mass is unitary and is given
by 
\begin{equation}\label{u-expansion}
U=
\left(
\begin{array}{cc}
\left(I-\frac{1}{2}m_D^*(M_R^*)^{-1}\,M_R^{-1}m_D^T\right)V_1 & m_D^*(M_R^*)^{-1}\,V_2 \\
 -M_R^{-1}\,m_D^T \,V_1& \left(I-\frac{1}{2}M_R^{-1}\,m_D^T\,m_D^*\,(M_R^*)^{-1}\right)V_2
\end{array}
\right)+O(\epsilon^3),
\end{equation}
where $V_1$ and $V_2$ are the unitary matrices that diagonalizes the
light and heavy sub-block respectively.  From Eq.~(\ref{u-expansion})
one sees that the active $3\times 3$ sub-block is no longer unitary
and the deviation from unitary is of the order of $\epsilon^2 \sim
(m_D/M_R)^2$.  The expansion parameter $\epsilon$ is very small if the
scale of new physics is at the GUT scale so the induced \lfv processes
are suppressed.  In this case there are no detectable direct
production signatures at colliders nor LFV processes.
This follows from the well know type-I seesaw relation
$$m_\nu\sim m_D^2/M_{\rm messenger}$$ where $M_{\rm messenger} = M_R$ implying that 
\begin{equation}
  \label{eq:expa}
\epsilon^2 \sim m_\nu/M_R  
\end{equation}
is suppressed by the neutrino mass, hence negligible regardless of
whether the messenger scale $M_R$ lies in the TeV scale~\footnote{Weak
  universality tests as well as searches at LEP and previous colliders
  rule out lower messenger mass
  scales~\cite{abreu:1996pa,Beringer:1900zz}.}.
As a result there is a decoupling of the effects of the messengers at
low energy, other than providing neutrino masses. This includes for
example \lfv effects in both type-I and type-III seesaw mechanism.
Regarding direct signatures at collider experiments these require TeV
scale messengers which can be artificially implemented in both type-I
and type-III cases by assuming the Dirac-type Yukawa couplings to be
tiny. This makes messenger production at colliders totally hopeless in
type-I seesaw, but does not affect the production rate in type-III
seesaw mechanism, since it proceeds with gauge
strength~\cite{Franceschini:2008pz}.

Coming to the type-II scheme, neutrino masses are proportional to the
vev of the neutral component of a scalar electroweak triplet
$\Delta^0$ and we have
\begin{equation}
m_\nu=y_\nu\, v_T\,, \quad \mbox{with} \quad v_T= \frac{\mu_T\, v^2}{M^2_T}\,,
\end{equation}
where $v$ is the vev of the \sm Higgs, $M_T$ is the mass of the scalar
triplet $\Delta$, $y_\nu$ is the coupling of the neutrino with the
scalar triplet and $\mu_T$ is the coupling (with mass dimension) of
the trilinear term between the \sm Higgs boson and the scalar triplet
$H^T\Delta H$. Assuming $y_\nu$ of order one, in order to have light
neutrino mass there are two possibilities: either $M_T$ is large or
$\mu_T$ is small.  The first case is the standard type-II seesaw where
all the parameters of the model are naturally of order one.

In such high scale type-I and type-III seesaw varieties neutrino mass
messengers are above the energy reach of any conceivable accelerator,
while \lfv effects arising from messenger exchange are also highly
suppressed. Should \lfv ever be observed in nature, such schemes would
suggest the existence of an alternative \lfv mechanism.
A celebrated example of the latter is provided the exchange of scalar
leptons in supersymmetric
models~\cite{borzumati:1986qx,hirsch:2008dy,Esteves:2010si}.

In contrast, if type-II seesaw schemes are chosen to lie at the TeV
scale, then \lfv effects as well as same-sign di-lepton signatures at
colliders remain~\cite{perez:2009mu}, see below.
Obviously supersymmetrized ``low-scale'' type-II seesaw have an even richer
phenomenology~\cite{AristizabalSierra:2003ix,esteves:2009vg}.

\subsection{Low scale type-I seesaw}

The most general approach to the seesaw mechanism is that provided by
the standard \SM gauge group structure which holds at low
energies. Within this framework one can construct seesaw theories with
an arbitrary number of right handed neutrinos,
$m$~\cite{Schechter:1980gr}, since gauge singlets carry no
anomalies. In fact the same trick can be upgraded to other extended
gauge groups, such as \lr or Pati-Salam and also unified groups such
as SO(10) \cite{Dev:2009aw,Dev:2013oxa} or E$_6$.  This opens the door
to genuine low-scale realizations of the seesaw mechanism.

\begin{figure}[t]
\centering
\includegraphics[clip,width=0.80\textwidth]{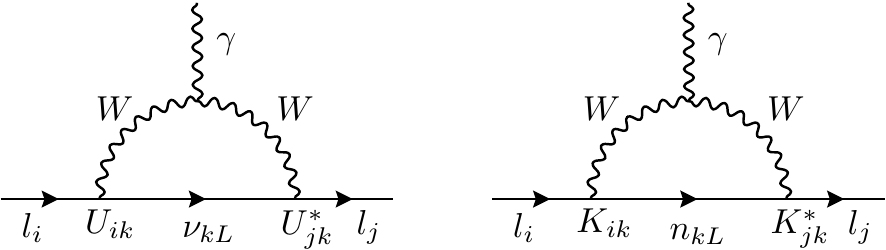}
\caption{Radiative decays $\ell_i \to \ell_j \gamma$ in the \sm with
  massive light neutrinos (left) and heavy neutrinos (right).}
\label{fig:LFV}
\end{figure}

Before turning to the description of specific low scale type-I seesaw
schemes lets us briefly note their basic phenomenological feature,
namely that in genuine low-scale seesaw schemes Eq.~(\ref{eq:expa})
does not hold so that, for light enough messengers, one can have \lfv
processes~\cite{deppisch:2005zm,Deppisch:2004fa,Forero:2011pc}.
For example, radiative decays $\ell_i \to \ell_j \gamma$ proceed
through the exchange of light (left panel) as well as heavy neutrinos
(right) in Fig.~(\ref{fig:LFV}).
Clearly expected \lfv rates such as that for the $\mu\to e\gamma$
process are too small to be of interest. Another important conceptual
feature of phenomenological importance is that \lfv survives even in
the limit of strictly massless neutrinos (i.e. $\mu\to 0$, see text below)~\cite{bernabeu:1987gr,gonzalez-garcia:1992be}.

\subsubsection{Inverse type-I seesaw}
\label{sec:inv}

In its simplest realization the inverse seesaw extends the \sm by
means of two sets of electroweak two-component singlet fermions
${N_R}_i$ and ${S_L}_j$~\cite{Mohapatra:1986bd}.  The lepton number
$L$ of the two sets of fields $N_R$ and $S_L$ can be assigned as
$L({N_R})=+1$ and $L({S_L})=+1$.  One assumes that the fermion pairs
are added sequentially, i.e. $i,j=1,2,3$, though other variants are
possible.
After electroweak symmetry breaking the Lagrangian is given by
\begin{equation}
\mathcal{L}= m_D \overline{\nu}_L N_R + M\, \overline{N}_R S_L +\mu\, \tilde{S}_L S_L + \mathrm{h.c.},
\end{equation}
We define $\tilde{S}_L \equiv S_L^TC^{-1}$ where $C$ is the  charge conjugation matrix, $m_D$ and $M$ are arbitrary
$3\times 3 $ Dirac mass matrices and $\mu $ is a Majorana $3\times 3 $
matrix. We note that the lepton number is violated by the $\mu$ mass
term here.
The full neutrino mass matrix can be written as a $9\times 9 $ matrix
instead of $6\times 6$ as in the typical type-I seesaw and is given by
(in the basis $\nu_L$, $N_R$ and $S_L$)
\begin{equation}\label{eq:iss}
M_\nu=
\left(
\begin{array}{ccc}
0 & m_D^T & 0\\
m_D & 0&M^T\\
0&M & \mu
\end{array}
\right)
\end{equation}
The entry $\mu$ may be generated from the spontaneous breaking of
lepton number through the vacuum expectation value of a gauge singlet
scalar boson carrying $L=2$~\cite{gonzalezgarcia:1988rw}.

It is easy to see that in the limit where $\mu\to 0$ the exact U(1)
symmetry associated to total lepton number conservation holds, so the
light neutrinos are strictly massless. However individual symmetries
are broken hence flavor is violated, despite neutrinos being
massless~\cite{bernabeu:1987gr,gonzalez-garcia:1992be}. For complex
couplings, one can also show that CP is violated despite the fact that
light neutrinos are strictly
degenerate~\cite{branco:1989bn,rius:1989gk}. The fact that flavor and
CP are violated in the massless limit implies that the attainable
rates for the corresponding processes are unconstrained by the
observed smallness of neutrino masses, and are potentially large.

\begin{figure}[t]
\centering
\includegraphics[width=0.50\textwidth,keepaspectratio=true,clip=true]{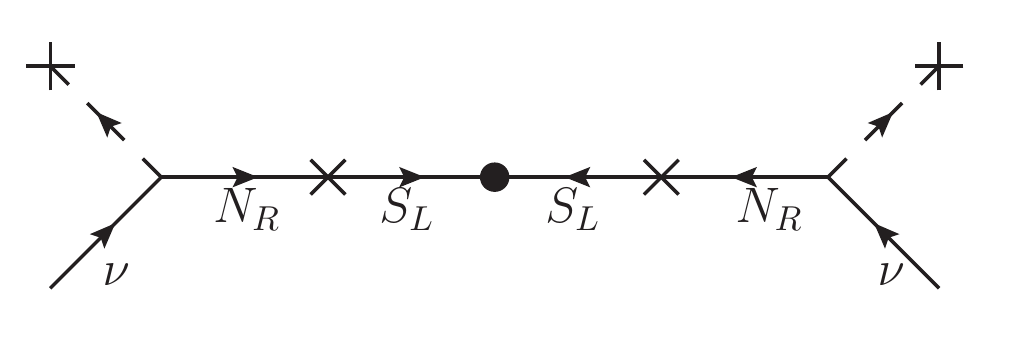}
\caption{Neutrino mass generation in the type-I inverse seesaw.}
\label{fig:invseesaw}
\end{figure}

This feature makes this scenario conceptually and phenomenologically
interesting and is a consequence of the fact that the lepton number is
conserved.  However when $\mu \ne 0$ light neutrino masses are
generated,
see Fig.~(\ref{fig:invseesaw}).
In particular in the limit where $\mu, m_D \lsim M$~\footnote{On the
  other hand, the opposite limit $\mu\gg M$ is called double
  seesaw. In contrast to the inverse seesaw, the double seesaw brings
  no qualitative differences with respect to standard seesaw and will
  not be considered here.}  the light neutrino $3\times 3 $ mass matrix
is given by
\begin{equation}
m_\nu\simeq m_D\frac{1}{M}\mu\frac{1}{M^T}m_D^T\,.
\end{equation}
It is clear from this formula that for ``reasonable'' Yukawa strength
or $m_D$ values, $M$ of the order of TeV, and suitably small $\mu$
values one can account for the required light neutrino mass scale at
the eV scale. 
There are two new physics scales, $M$ and $\mu$, the last of which is
very small.
Therefore it constitutes an extension of the \sm from below, rather
than from above. For this reason, it has been called inverse seesaw:
in contrast with the standard type-I seesaw mechanism, neutrino masses
are suppressed by a small parameter, instead of the inverse of a large
one.
The smallness of the scale $\mu$ is {\it natural} in t'Hooft's sense,
namely in the limit $\mu\to 0$, a symmetry the symmetry is enhanced
since lepton number is recovered~\footnote{There are realizations
  where the low scale of $\mu$ is radiatively calculable. As examples
  see the supersymmetry framework given in~\cite{Bazzocchi:2009kc}, or
  the \sm extension suggested in~\cite{Bazzocchi:2010dt}.}.
 
In this case the seesaw expansion parameter $\epsilon \sim m_D/M$ also
characterizes the strength of unitarity and universality violation and
can be of order of percent or
so~\cite{Hettmansperger:2011bt,Forero:2011pc}, leading to sizable \lfv
rates, close to future experimental sensitivities.
For example, with $m_D=30~$GeV, $M=300~$GeV and $\mu=10~$eV we have that
$\epsilon^2\sim 10^{-2}$.  The deviation from the unitary is typically
of order $\epsilon^2$.
\begin{figure*}[t!] 
\label{fig:meg}
  \centering 
\includegraphics[height=50mm,keepaspectratio=true,width=.52\linewidth]{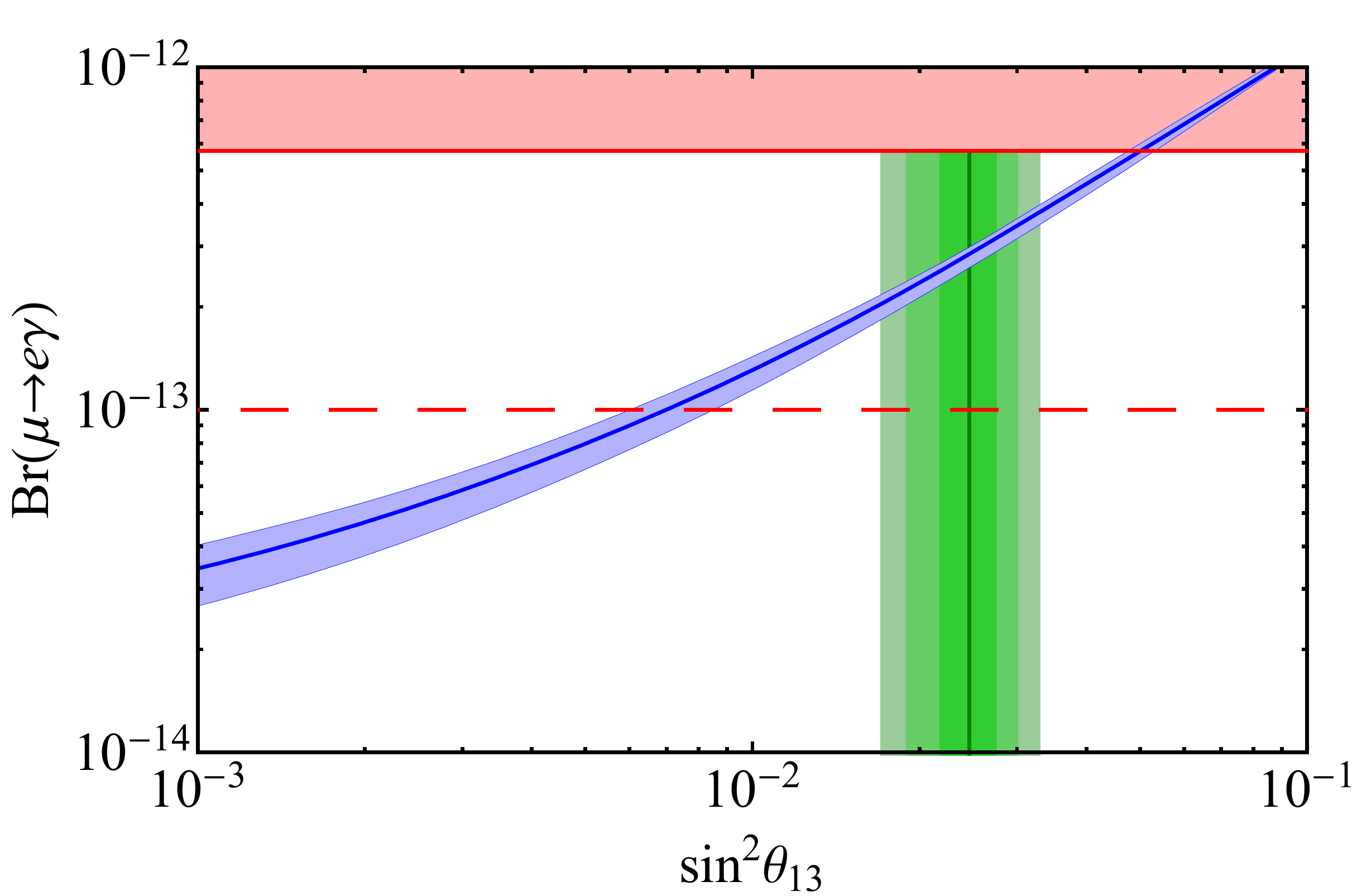}
\caption{\label{fig:mueg2} Branching ratios \(Br(\mu\to e\gamma)\) in
  the inverse seesaw model of neutrino mass~\cite{Deppisch:2004fa}. }
\end{figure*}
As mentioned above, typical expected \lfv rates in the inverse seesaw
model can be potentially large. For example, the rates for the classic
$\mu\to e\gamma$ process are illustrated in Fig.~\ref{fig:meg}.
The figure gives the predicted branching ratios \(Br(\mu\to e\gamma)\)
in terms of the small neutrino mixing angle $\theta_{13}$, for
different values of the remaining oscillation parameters, with the
solar mixing parameter $\sin^2\theta_{12}$ within its $3\sigma$
allowed range and fixing the inverse seesaw parameters as: $M=1$~TeV
and $\mu=3$~KeV.  The vertical band corresponds to the $3\sigma$
allowed $\theta_{13}$ range.

Regarding direct production at colliders, although kinematically
possible, the associated signatures are not easy to catch given the
low rates as the right handed neutrinos are gauge singlets and due to
the expected backgrounds (see for instance \cite{Das:2012ze}).

The way out is by embedding the model within an extended gauge
structure that can hold at TeV energies, such as an extra U(1) coupled
to $B-L$ which may arise from SO(10)~\cite{Malinsky:2005bi}. Viable
scenarios may also have TeV-scale \lr or Pati-Salam intermediate
symmetries~\cite{DeRomeri:2011ie}.
In this case the right handed messengers can be produced through a new
charged~\cite{Nath:2010zj,Das:2012ii,AguilarSaavedra:2012fu} or
neutral gauge boson~\cite{Deppisch:2013cya}.
In fact one has the fascinating additional possibility of detectable
\lfv taking place at the large energies now accessible at the
LHC~\cite{Deppisch:2013cya}.

\subsubsection{Linear type-I seesaw}

This variant of low-scale seesaw was first studied in the context of
\lr theories~\cite{Akhmedov:1995ip,Akhmedov:1995vm} and subsequently
demonstrated to arise naturally within the SO(10) framework in the
presence of gauge singlets~\cite{Malinsky:2005bi}. The lepton number
assignment is as follows: $L({\nu_L}_i) = +1$, $L({N_R}_i) = 1$ and
$L({S_L}_i) = +1$ so that, after electroweak symmetry breaking the
Lagrangian is given by
\begin{equation}
  \mathcal{L}= m_D \overline{\nu}_L N_R + M_R\, \overline{N}_R S_L + M_L \nu_L {\tilde{S}}_L + \mathrm{h.c.}
\end{equation}
Notice that the lepton number is broken by the mass term proportional to $M_L$. This corresponds to the neutrino mass matrix in the basis
$\nu_L$, $N_R$ and $S_L$ given as
\begin{equation}
M_\nu=
\left(
\begin{array}{ccc}
0 & m_D^T & M_L\\
m_D & 0&M_R\\
M_L&M_R & 0
\end{array}
\right)
\end{equation}
If $m_D\ll M_{L,R}$ then the effective light neutrino mass matrix is
given by
\begin{equation}\label{lss}
m_\nu = m_D\,M_L \frac{1}{M_R}+{\rm Transpose}.
\end{equation}
Note that, in contrast with other seesaw varieties which lead to
$m_\nu \propto m_D^2$, this relation is linear in the Dirac mass
entry, hence the origin of the name ``linear seesaw''.
Clearly neutrino masses will be suppressed by the small value of $M_L$
irrespective of how low is the $M_R$ scale characterizing the heavy
messengers. For example, if one takes the SO(10) unification
framework~\cite{Malinsky:2005bi}, natural in this context, one finds
that the scale of $M_L$, i.e $v_L$, is related to the scale of $M_R$,
i.e. $v_R$ through
\begin{equation}\label{vLvR}
v_L \sim \frac{v_R\, v}{M_{\rm GUT}}\,,
\end{equation}
where $M_{\rm GUT}$ is the unification scale of the order of
$\mathcal{O}(10^{16}\, \mathrm{GeV})$ and $v$ is the electroweak breaking
scale of the order of $\mathcal{O}(100\, \mathrm{GeV})$. Replacing the relation
(\ref{vLvR}) in Eq.~(\ref{lss}) the new physics scale drops out
and can be very light, of the order of TeV.

Neutrino mass messengers are naturally accessible at colliders, like
the LHC, since the right handed neutrinos can be produced through the
$Z^\prime$ ``portal'', as light as few TeV. The scenario has been shown
to be fully consistent with the required smallness of neutrino mass as
well as with the requirement of gauge coupling
unification~\cite{Malinsky:2005bi}. Other \lr and Pati-Salam
implementations also been studied in~\cite{DeRomeri:2011ie}.

Similarly to the inverse type-I seesaw scheme, we also have here
potentially large unitarity violation in the effective lepton mixing
matrix governing the couplings of the light neutrinos. This gives rise
to \lfv effects similar to the inverse seesaw case. Finally we note
that, in general, a left-right symmetric linear seesaw construction
also contains the lepton number violating Majorana mass term
$\tilde{S}_L S_L $ considered previously.

\subsection{Low scale type-III seesaw}

Here we consider a variant of the low scale type-III seesaw model introduced in~\cite{Ma:2009kh} based on the
inverse seesaw mechanism~\cite{Mohapatra:1986bd} but replacing the
$N_R$ lepton field with the neutral component $\Sigma^0$ of a fermion
triplet under SU(2)$_L$ with hypercharge zero~\cite{Ibanez:2009du}, $$\Sigma=(\Sigma^+,\Sigma^0,\Sigma^-).$$
As in the the inverse type-I seesaw one introduces an extra set of
gauge singlet fermions $S_L$ with lepton number $L(S_L)=+1$ and
$L(\Sigma^0)=+1$. The mass Lagrangian is given by
\begin{equation}
\mathcal{L}= m_D \overline{\nu}_L \Sigma^0 + M\, \overline{\Sigma^0}S_L +\mu\, \tilde{S}_L S_L -\frac{1}{2} m_\Sigma \rm{Tr}(\overline{\Sigma} \Sigma^c) + \mathrm{h.c.}
\end{equation}
In the basis $(\nu,\,\Sigma^0,\,S_L)$ the neutrino mass matrix is
given by
\begin{equation}\label{eq:iss}
M_\nu=
\left(
\begin{array}{ccc}
0 & m_D^T & 0\\
m_D & m_\Sigma&M^T\\
0&M & \mu
\end{array}
\right).
\end{equation}

As in the inverse seesaw case, in the limit $\mu=0$ the light neutrinos are massless at tree level even if the mass term $m_\Sigma$ breaks lepton number.
And for a small $\mu\ne0$ neutrinos get mass. Again, the scale of new physics is naturally small leading to sizable \lfv rates. 

On the other hand the charged component of the fermion triplet
$\Sigma^\pm $ gives also a contribution to the charged lepton mass
matrix
\begin{equation}
\label{eq:mlep}
M_{\rm ch. lep}=\left(
\begin{array}{ccc}
M_l&m_D\\
0 & m_{\Sigma}
\end{array}
\right),
\end{equation}
leading to a violation of the Glashow-Iliopoulos-Maiani
mechanism~\cite{glashow:1970gm} in the charged lepton sector, leading
to tree level contributions to $\mu\to eee$ and similar tau decay
processes.

\begin{figure}[h!]
\begin{center}
\includegraphics[angle=0,height=5cm,keepaspectratio=true,width=0.5\textwidth]{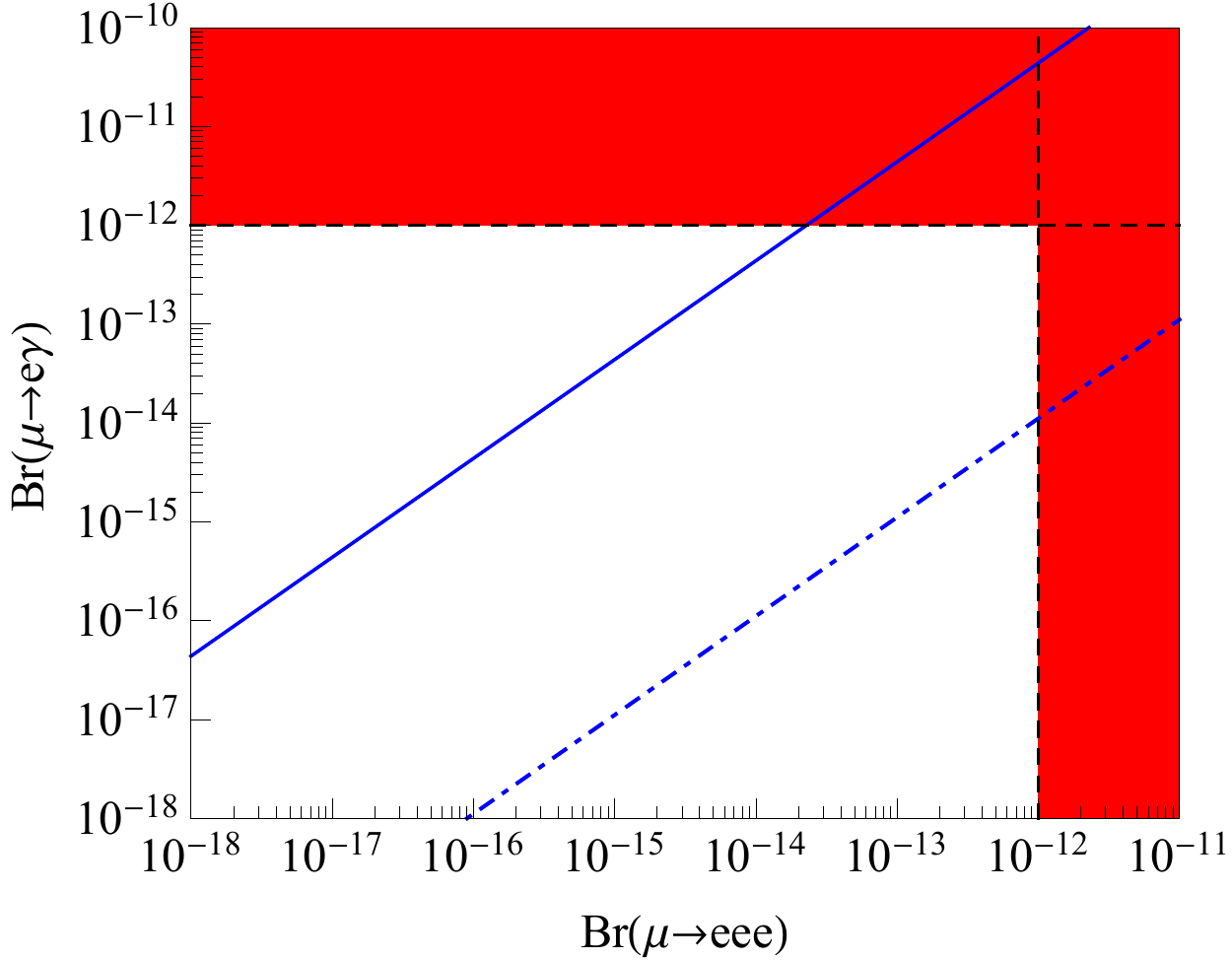}
\caption{Branching of $\mu$ decay into $3e$ vs the branching of
  $\mu\to e \gamma$ varying the parameter $\mu$ parameter for
  different values of the mixing between the $\Sigma^0$ and $S$ fields, 0.5 (continuous)
  and 0.1 (dashed) and with $M_{}$ is fixed at 1 TeV.  }
\label{fig:3e}
\end{center}
\end{figure}

As in the standard type-III seesaw mechanism~\cite{Foot:1988aq},
universality violation is also present here. However, in contrast to
the standard case, here its amplitude is of the order $$\epsilon^2
\sim (m_D / m_\Sigma)^2,$$ which need not be neutrino mass suppressed.
Indeed, in the inverse type-III seesaw scheme neutrino masses are
proportional to the parameter $\mu$. As a result there are sizeable
\lfv processes such as $\mu\to e\gamma$ and $\mu\to eee$, whose
attainable branching ratios are shown in Fig.~(\ref{fig:3e}). %

Finally, to conclude this discussion, we stress that, in contrast with
the inverse type-I seesaw mechanism, here the neutrino mass messenger
$\Sigma^0$, being an isotriplet member, has gauge interactions. Hence,
if kinematically allowed it will be copiously produced in collider
experiments like the LHC~\cite{Franceschini:2008pz}. 

In short this scheme is a very interesting one both from the point of
view of the detectability of collider signatures at the LHC as well as
\lfv phenomenology.

\begin{table}
\begin{center}
\begin{tabular}{|l|c|c|c|c|}
\hline
Model & Scalars & Fermions & LFV & LHC \\
\hline
\hline
Type-I &   & $({\bf 1},{\bf 1},{ 0})_{+1}$ &  \no  & \no \\
\hline
Type-II & $({\bf 1},{\bf 3},{ 2})_{+2}$  & &   \yes  & \yes \\
\hline
Type-III &  & $({\bf 1},{\bf 3},{ 0})_{+1}$ & \no  & \yes \\
\hline
Inverse &  & $({\bf 1},{\bf 1},{ 0})_{+1}$ &  \yes  & \no \\
\hline
Linear & & $({\bf 1},{\bf 1},{ 0})_{+1}$ &  \yes & \no \\
\hline
Inverse Type-III &  & $({\bf 1},{\bf 3},{ 0})_{+1},\,({\bf 1},{\bf 1},{ 0})_{+1}$ & \yes & \yes \\
\hline
\end{tabular}
\caption{Phenomenological implications of low-scale \SM seesaw
  models together with their particle content. The subscript in the representations is lepton number. ``\no"~would change to ``\yes"~in the presence of new gauge
  bosons or supersymmetry, as explained in the text. }
\end{center}
\label{tab:seesaw}
\end{table}

\subsection{Low scale type-II seesaw}

We now turn to the so-called type-II seesaw
mechanism~\cite{Schechter:1980gr,Cheng:1980qt,Magg:1980ut,Mohapatra:1980yp}
which, though normally assumed to involve new physics at high energy
scales, typically close to the unification scale, may also be
considered (perhaps articially) as a low scale construction, provided
one adopts a tiny value for the trilinear mass parameter $$\mu_T\sim 10^{-8} {\rm GeV},$$ in the scalar potential, then the triplet mass
$M_T$ can be assumed to lie around the TeV scale.
Barring naturalness issues, such a scheme could be a possibility
giving rise to very interesting phenomenological implications. In fact, in
this case, if kinematically allowed, the scalar triplet $\Delta$ will
be copiously produced at the LHC because it interacts with gauge
bosons.

Moreover the couplings $y_\nu$ that mediate \lfv processes are of
order one and therefore such processes are not neutrino mass
suppressed, as in the standard type-I seesaw.
Indeed, from the upper limit ${\rm Br}(\mu\to 3 e) < 10^{-12}$ it
follows that (see Ref.\,\cite{Perez:2008ha})
\begin{equation}
y_\nu^2 < 1.4 \times 10^{-5}\, \left(\frac{m_\Delta}{1\,\rm TeV}\right)\,,
\end{equation}
implying a sizeable triplet Yukawa coupling. With $y_\nu\sim 10^{-2}$,
in order to get adequate neutrino mass values, one needs
\begin{equation}
  \label{eq:triplet-vev}
  v_T\sim 10^{-7}\, {\rm GeV} \,,
\end{equation}
which restricts the scalar triplet vacuum expectation value (vev). For
such small value of the vev, the decay of the $\Delta^{++}$ is mainly
into a pair of leptons with the same charge, while for $v_T>
10^{-4}\, \rm GeV$, the $\Delta^{++}$ decays mainly into a sam-sign $WW$ pair,
see Ref.~\cite{Perez:2008ha}.

Note that the tiny parameter $\mu_T$ controls the neutrino mass scale
but does not enter in the couplings with fermions.  This is why the
\lfv rates can be sizable in this case. For detailed phenomenological
studies of low energy type-II seesaw see, for example,
Ref.~\cite{Perez:2008ha,delAguila:2008cj,Nath:2010zj}

Before reviewing the models based on radiative generation
mechanisms for neutrino masses, we summarize the
phenomenological implications of low scale seesaw models,
together with their particle content, in Tab.~(\ref{tab:seesaw}).
\vskip3.mm
\section{Radiative neutrino masses}

In the previous sections we reviewed mechanisms ascribing the
smallness of neutrino masses to the small coefficient in front of
Weinberg's dimension-five operator. This was generated either through
tree-level exchange of super-heavy messengers, with mass associated to
high-scale symmetry breaking, or conversely, because of symmetry
breaking at low scale.
In what follows we turn to radiatively induced neutrino masses, a
phenomenologically attractive way to account for neutrino masses. In
such scenarios the smallness of the neutrino mass follows from loop
factor(s) suppression. From a purely phenomenological perspective,
radiative models are perhaps quite interesting as they rely on new
particles that typically lie around the TeV scale, hence accessible to
collider searches.

Unlike seesaw models, radiative mechanisms can go beyond the effective
$\Delta L=2$ dimension-five operator in Eq.~(\ref{eq:d5}) and generate
the neutrino masses at higher order. This leads to new operators and
to further mass suppression.
Such an approach has been reviewed in Refs.~\cite{Babu:2001ex,Gouvea:2007xp,Angel:2012ug,Farzan:2012ev,Law:2013dya}.  In what follows we'll survey some representative
underlying models up to the third loop level.

\subsection{One--loop schemes}

A general survey of one-loop neutrino mass operators leading to
neutrino mass has been performed in \cite{Bonnet:2012kz}. Neutrino
mass models in extensions of the SM with singlet right-handed
neutrinos have been systematically analyzed in
\cite{Pilaftsis:1991ug,Dev:2012sg} and for higher representations in
\cite{FileviezPerez:2009ud}.  Here we review the most representative
model realizations.

\subsubsection{Zee Model}

\begin{figure}[!h]
\centering
\includegraphics[width=0.35\textwidth,keepaspectratio=true,clip=true]{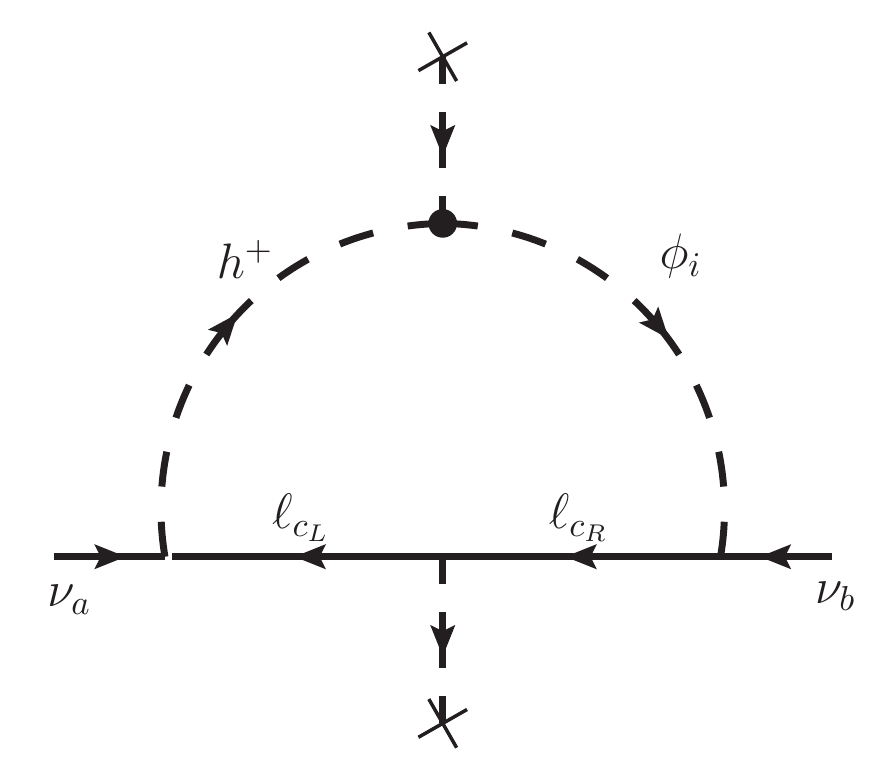}
\caption{Neutrino mass generation in the Zee model.}
\label{fig:zee}
\end{figure}

The Zee Model \cite{Zee:1980ai} extends the standard \SM model with
the following fields
\begin{equation}
  \label{zee-content}
  h^+ \, \sim \, (\oneR,\oneR,+1)_{-2} \quad ,\quad \phi_{1,2} \, \sim\, (\oneR,\twoR,+1/2)_0  \, ,
\end{equation}
where the subscript denotes lepton number. Given this particle content neutrino mass are one-loop calculable. The
relevant terms are given by
\begin{equation}
\label{zee-lag} 
\mathcal{L} = y_i^{ab}\, \overline{L}_a \phi_i \ell_{b R}
+ f^{ab} \tilde{L}_a \,i \tau_2\, L_b\, h^+ - \mu\, \phi^\dagger_1\,i
\tau_2\,\phi_2^* \, h^+ + \mathrm{h.c.}   \,,
\end{equation}
where $a,b$ indicate the flavor indices, i.e.~$a,b=e,\mu,\tau$,
$\tilde{L} \equiv L^T C^{-1}$ and $\tau_2$ is the second Pauli
matrix. Notice that the matrix $f$ must be anti-symmetric in
generation indices. The violation of lepton number, required to
generate a Majorana mass term for neutrinos, resides in the
coexistence of the two Higgs doublets in the $\mu$ term. The one-loop radiative diagram is 
shown in Fig.~(\ref{fig:zee}).
The model has been extensively studied in the literature~
\cite{AristizabalSierra:2006ri,Smirnov:1996bv,Jarlskog:1998uf,Frampton:1999yn,Joshipura:1999xe,McLaughlin:1999un,Cheung:1999az,
  Chang:1999hga,Dicus:2001ph,Balaji:2001ex,Mitsuda:2001vh,Ghosal:2001ep,Koide:2001xy,Brahmachari:2001rn,Kitabayashi:2001it,Koide:2002uk,Cheng:2002hr,
  He:2003rm,Hasegawa:2003by,Kanemura:2005hr,Brahmachari:2006qm,Sahu:2008aw,Fukuyama:2010ff,delAguila:2012nu},
particularly in the Zee-Wolfenstein limit where only $\phi_1$ couples
to leptons due to a
$\mathbb{Z}_2$ symmetry~\cite{Wolfenstein:1980sy}. \\[-.2cm]

This particular simplification forbids tree-level Higgs-mediated flavor-changing neutral currents (FCNC),
although it is now disfavored by neutrino oscillation data
\cite{Koide:2001xy, Frampton:2001eu}.
However the general Zee model is still valid phenomenologically
\cite{Balaji:2001ex} and is in testable with FCNC experiments. For
instance the exchange of the Higgs bosons leads to tree level decays
of the form $\ell_i \to \ell_j\ell_k\bar{\ell}_k$, in particular $\tau
\to \mu\mu\bar{\mu}, \mu e \bar{e}$ (see for instance
\cite{He:2003ih}). Collider phenomenology has been studied in
\cite{Kanemura:2000bq,Assamagan:2002kf}.

Recently, a variant of the Zee model have been considered in
\cite{Babu:2013pma} (see also \cite{Aranda:2011rt}) by imposing a family-dependent $\mathbb{Z}_4$
symmetry acting on the leptons, thereby reducing the number of
effective free parameters to four. The model predicts inverse
hierarchy spectrum in addition to correlations among the mixing
angles.

\subsubsection{Radiative seesaw model}

\begin{figure}[t]
\centering
\includegraphics[width=0.35\textwidth,keepaspectratio=true,clip=true]{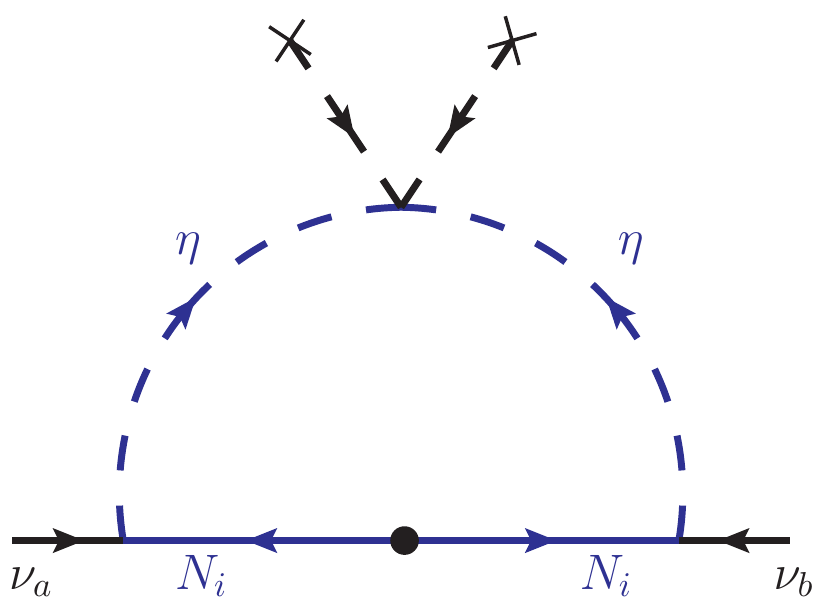}
\caption{Neutrino mass generation in the radiative seesaw model. The blue color represents the potential dark matter candidates.}
\label{fig:ma}
\end{figure}

Another one-loop scenario was suggested by Ma~\cite{Ma:2006km}.
Besides the \sm fields, three right handed Majorana fermions $N_i$
($i=1,2,3$) and a Higgs doublet are added to the \SM model,
\begin{equation}
\label{ma-content} N_i \, \sim \, (\oneR,\oneR,0)_{+1} \quad ,\quad
\eta \, \sim\, (\oneR,\twoR,+1/2)_0 \,.
\end{equation}
In addition, a parity symmetry acting only on the new fields is
postulated. This $\mathbb{Z}_2$ is imposed in order to forbid Dirac
neutrino mass terms. The relevant interactions of this model are given
by
\begin{equation}
\label{malag}
{\mathcal L} = y_{ab}\,\overline{L}_a \, i \tau_2 \eta^* N_b -  \frac{M_{N_{i}}}{2} \tilde{N}_i N_i + \mathrm{h.c.}
\end{equation} 
In the scalar potential a quartic scalar term of the form $(H^\dagger
\eta)^2$ is allowed.  The one-loop radiative diagram is shown in
Fig.~(\ref{fig:ma})
and generates calculable ${\cal M}_\nu$ if $\vev{\eta}=0$, which
follows from the assumed symmetry.
The neutrino masses are given by
\begin{equation}
\label{ma-mat}
({M_\nu})_{a b}=  \sum_i {y_{a i} y_{b i}\, M_{N_i} \over 16 \pi^{2}} \left[ {m_R^{2} \over m_R^{2}-M_{N_i}^{2}} \ln {m_R^{2} 
\over M_{N_i}^{2}} - {m_I^{2} \over m_I^{2}-M_{N_i}^{2}} \ln {m_I^{2} \over 
M_i^{2}} \right],
\end{equation}
where $m_{R}$ ($m_I$) is the mass of the real (imaginary) part of the
neutral component of $\eta$.

Thanks to its simplicity and rich array of predictions, the model has
become very popular and an extensive literature has been devoted to
its phenomenological consequences. As is generally the case with
multi-Higgs \sm extensions, the induced \lfv effects such as $\mu \to
e\gamma$ provide a way to probe the model parameters. In particular
the \lfv phenomenology has been studied in
\cite{Kubo:2006yx,Sierra:2008wj,Suematsu:2009ww,Adulpravitchai:2009gi,Toma:2013zsa,Schmidt:2012yg}. The
effect of corrections induced by renormalization group running have
also been considered \cite{Bouchand:2012dx}, showing that highly
symmetric patterns such as the bimaximal lepton mixing structure can
still be valid at high-energy but modified by the running to correctly
account for the parameters required by the neutrino oscillation
measurements~\cite{Tortola:2012te}. Collider signatures have also been
investigated in \cite{Aoki:2010tf,Aoki:2013lhm,Ho:2013spa,Ho:2013hia}.

A remarkable feature of this model is the natural inclusion of a WIMP
(weakly interacting massive particle) dark matter candidate. Indeed,
the same parity that makes the neutrino mass calculable, also
stabilizes $N_i$ and the neutral component of $\eta$. The lightest
$\mathbb{Z}_2$-odd particle, either a boson or a fermion, can play the
role of WIMP cold dark matter candidate
\cite{Kubo:2006yx,Kajiyama:2006ww, Kajiyama:2006ww,
  Suematsu:2009ww,Suematsu:2010gv,Kajiyama:2011fe,
  Kajiyama:2011fx,Klasen:2013jpa, Schmidt:2012yg}. There is also the
interesting possibility of the dark matter being warm in this setup
\cite{Sierra:2008wj,Hu:2012az}. Various extensions of the model have
also been considered, for e.g. \cite{Hirsch:2013ola,Brdar:2013iea}.
For a review on models with one--loop radiative neutrino masses and
viable dark matter candidates we refer the reader to the complete
classification given in \cite{Restrepo:2013aga,Law:2013saa}.

\subsection{Two--loop schemes}

\begin{figure}[t]
\centering
\includegraphics[width=0.35\textwidth,keepaspectratio=true,clip=true]{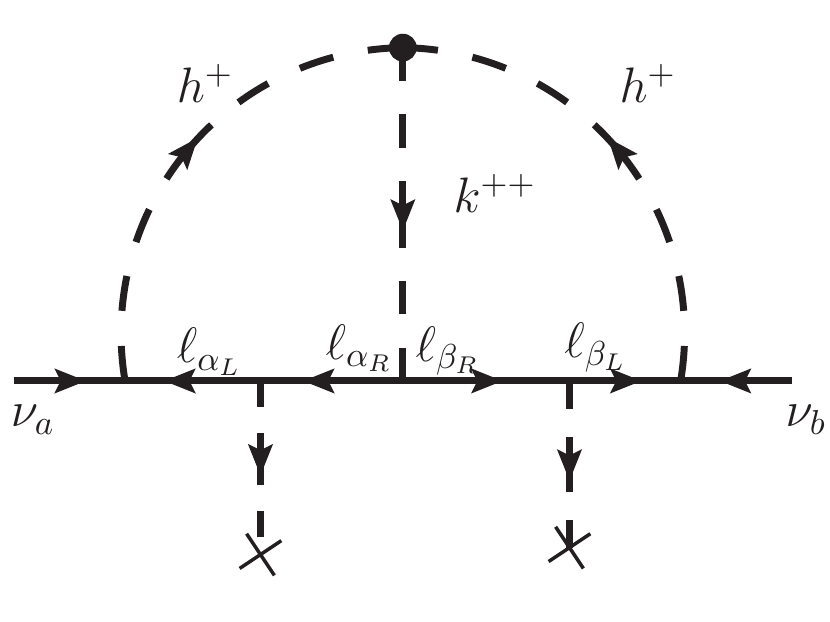}
\caption{Neutrino mass generation in the Zee-Babu model.}
\label{fig:zeebabu}
\end{figure}

As a prototype two--loop scheme we consider the model proposed by
Zee~\cite{Zee:1985id} and Babu~\cite{Babu:1988ki} (which first
appeared in \cite{Cheng:1980qt}), that leads to neutrino masses at
two-loop level by extending the \sm with two complex singly and
doubly~\cite{Schechter:1982bd} charged SU(2)$_L$ singlet scalars
\begin{equation}
\label{zeebabu-content} 
h^{+}\, \sim \, (\oneR,\oneR,+1)_{-2} \quad
,\quad k^{++} \, \sim\, (\oneR,\oneR,+2)_{-2} \,.
\end{equation}
The relevant terms in the Lagrangian are therefore
\begin{equation}
\label{zeebabu-lag}
\mathcal{L} = f_{a b}\, \tilde{L}_a\, i \tau_2\, L_b h^{+} + g_{a b} \tilde{\ell}_{a R}\, \ell_{b R} k^{++} - \mu\, h^- h^- k^{++} + \mathrm{h.c.}
\end{equation}
The trilinear $\mu$ term in the scalar potential~\footnote{This term can arise spontaneously through the vev of an extra gauge singlet scalar
  boson~\cite{peltoniemi:1993pd}.} provides lepton number violation
and leads to a calculable Majorana neutrino mass generated at the second loop order, as shown in Fig.~(\ref{fig:zeebabu}) and given by

\begin{equation}
  \label{zeebabu-mass} (M_\nu)_{a b} \sim \mu\, \frac{1}{(16
    \pi)^2}\frac{1}{M} \frac{16 \pi^2}{3} f_{a c} m_c g^*_{c d} m_d f_{b
    d} \end{equation} 
where $M=\mathrm{max}(M_{k^{++}},M_{h^+})$ and $m_a$ are
charged lepton masses \cite{Nebot:2007bc}. As in the Zee model, the
matrix $f$ is anti-symmetric. Therefore the determinant of $m_\nu$ 
vanishes and, as a result, one of the light neutrinos must be massless.

The Zee--Babu model is constrained by a variety of \lfv processes
among which the tree-level \lfv $\ell_i\to \ell_j\ell_k\bar{\ell}_l$ decays
induced by $k^{++}$ exchange and the radiative decays
$\ell_i\to\ell_j\gamma$ mediated by the charged scalars $h^+$ and
$k^{++}$.  Weak universality is also violated since the $h^+$ exchange
induces new contributions for muon
decay~\cite{Schmidt:2014zoa,Babu:2002uu, Nebot:2007bc,
  AristizabalSierra:2006gb}. Both \lfv and weak universality tests
constrain the model parameters.
Combining \lfv and universality constraints \cite{Schmidt:2014zoa}
pushes the mass of $h^+$ and $k^{++}$ above the TeV scale, for both
inverted and normal hierarchies, making it a challenge to probe the
model at the LHC. The collider phenomenology of the model have been
considered in
\cite{Nebot:2007bc,Schmidt:2014zoa,Herrero-Garcia:2014hfa}.

\subsection{Three--loop schemes}

\begin{figure}[!h]
\centering
\includegraphics[width=0.50\textwidth,keepaspectratio=true,clip=true]{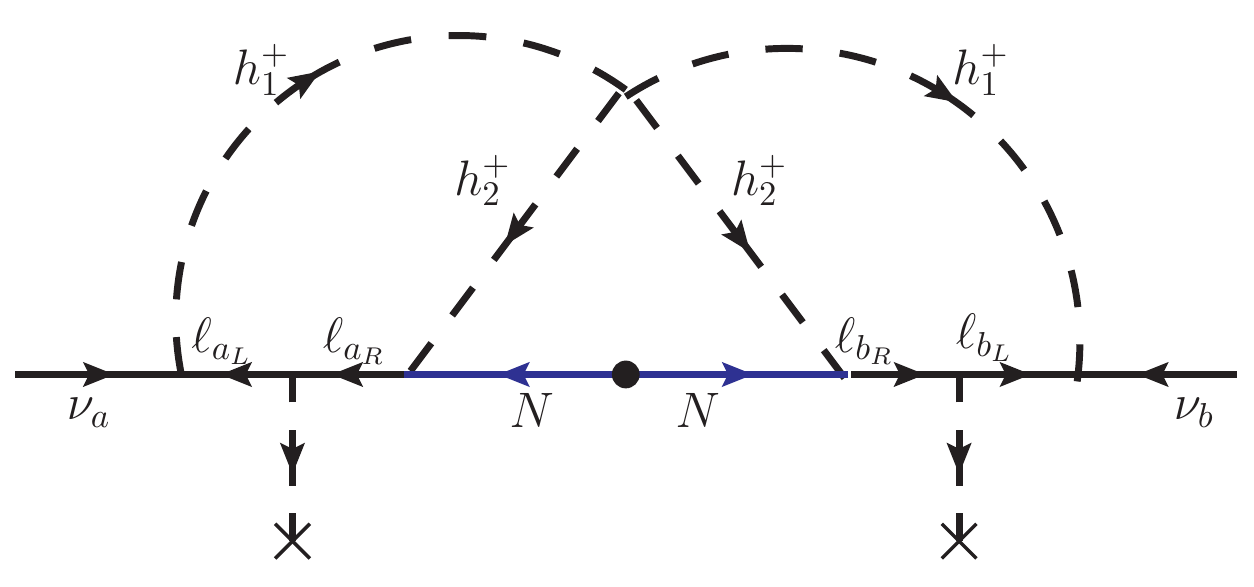}
\caption{Neutrino mass generation in the KNT model.}
\label{fig:nasri}
\end{figure}

Of the possible three--loop schemes we will focus on the one suggested
by Krauss-Nasri-Trodden (KNT)~\cite{Krauss:2002px}. 
These authors considered an extension of the \sm with two charged
scalar singlets $h_1$ and $h_2$ and one right handed neutrino $N$.
\begin{equation}
\label{nasri-content}
h_{1,2}^{+}\, \sim \, (\oneR,\oneR,+1)_{-2} \quad ,\quad N \, \sim\, (\oneR,\oneR,0)_{+1}\,.
\end{equation}
As usual in radiative neutrino mass models that include gauge singlet
Majorana fermions, an additional $\mathbb{Z}_2$ symmetry is imposed,
under which the \sm fields as well as $h_1$ transform trivially, while $N$ and $h_2$ are odd.  The most general
renormalizable terms that may be added to the \sm fermion Lagrangian
are
\begin{equation}
\label{nasrilag}
\mathcal L = f_{a b}  \tilde{L}_{a}\, i\tau_2 L_{b}\, h_1^{+} +g_{a} N h_2^{+} \ell_{{a}_R} - \frac{M_N}{2} \tilde{N} N + {\rm h.c.} 
\end{equation}
Note that the scalar potential contains a term of the form $(h_1
h_2^{*})^2$, which makes possible the diagram of Fig.~(\ref{fig:nasri}) possible.
Hence neutrinos acquire Majorana masses induced only at the 3-loop
level. Such strong suppression allows for sizable couplings of the
TeV scale singlet messenger states.

In addition to neutrino masses, the model also includes a WIMP dark
matter candidate. Indeed for the choice of parameters $M_{h_2} > M_N$,
$N$ is stable and can be thermally produced in the early universe,
leading naturally to the correct dark matter abundance.

A very similar model with the same loop topology has been proposed in
\cite{Ng:2013xja}, replacing the neutral gauge singlets by new colored
fields and the charged leptons by quarks and in \cite{Ahriche:2014cda}
the triplet variant of the model has been introduced. These variations
makes the model potentially testable at hadron colliders. Other three
loop mass models have also been considered more recently, for instance
in ~\cite{Aoki:2008av,Aoki:2009vf,Gustafsson:2012vj,Ahriche:2014cda}.
A systematic study generalizing the KNT model was presented in
\cite{Chen:2014ska}.
\begin{table}[h!]
\label{tab:rad}
\begin{center}
\begin{tabular}{|l|c|c|c|c|c|c|}
\hline
& Model & Scalars & Fermions & LFV & DM & LHC\\
\hline
\hline
1--Loop \, & Zee  &  $(\oneR,\oneR,+1)_{-2} \, ,\, (\oneR,\twoR,-1/2)_0$ & &  \yes & \no & \yes \\
 	& Ma  &   $(\oneR,\twoR,+1/2)_0$ & $(\oneR,\oneR,0)_{+1}$  &  \yes & \yes & \yes\\
\hline
2--Loops \, & Zee-Babu  & $(\oneR,\oneR,+1)_{-2}\, , \, (\oneR,\oneR,+2)_{-2}$ & & \yes & \no &\yes\\
\hline
3--Loops \, & KNT  & $(\oneR,\oneR,+1)_{-2}$ & $(\oneR,\oneR,0)_{+1}$ & \yes & \yes & \no\\
\hline
\end{tabular}
\caption{Phenomenological implications of radiative \SM neutrino mass
  models discussed in this review. Representations are labelled as in
  the rest of the paper. }
\end{center}
\end{table}
\vskip3.mm

We summarize the models discussed in this section and
their phenomenological implications in Tab.~(\ref{tab:rad}).

\section{Supersymmetry as the origin of neutrino masses}

The standard formulation of supersymmetry assumes the conservation of
a discrete symmetry called R--parity ($R_p$), under which all the
standard model states are R-even, while their superpartners are
R-odd~\cite{martin:1997ns}.  $R_p$ is related to the spin (S), total
lepton (L), and baryon (B) number as
$$R_p=(-1)^{(3B+L+2S)}.$$
Hence requiring baryon and lepton number conservation implies $R_p$
conservation. In this case the supersymmetric states must be produced
in pairs, while the lightest of them is absolutely stable.

On general grounds, however, neither gauge invariance nor
supersymmetry require $R_p$ conservation and many implications can be
associated to R-parity violation~\cite{barbier:2004ez}. The most
general supersymmetric \sm extension contains explicit $R_p$ violating
interactions.  Constraints on the relevant parameters and their
possible signals have been analysed~\cite{barger:1989rk,Baer:2006rs}.
In general, there are too many independent couplings, some of which
must be set to zero in order to avoid too fast the proton decay.
For these reasons we focus our attention to the possibility that $R_p$
can be an exact symmetry of the Lagrangian, broken spontaneously
through the Higgs mechanism~\cite{Nogueira:1990wz,masiero:1990uj}.
This may occur via nonzero vacuum expectation values for scalar
neutrinos, such as
\begin{equation}
v_R = \vev {\tilde{\nu}_{R\tau}}\qquad ;
\qquad v_L = \vev {\tilde{\nu}_{L\tau}}\ .
\end{equation}

Here we consider the simplest prototype scheme where supersymmetry
seeds neutrino masses in an essential way. The idea is to take the
simplest effective description of the above picture, namely, bilinear
R-parity violation~\cite{Hall:1983id,diaz:1997xc,romao:1997xf}.
This is the minimal way to incorporate lepton number and R-parity
violation to the Minimal Supersymmetric Standard Model (MSSM),
providing a simple way to accommodate neutrino masses in
supersymmetry. The superpotential is
\begin{equation}\label{brpv-superpotential}
W=W^{MSSM} + \epsilon_a \widehat{L}_a \widehat{H}_u .
\end{equation}
The three $\epsilon_a = (\epsilon_e,\epsilon_\mu,\epsilon_\tau)$
parameters have dimensions of mass and explicitly break lepton number
by $\Delta L=1$. Their size and origin can be naturally explained in extended
models where the breaking of lepton number is
spontaneous~\cite{masiero:1990uj,Nogueira:1990wz,romao:1997xf}.
These parameters are constrained to be small ($\epsilon_a \ll m_W$) so
as to account for the small neutrino masses. Furthermore, the presence
of the new superpotential terms implies new soft supersymmetry
breaking terms as well
\begin{equation}\label{brpv-soft}
V_{soft}^{
} = V^{MSSM}_{soft} + B_a {\epsilon_a} \tilde{L}_a {H}_u \,,
\end{equation}
where the $B_a$ are parameters with units of mass.

In this scheme, neutrinos get tree-level mass by mixing with the
neutralino sector~\cite{Hirsch:2000ef,Diaz:2003as,Chun:1999bq}. In the
basis $(\psi^0)^T=
(-i\tilde{B}^0,-i\tilde{W}_3^0,\widetilde{H}_d^0,\widetilde{H}_u^0,
\nu_{e}, \nu_{\mu}, \nu_{\tau} )$ the neutral fermion mass matrix
$M_N$ this matrix is given by
\begin{equation} \label{brpvmat}
M_{N\chi}=\left(
\begin{array}{cc}
{\cal M}_{\chi^0}& m^T \cr 
m & 0 \cr
\end{array}
\right),
\end{equation}
where ${\cal M}_{\chi^0}$ is the usual neutralino mass matrix and
\begin{equation}
m=\left(
\begin{array}{cccc}
-\frac 12g^{\prime }v_{L_e} & \frac 12g v_{L_e} & 0 & \epsilon_e \cr 
-\frac 12g^{\prime }v_{L_\mu} & \frac 12g v_{L_\mu} & 0 & \epsilon_\mu  \cr
 -\frac 12g^{\prime }v_{L_\tau} & \frac 12g v_{L_\tau} & 0 & \epsilon_\tau
\cr
\end{array}
\right),
\end{equation}
is the matrix describing R-parity violation. Here $v_{L_a}$ are the
vevs of sneutrinos induced by the presence of $\epsilon_i$ and
$B_i$. The smallness of the R-parity violating parameters implies that
the components of $m$ are suppressed with respect to those in ${\cal
  M}_{\chi^0}$. Hence the resulting $M_N$ matrix has a type-I seesaw
structure so the effective light neutrino mass matrix can be obtained
from the usual formula $m_\nu^{0} = - m \cdot {\cal M}_{\chi^0}^{-1}
\cdot m^T$, which can be expanded to give
\begin{equation} \label{meff-lambda-brpv}
\left( M_\nu\right)_{ab} = \alpha \Lambda_a \Lambda_b ,
\end{equation}
where $\alpha$ is a combination of SUSY parameters, while $\Lambda_a =
\mu v_{L_a} + v_d \epsilon_a$ are known as the \emph{alignment
  parameters}.  The above matrix is projective and has two zero
eigenvalues, therefore only one neutrino is massive at tree level. A
natural choice is to ascribe this eigenvalue to the atmospheric scale
whereas the solar mass scale, $\Delta m_{sol}^2 \ll \Delta m_{atm}^2$,
arises from quantum corrections calculable at the one-loop level of
the neutrino mass matrix in Eq.~(\ref{meff-lambda-brpv}).
Detailed computations of the one-loop contributions to the neutrino
mass matrix are given in Refs.~\cite{Hirsch:2000ef,Diaz:2003as}. The
corrections are of the type
\begin{equation}\label{1loop}
\left( m_\nu^{\rm rad} \right)_{ab} \approx  \alpha^{(\rm rad)} \Lambda_a \Lambda_b + \beta^{(rad)} (\Lambda_a \epsilon_b + \Lambda_a \epsilon_b) + 
\gamma^{(\rm rad)} \epsilon_a \epsilon_b ,
\end{equation}
where the coefficients $\alpha^{(\rm rad)}, \beta^{(\rm rad)},
\gamma^{(\rm rad)}$ are complicated functions of the SUSY
parameters. These corrections generate a second non-zero mass
eigenstate associated with the solar scale, and the corresponding
mixing angle~\footnote{The neutrino mixing angles are determined as
  ratios of R-parity violating parameters $\epsilon_i$ and
  $\Lambda_i$.}  $\theta_{12}$.

The bilinear R-parity breaking model offers a hybrid mechanism
combining seesaw-type and radiative contributions, thereby providing
an explanation for the observed smallness of the solar squared mass
splitting with respect to the atmospheric one.

The above scheme is both well motivated and testable at colliders.
Indeed in the absence of R-parity, the lightest supersymmetric
particle (LSP) is no longer protected and decays to \sm particles. The
smallness of the breaking strength, required to account for neutrino
masses, makes the lifetime of the LSP long enough so that it may decay
within the detector with displaced vertices.
Since LSP decays and neutrino masses have a common origin, one can
show that ratios of LSP decay branching ratios correlate with the
neutrino mixing angles measured at low energies
\cite{Mukhopadhyaya:1998xj}.  
This provides a remarkable connection which allows one to use neutrino
oscillation data to test the model at the LHC
see~e.~g.~\cite{deCampos:2012pf,DeCampos:2010yu}.

\section{Summary and outlook}

We have given a brief overview of the low-scale \SM approach to
neutrino mass generation. To chart out directions within such a broad
neutrino landscape we used their possible phenomenological potential
as a guide.
We analyzed signatures associated to direct neutrino mass messenger
production at the LHC, as well as messenger-induced \lfv processes.
We have considered seesaw-based schemes as well as those with
radiative or supersymmetric origin for the neutrino mass.
We summarize our conclusions in table\,\ref{tab1}.
We stressed the phenomenological interest on radiative models, low
scale seesaw schemes as well as the type-II seesaw "tuned" to lie at
the low scale.
We also briefly comment on the presence of WIMP cold dark matter
candidates.


\begin{table}[h!]
\begin{center}
\begin{tabular}{|l|c|c|c|c|c|c|c|}
\hline
     & \,\,\, Type-I \,\,\, &   \,\,  Type-II  \,\, &  \,\,   Type-III \,\,   &  \,\, Inverse \,\,   &  \,\, Linear \,\,  & \, Invers Type-III \, &
\, Radiative \,  \\
\hline
LHC  \,\, & \no& \yes & \yes & \no& \no& \yes& \yes \\
\hline
LFV \,\, & \no & \yes& \no &\yes &  \yes & \yes& \yes\\
\hline
\end{tabular}
\caption{Neutrino mass models in terms of their phenomenological
potential at the LHC and/or the sizable presence of \lfv phenomena  
where we use the same labeling convention as in the text. As we have
explained in the text, ``\no" could change to ``\yes" in the presence of 
new gauge bosons or supersymmetry.}
\label{tab1}
\end{center}
\end{table}

\noindent
In conclusion if the messengers responsible for the light neutrino
masses lie at a very high scale, like in type-I seesaw, it will be very
difficult if not impossible to have any detectable signal within the
non-supersymmetric \SM seesaw framework.
In contrast, within the low scale approach to neutrino mass we can
have very interesting phenomenological implications. They can give
rise to signatures at high energy collider experiments, as well as
\lfv rates close to the sensitivity of planned experiments.
In some of the schemes there is a natural WIMP dark matter candidate.
In short, these scenarios may help reconstructing the neutrino mass
from a variety of potentially over-constrained set of observables.


\section*{Acknowledgments}
\noindent Work supported by the Spanish MINECO under grants
FPA2011-22975 and MULTIDARK CSD2009-00064 (Consolider-Ingenio 2010
Programme).  S.M. thanks the DFG grants WI 2639/3-1 and WI 2639/4-1
for financial support.

\bibliography{merged,extraref}

\begin{thebibliography}{159}%
\makeatletter
\providecommand \@ifxundefined [1]{%
 \@ifx{#1\undefined}
}%
\providecommand \@ifnum [1]{%
 \ifnum #1\expandafter \@firstoftwo
 \else \expandafter \@secondoftwo
 \fi
}%
\providecommand \@ifx [1]{%
 \ifx #1\expandafter \@firstoftwo
 \else \expandafter \@secondoftwo
 \fi
}%
\providecommand \natexlab [1]{#1}%
\providecommand \enquote  [1]{``#1''}%
\providecommand \bibnamefont  [1]{#1}%
\providecommand \bibfnamefont [1]{#1}%
\providecommand \citenamefont [1]{#1}%
\providecommand \href@noop [0]{\@secondoftwo}%
\providecommand \href [0]{\begingroup \@sanitize@url \@href}%
\providecommand \@href[1]{\@@startlink{#1}\@@href}%
\providecommand \@@href[1]{\endgroup#1\@@endlink}%
\providecommand \@sanitize@url [0]{\catcode `\\12\catcode `\$12\catcode
  `\&12\catcode `\#12\catcode `\^12\catcode `\_12\catcode `\%12\relax}%
\providecommand \@@startlink[1]{}%
\providecommand \@@endlink[0]{}%
\providecommand \url  [0]{\begingroup\@sanitize@url \@url }%
\providecommand \@url [1]{\endgroup\@href {#1}{\urlprefix }}%
\providecommand \urlprefix  [0]{URL }%
\providecommand \Eprint [0]{\href }%
\providecommand \doibase [0]{http://dx.doi.org/}%
\providecommand \selectlanguage [0]{\@gobble}%
\providecommand \bibinfo  [0]{\@secondoftwo}%
\providecommand \bibfield  [0]{\@secondoftwo}%
\providecommand \translation [1]{[#1]}%
\providecommand \BibitemOpen [0]{}%
\providecommand \bibitemStop [0]{}%
\providecommand \bibitemNoStop [0]{.\EOS\space}%
\providecommand \EOS [0]{\spacefactor3000\relax}%
\providecommand \BibitemShut  [1]{\csname bibitem#1\endcsname}%
\let\auto@bib@innerbib\@empty
\bibitem [{1234547()}]{ATLASandCMSCollaborations:2013pga}%
  \BibitemOpen
  \bibfield  {author} {1234547,\ }\href@noop {} {\bibfield  {journal} {\bibinfo
   {journal} {CERN Cour.}\ }\textbf {\bibinfo {volume} {53}},\ \bibinfo {pages}
  {21} (\bibinfo {year} {2013})}\BibitemShut {NoStop}%
\bibitem [{\citenamefont {Aad}\ \emph {et~al.}(2012)\citenamefont {Aad} \emph
  {et~al.}}]{Aad:2012tfa}%
  \BibitemOpen
  \bibfield  {author} {\bibinfo {author} {\bibfnamefont {G.}~\bibnamefont
  {Aad}} \emph {et~al.} (\bibinfo {collaboration} {ATLAS Collaboration}),\
  }\href@noop {} {\bibfield  {journal} {\bibinfo  {journal} {Phys.Lett.}\
  }\textbf {\bibinfo {volume} {B716}},\ \bibinfo {pages} {1} (\bibinfo {year}
  {2012})},\ \Eprint {http://arxiv.org/abs/1207.7214} {arXiv:1207.7214
  [hep-ex]} \BibitemShut {NoStop}%
\bibitem [{\citenamefont {Chatrchyan}\ \emph {et~al.}(2012)\citenamefont
  {Chatrchyan} \emph {et~al.}}]{Chatrchyan:2012ufa}%
  \BibitemOpen
  \bibfield  {author} {\bibinfo {author} {\bibfnamefont {S.}~\bibnamefont
  {Chatrchyan}} \emph {et~al.} (\bibinfo {collaboration} {CMS Collaboration}),\
  }\href@noop {} {\bibfield  {journal} {\bibinfo  {journal} {Phys.Lett.}\
  }\textbf {\bibinfo {volume} {B716}},\ \bibinfo {pages} {30} (\bibinfo {year}
  {2012})},\ \Eprint {http://arxiv.org/abs/1207.7235} {arXiv:1207.7235
  [hep-ex]} \BibitemShut {NoStop}%
\bibitem [{\citenamefont {Weinberg}(1979)}]{Weinberg:1979sa}%
  \BibitemOpen
  \bibfield  {author} {\bibinfo {author} {\bibfnamefont {S.}~\bibnamefont
  {Weinberg}},\ }\href@noop {} {\bibfield  {journal} {\bibinfo  {journal}
  {Phys. Rev. Lett.}\ }\textbf {\bibinfo {volume} {43}},\ \bibinfo {pages}
  {1566} (\bibinfo {year} {1979})}\BibitemShut {NoStop}%
\bibitem [{\citenamefont {Bonnet}\ \emph {et~al.}(2013)\citenamefont {Bonnet},
  \citenamefont {Hirsch}, \citenamefont {Ota},\ and\ \citenamefont
  {Winter}}]{Bonnet:2012kh}%
  \BibitemOpen
  \bibfield  {author} {\bibinfo {author} {\bibfnamefont {F.}~\bibnamefont
  {Bonnet}}, \bibinfo {author} {\bibfnamefont {M.}~\bibnamefont {Hirsch}},
  \bibinfo {author} {\bibfnamefont {T.}~\bibnamefont {Ota}}, \ and\ \bibinfo
  {author} {\bibfnamefont {W.}~\bibnamefont {Winter}},\ }\href@noop {}
  {\bibfield  {journal} {\bibinfo  {journal} {JHEP}\ }\textbf {\bibinfo
  {volume} {1303}},\ \bibinfo {pages} {055} (\bibinfo {year} {2013})},\ \Eprint
  {http://arxiv.org/abs/1212.3045} {arXiv:1212.3045 [hep-ph]} \BibitemShut
  {NoStop}%
\bibitem [{\citenamefont {Bonnet}\ \emph {et~al.}(2012)\citenamefont {Bonnet},
  \citenamefont {Hirsch}, \citenamefont {Ota},\ and\ \citenamefont
  {Winter}}]{Bonnet:2012kz}%
  \BibitemOpen
  \bibfield  {author} {\bibinfo {author} {\bibfnamefont {F.}~\bibnamefont
  {Bonnet}}, \bibinfo {author} {\bibfnamefont {M.}~\bibnamefont {Hirsch}},
  \bibinfo {author} {\bibfnamefont {T.}~\bibnamefont {Ota}}, \ and\ \bibinfo
  {author} {\bibfnamefont {W.}~\bibnamefont {Winter}},\ }\href {\doibase
  10.1007/JHEP07(2012)153} {\bibfield  {journal} {\bibinfo  {journal} {JHEP}\
  }\textbf {\bibinfo {volume} {1207}},\ \bibinfo {pages} {153} (\bibinfo {year}
  {2012})},\ \Eprint {http://arxiv.org/abs/1204.5862} {arXiv:1204.5862
  [hep-ph]} \BibitemShut {NoStop}%
\bibitem [{\citenamefont {de~Gouvea}\ and\ \citenamefont
  {Jenkins}(2008)}]{Gouvea:2007xp}%
  \BibitemOpen
  \bibfield  {author} {\bibinfo {author} {\bibfnamefont {A.}~\bibnamefont
  {de~Gouvea}}\ and\ \bibinfo {author} {\bibfnamefont {J.}~\bibnamefont
  {Jenkins}},\ }\href {\doibase 10.1103/PhysRevD.77.013008} {\bibfield
  {journal} {\bibinfo  {journal} {Phys.Rev.}\ }\textbf {\bibinfo {volume}
  {D77}},\ \bibinfo {pages} {013008} (\bibinfo {year} {2008})},\ \Eprint
  {http://arxiv.org/abs/0708.1344} {arXiv:0708.1344 [hep-ph]} \BibitemShut
  {NoStop}%
\bibitem [{\citenamefont {Bonnet}\ \emph {et~al.}(2009)\citenamefont {Bonnet},
  \citenamefont {Hernandez}, \citenamefont {Ota},\ and\ \citenamefont
  {Winter}}]{Bonnet:2009ej}%
  \BibitemOpen
  \bibfield  {author} {\bibinfo {author} {\bibfnamefont {F.}~\bibnamefont
  {Bonnet}}, \bibinfo {author} {\bibfnamefont {D.}~\bibnamefont {Hernandez}},
  \bibinfo {author} {\bibfnamefont {T.}~\bibnamefont {Ota}}, \ and\ \bibinfo
  {author} {\bibfnamefont {W.}~\bibnamefont {Winter}},\ }\href@noop {}
  {\bibfield  {journal} {\bibinfo  {journal} {JHEP}\ }\textbf {\bibinfo
  {volume} {10}},\ \bibinfo {pages} {076} (\bibinfo {year} {2009})},\ \Eprint
  {http://arxiv.org/abs/0907.3143} {arXiv:0907.3143 [hep-ph]} \BibitemShut
  {NoStop}%
\bibitem [{\citenamefont {Krauss}\ \emph {et~al.}(2013)\citenamefont {Krauss},
  \citenamefont {Meloni}, \citenamefont {Porod},\ and\ \citenamefont
  {Winter}}]{Krauss:2013gy}%
  \BibitemOpen
  \bibfield  {author} {\bibinfo {author} {\bibfnamefont {M.~B.}\ \bibnamefont
  {Krauss}}, \bibinfo {author} {\bibfnamefont {D.}~\bibnamefont {Meloni}},
  \bibinfo {author} {\bibfnamefont {W.}~\bibnamefont {Porod}}, \ and\ \bibinfo
  {author} {\bibfnamefont {W.}~\bibnamefont {Winter}},\ }\href {\doibase
  10.1007/JHEP05(2013)121} {\bibfield  {journal} {\bibinfo  {journal} {JHEP}\
  }\textbf {\bibinfo {volume} {1305}},\ \bibinfo {pages} {121} (\bibinfo {year}
  {2013})},\ \Eprint {http://arxiv.org/abs/1301.4221} {arXiv:1301.4221
  [hep-ph]} \BibitemShut {NoStop}%
\bibitem [{\citenamefont {Nunokawa}\ \emph {et~al.}(2008)\citenamefont
  {Nunokawa}, \citenamefont {Parke},\ and\ \citenamefont
  {Valle}}]{Nunokawa:2007qh}%
  \BibitemOpen
  \bibfield  {author} {\bibinfo {author} {\bibfnamefont {H.}~\bibnamefont
  {Nunokawa}}, \bibinfo {author} {\bibfnamefont {S.~J.}\ \bibnamefont {Parke}},
  \ and\ \bibinfo {author} {\bibfnamefont {J.~W.~F.}\ \bibnamefont {Valle}},\
  }\href@noop {} {\bibfield  {journal} {\bibinfo  {journal} {Prog. Part. Nucl.
  Phys.}\ }\textbf {\bibinfo {volume} {60}},\ \bibinfo {pages} {338} (\bibinfo
  {year} {2008})}\BibitemShut {NoStop}%
\bibitem [{\citenamefont {Forero}\ \emph {et~al.}(2012)\citenamefont {Forero},
  \citenamefont {Tortola},\ and\ \citenamefont {Valle}}]{Tortola:2012te}%
  \BibitemOpen
  \bibfield  {author} {\bibinfo {author} {\bibfnamefont {D.}~\bibnamefont
  {Forero}}, \bibinfo {author} {\bibfnamefont {M.}~\bibnamefont {Tortola}}, \
  and\ \bibinfo {author} {\bibfnamefont {J.~W.~F.}\ \bibnamefont {Valle}},\
  }\href@noop {} {\bibfield  {journal} {\bibinfo  {journal} {Phys.Rev.}\
  }\textbf {\bibinfo {volume} {D86}},\ \bibinfo {pages} {073012} (\bibinfo
  {year} {2012})},\ \Eprint {http://arxiv.org/abs/arXiv:1205.4018}
  {arXiv:arXiv:1205.4018 [hep-ph]} \BibitemShut {NoStop}%
\bibitem [{\citenamefont {Osipowicz}\ \emph {et~al.}(2001)\citenamefont
  {Osipowicz} \emph {et~al.}}]{Osipowicz:2001sq}%
  \BibitemOpen
  \bibfield  {author} {\bibinfo {author} {\bibfnamefont {A.}~\bibnamefont
  {Osipowicz}} \emph {et~al.} (\bibinfo {collaboration} {KATRIN
  collaboration}),\ }\href@noop {} {\  (\bibinfo {year} {2001})},\ \Eprint
  {http://arxiv.org/abs/hep-ex/0109033} {hep-ex/0109033} \BibitemShut {NoStop}%
\bibitem [{\citenamefont {Lesgourgues}\ \emph {et~al.}(2013)\citenamefont
  {Lesgourgues}, \citenamefont {Mangano}, \citenamefont {Miele},\ and\
  \citenamefont {Pastor}}]{pastor-book}%
  \BibitemOpen
  \bibfield  {author} {\bibinfo {author} {\bibfnamefont {J.}~\bibnamefont
  {Lesgourgues}}, \bibinfo {author} {\bibfnamefont {G.}~\bibnamefont
  {Mangano}}, \bibinfo {author} {\bibfnamefont {G.}~\bibnamefont {Miele}}, \
  and\ \bibinfo {author} {\bibfnamefont {S.}~\bibnamefont {Pastor}},\
  }\href@noop {} {\emph {\bibinfo {title} {{Neutrino Cosmology}}}}\ (\bibinfo
  {publisher} {Cambridge Univ Pr},\ \bibinfo {year} {2013})\BibitemShut
  {NoStop}%
\bibitem [{\citenamefont {Barabash}(2011)}]{Barabash:2011fn}%
  \BibitemOpen
  \bibfield  {author} {\bibinfo {author} {\bibfnamefont {A.}~\bibnamefont
  {Barabash}},\ }\href@noop {} {\  (\bibinfo {year} {2011})},\ \bibinfo {note}
  {{75 years of double beta decay: yesterday, today and tomorrow}},\ \Eprint
  {http://arxiv.org/abs/1101.4502} {arXiv:1101.4502 [nucl-ex]} \BibitemShut
  {NoStop}%
\bibitem [{\citenamefont {Ma}(1998)}]{Ma:1998dn}%
  \BibitemOpen
  \bibfield  {author} {\bibinfo {author} {\bibfnamefont {E.}~\bibnamefont
  {Ma}},\ }\href@noop {} {\bibfield  {journal} {\bibinfo  {journal} {Phys. Rev.
  Lett.}\ }\textbf {\bibinfo {volume} {81}},\ \bibinfo {pages} {1171} (\bibinfo
  {year} {1998})},\ \Eprint {http://arxiv.org/abs/hep-ph/9805219}
  {arXiv:hep-ph/9805219} \BibitemShut {NoStop}%
\bibitem [{\citenamefont {Minkowski}(1977)}]{Minkowski:1977sc}%
  \BibitemOpen
  \bibfield  {author} {\bibinfo {author} {\bibfnamefont {P.}~\bibnamefont
  {Minkowski}},\ }\href@noop {} {\bibfield  {journal} {\bibinfo  {journal}
  {Phys. Lett.}\ }\textbf {\bibinfo {volume} {B67}},\ \bibinfo {pages} {421}
  (\bibinfo {year} {1977})}\BibitemShut {NoStop}%
\bibitem [{\citenamefont {Yanagida}(1979)}]{Yanagida:1979as}%
  \BibitemOpen
  \bibfield  {author} {\bibinfo {author} {\bibfnamefont {T.}~\bibnamefont
  {Yanagida}},\ }\href@noop {} {\bibfield  {journal} {\bibinfo  {journal}
  {Conf.Proc.}\ }\textbf {\bibinfo {volume} {C7902131}},\ \bibinfo {pages} {95}
  (\bibinfo {year} {1979})}\BibitemShut {NoStop}%
\bibitem [{\citenamefont {Gell-Mann}\ \emph {et~al.}(1979)\citenamefont
  {Gell-Mann}, \citenamefont {Ramond},\ and\ \citenamefont
  {Slansky}}]{GellMann:1980vs}%
  \BibitemOpen
  \bibfield  {author} {\bibinfo {author} {\bibfnamefont {M.}~\bibnamefont
  {Gell-Mann}}, \bibinfo {author} {\bibfnamefont {P.}~\bibnamefont {Ramond}}, \
  and\ \bibinfo {author} {\bibfnamefont {R.}~\bibnamefont {Slansky}},\
  }\href@noop {} {\bibfield  {journal} {\bibinfo  {journal} {Conf.Proc.}\
  }\textbf {\bibinfo {volume} {C790927}},\ \bibinfo {pages} {315} (\bibinfo
  {year} {1979})},\ \Eprint {http://arxiv.org/abs/1306.4669} {arXiv:1306.4669
  [hep-th]} \BibitemShut {NoStop}%
\bibitem [{\citenamefont {Schechter}\ and\ \citenamefont
  {Valle}(1980)}]{Schechter:1980gr}%
  \BibitemOpen
  \bibfield  {author} {\bibinfo {author} {\bibfnamefont {J.}~\bibnamefont
  {Schechter}}\ and\ \bibinfo {author} {\bibfnamefont {J.~W.~F.}\ \bibnamefont
  {Valle}},\ }\href@noop {} {\bibfield  {journal} {\bibinfo  {journal} {Phys.
  Rev.}\ }\textbf {\bibinfo {volume} {D22}},\ \bibinfo {pages} {2227} (\bibinfo
  {year} {1980})}\BibitemShut {NoStop}%
\bibitem [{\citenamefont {Mohapatra}\ and\ \citenamefont
  {Senjanovic}(1980)}]{Mohapatra:1979ia}%
  \BibitemOpen
  \bibfield  {author} {\bibinfo {author} {\bibfnamefont {R.~N.}\ \bibnamefont
  {Mohapatra}}\ and\ \bibinfo {author} {\bibfnamefont {G.}~\bibnamefont
  {Senjanovic}},\ }\href {\doibase 10.1103/PhysRevLett.44.912} {\bibfield
  {journal} {\bibinfo  {journal} {Phys.Rev.Lett.}\ }\textbf {\bibinfo {volume}
  {44}},\ \bibinfo {pages} {912} (\bibinfo {year} {1980})}\BibitemShut
  {NoStop}%
\bibitem [{\citenamefont {Schechter}\ and\ \citenamefont
  {Valle}(1982{\natexlab{a}})}]{Schechter:1981cv}%
  \BibitemOpen
  \bibfield  {author} {\bibinfo {author} {\bibfnamefont {J.}~\bibnamefont
  {Schechter}}\ and\ \bibinfo {author} {\bibfnamefont {J.~W.~F.}\ \bibnamefont
  {Valle}},\ }\href@noop {} {\bibfield  {journal} {\bibinfo  {journal} {Phys.
  Rev.}\ }\textbf {\bibinfo {volume} {D25}},\ \bibinfo {pages} {774} (\bibinfo
  {year} {1982}{\natexlab{a}})}\BibitemShut {NoStop}%
\bibitem [{\citenamefont {Cheng}\ and\ \citenamefont
  {Li}(1980)}]{Cheng:1980qt}%
  \BibitemOpen
  \bibfield  {author} {\bibinfo {author} {\bibfnamefont {T.~P.}\ \bibnamefont
  {Cheng}}\ and\ \bibinfo {author} {\bibfnamefont {L.-F.}\ \bibnamefont {Li}},\
  }\href@noop {} {\bibfield  {journal} {\bibinfo  {journal} {Phys. Rev.}\
  }\textbf {\bibinfo {volume} {D22}},\ \bibinfo {pages} {2860} (\bibinfo {year}
  {1980})}\BibitemShut {NoStop}%
\bibitem [{\citenamefont {Magg}\ and\ \citenamefont
  {Wetterich}(1980)}]{Magg:1980ut}%
  \BibitemOpen
  \bibfield  {author} {\bibinfo {author} {\bibfnamefont {M.}~\bibnamefont
  {Magg}}\ and\ \bibinfo {author} {\bibfnamefont {C.}~\bibnamefont
  {Wetterich}},\ }\href {\doibase 10.1016/0370-2693(80)90825-4} {\bibfield
  {journal} {\bibinfo  {journal} {Phys.Lett.}\ }\textbf {\bibinfo {volume}
  {B94}},\ \bibinfo {pages} {61} (\bibinfo {year} {1980})}\BibitemShut
  {NoStop}%
\bibitem [{\citenamefont {Wetterich}(1981)}]{Wetterich:1981bx}%
  \BibitemOpen
  \bibfield  {author} {\bibinfo {author} {\bibfnamefont {C.}~\bibnamefont
  {Wetterich}},\ }\href {\doibase 10.1016/0550-3213(81)90279-0} {\bibfield
  {journal} {\bibinfo  {journal} {Nucl.Phys.}\ }\textbf {\bibinfo {volume}
  {B187}},\ \bibinfo {pages} {343} (\bibinfo {year} {1981})}\BibitemShut
  {NoStop}%
\bibitem [{\citenamefont {Mohapatra}\ and\ \citenamefont
  {Senjanovic}(1981)}]{Mohapatra:1980yp}%
  \BibitemOpen
  \bibfield  {author} {\bibinfo {author} {\bibfnamefont {R.~N.}\ \bibnamefont
  {Mohapatra}}\ and\ \bibinfo {author} {\bibfnamefont {G.}~\bibnamefont
  {Senjanovic}},\ }\href@noop {} {\bibfield  {journal} {\bibinfo  {journal}
  {Phys. Rev.}\ }\textbf {\bibinfo {volume} {D23}},\ \bibinfo {pages} {165}
  (\bibinfo {year} {1981})}\BibitemShut {NoStop}%
\bibitem [{\citenamefont {Foot}\ \emph {et~al.}(1989)\citenamefont {Foot},
  \citenamefont {Lew}, \citenamefont {He},\ and\ \citenamefont
  {Joshi}}]{Foot:1988aq}%
  \BibitemOpen
  \bibfield  {author} {\bibinfo {author} {\bibfnamefont {R.}~\bibnamefont
  {Foot}}, \bibinfo {author} {\bibfnamefont {H.}~\bibnamefont {Lew}}, \bibinfo
  {author} {\bibfnamefont {X.~G.}\ \bibnamefont {He}}, \ and\ \bibinfo {author}
  {\bibfnamefont {G.~C.}\ \bibnamefont {Joshi}},\ }\href@noop {} {\bibfield
  {journal} {\bibinfo  {journal} {Z. Phys.}\ }\textbf {\bibinfo {volume}
  {C44}},\ \bibinfo {pages} {441} (\bibinfo {year} {1989})}\BibitemShut
  {NoStop}%
\bibitem [{\citenamefont {Kersten}\ and\ \citenamefont
  {Smirnov}(2007)}]{Kersten:2007vk}%
  \BibitemOpen
  \bibfield  {author} {\bibinfo {author} {\bibfnamefont {J.}~\bibnamefont
  {Kersten}}\ and\ \bibinfo {author} {\bibfnamefont {A.~Y.}\ \bibnamefont
  {Smirnov}},\ }\href@noop {} {\bibfield  {journal} {\bibinfo  {journal} {Phys.
  Rev.}\ }\textbf {\bibinfo {volume} {D76}},\ \bibinfo {pages} {073005}
  (\bibinfo {year} {2007})},\ \Eprint {http://arxiv.org/abs/0705.3221}
  {arXiv:0705.3221 [hep-ph]} \BibitemShut {NoStop}%
\bibitem [{\citenamefont {Gu}\ \emph {et~al.}(2009)\citenamefont {Gu} \emph
  {et~al.}}]{Gu:2008yj}%
  \BibitemOpen
  \bibfield  {author} {\bibinfo {author} {\bibfnamefont {P.-H.}\ \bibnamefont
  {Gu}} \emph {et~al.},\ }\href@noop {} {\bibfield  {journal} {\bibinfo
  {journal} {Phys. Rev.}\ }\textbf {\bibinfo {volume} {D79}},\ \bibinfo {pages}
  {033010} (\bibinfo {year} {2009})},\ \Eprint {http://arxiv.org/abs/0811.0953}
  {arXiv:0811.0953 [hep-ph]} \BibitemShut {NoStop}%
\bibitem [{\citenamefont {Dev}\ \emph {et~al.}(2014)\citenamefont {Dev},
  \citenamefont {Pilaftsis},\ and\ \citenamefont {Yang}}]{Dev:2013wba}%
  \BibitemOpen
  \bibfield  {author} {\bibinfo {author} {\bibfnamefont {P.~S.~B.}\
  \bibnamefont {Dev}}, \bibinfo {author} {\bibfnamefont {A.}~\bibnamefont
  {Pilaftsis}}, \ and\ \bibinfo {author} {\bibfnamefont {U.-k.}\ \bibnamefont
  {Yang}},\ }\href {\doibase 10.1103/PhysRevLett.112.081801} {\bibfield
  {journal} {\bibinfo  {journal} {Phys.Rev.Lett.}\ }\textbf {\bibinfo {volume}
  {112}},\ \bibinfo {pages} {081801} (\bibinfo {year} {2014})},\ \Eprint
  {http://arxiv.org/abs/1308.2209} {arXiv:1308.2209 [hep-ph]} \BibitemShut
  {NoStop}%
\bibitem [{\citenamefont {Mohapatra}\ and\ \citenamefont
  {Valle}(1986)}]{Mohapatra:1986bd}%
  \BibitemOpen
  \bibfield  {author} {\bibinfo {author} {\bibfnamefont {R.~N.}\ \bibnamefont
  {Mohapatra}}\ and\ \bibinfo {author} {\bibfnamefont {J.~W.~F.}\ \bibnamefont
  {Valle}},\ }\href@noop {} {\bibfield  {journal} {\bibinfo  {journal} {Phys.
  Rev.}\ }\textbf {\bibinfo {volume} {D34}},\ \bibinfo {pages} {1642} (\bibinfo
  {year} {1986})}\BibitemShut {NoStop}%
\bibitem [{\citenamefont {Akhmedov}\ \emph
  {et~al.}(1996{\natexlab{a}})\citenamefont {Akhmedov} \emph
  {et~al.}}]{Akhmedov:1995ip}%
  \BibitemOpen
  \bibfield  {author} {\bibinfo {author} {\bibfnamefont {E.}~\bibnamefont
  {Akhmedov}} \emph {et~al.},\ }\href@noop {} {\bibfield  {journal} {\bibinfo
  {journal} {Phys. Lett.}\ }\textbf {\bibinfo {volume} {B368}},\ \bibinfo
  {pages} {270} (\bibinfo {year} {1996}{\natexlab{a}})},\ \Eprint
  {http://arxiv.org/abs/hep-ph/9507275} {hep-ph/9507275} \BibitemShut {NoStop}%
\bibitem [{\citenamefont {Akhmedov}\ \emph
  {et~al.}(1996{\natexlab{b}})\citenamefont {Akhmedov} \emph
  {et~al.}}]{Akhmedov:1995vm}%
  \BibitemOpen
  \bibfield  {author} {\bibinfo {author} {\bibfnamefont {E.}~\bibnamefont
  {Akhmedov}} \emph {et~al.},\ }\href@noop {} {\bibfield  {journal} {\bibinfo
  {journal} {Phys. Rev.}\ }\textbf {\bibinfo {volume} {D53}},\ \bibinfo {pages}
  {2752} (\bibinfo {year} {1996}{\natexlab{b}})},\ \Eprint
  {http://arxiv.org/abs/hep-ph/9509255} {hep-ph/9509255} \BibitemShut {NoStop}%
\bibitem [{\citenamefont {Malinsky}\ \emph {et~al.}(2005)\citenamefont
  {Malinsky}, \citenamefont {Romao},\ and\ \citenamefont
  {Valle}}]{Malinsky:2005bi}%
  \BibitemOpen
  \bibfield  {author} {\bibinfo {author} {\bibfnamefont {M.}~\bibnamefont
  {Malinsky}}, \bibinfo {author} {\bibfnamefont {J.~C.}\ \bibnamefont {Romao}},
  \ and\ \bibinfo {author} {\bibfnamefont {J.~W.~F.}\ \bibnamefont {Valle}},\
  }\href@noop {} {\bibfield  {journal} {\bibinfo  {journal} {Phys. Rev. Lett.}\
  }\textbf {\bibinfo {volume} {95}},\ \bibinfo {pages} {161801} (\bibinfo
  {year} {2005})}\BibitemShut {NoStop}%
\bibitem [{\citenamefont {Babu}(1988)}]{Babu:1988ki}%
  \BibitemOpen
  \bibfield  {author} {\bibinfo {author} {\bibfnamefont {K.~S.}\ \bibnamefont
  {Babu}},\ }\href@noop {} {\bibfield  {journal} {\bibinfo  {journal} {Phys.
  Lett.}\ }\textbf {\bibinfo {volume} {B203}},\ \bibinfo {pages} {132}
  (\bibinfo {year} {1988})}\BibitemShut {NoStop}%
\bibitem [{\citenamefont {Masiero}\ and\ \citenamefont
  {Valle}(1990)}]{masiero:1990uj}%
  \BibitemOpen
  \bibfield  {author} {\bibinfo {author} {\bibfnamefont {A.}~\bibnamefont
  {Masiero}}\ and\ \bibinfo {author} {\bibfnamefont {J.}~\bibnamefont
  {Valle}},\ }\href {\doibase 10.1016/0370-2693(90)90935-Y} {\bibfield
  {journal} {\bibinfo  {journal} {Phys.Lett.}\ }\textbf {\bibinfo {volume}
  {B251}},\ \bibinfo {pages} {273} (\bibinfo {year} {1990})}\BibitemShut
  {NoStop}%
\bibitem [{\citenamefont {Romao}\ \emph {et~al.}(1992)\citenamefont {Romao},
  \citenamefont {Santos},\ and\ \citenamefont {Valle}}]{Romao:1992vu}%
  \BibitemOpen
  \bibfield  {author} {\bibinfo {author} {\bibfnamefont {J.~C.}\ \bibnamefont
  {Romao}}, \bibinfo {author} {\bibfnamefont {C.~A.}\ \bibnamefont {Santos}}, \
  and\ \bibinfo {author} {\bibfnamefont {J.~W.~F.}\ \bibnamefont {Valle}},\
  }\href@noop {} {\bibfield  {journal} {\bibinfo  {journal} {Phys. Lett.}\
  }\textbf {\bibinfo {volume} {B288}},\ \bibinfo {pages} {311} (\bibinfo {year}
  {1992})}\BibitemShut {NoStop}%
\bibitem [{\citenamefont {Abreu}\ \emph {et~al.}(1997)\citenamefont {Abreu}
  \emph {et~al.}}]{abreu:1996pa}%
  \BibitemOpen
  \bibfield  {author} {\bibinfo {author} {\bibfnamefont {P.}~\bibnamefont
  {Abreu}} \emph {et~al.} (\bibinfo {collaboration} {DELPHI}),\ }\href@noop {}
  {\bibfield  {journal} {\bibinfo  {journal} {Z. Phys.}\ }\textbf {\bibinfo
  {volume} {C74}},\ \bibinfo {pages} {57} (\bibinfo {year} {1997})}\BibitemShut
  {NoStop}%
\bibitem [{\citenamefont {Beringer}\ \emph {et~al.}(2012)\citenamefont
  {Beringer} \emph {et~al.}}]{Beringer:1900zz}%
  \BibitemOpen
  \bibfield  {author} {\bibinfo {author} {\bibfnamefont {J.}~\bibnamefont
  {Beringer}} \emph {et~al.} (\bibinfo {collaboration} {Particle Data Group}),\
  }\href@noop {} {\bibfield  {journal} {\bibinfo  {journal} {Phys.Rev.}\
  }\textbf {\bibinfo {volume} {D86}},\ \bibinfo {pages} {010001} (\bibinfo
  {year} {2012})}\BibitemShut {NoStop}%
\bibitem [{\citenamefont {Franceschini}\ \emph {et~al.}(2008)\citenamefont
  {Franceschini}, \citenamefont {Hambye},\ and\ \citenamefont
  {Strumia}}]{Franceschini:2008pz}%
  \BibitemOpen
  \bibfield  {author} {\bibinfo {author} {\bibfnamefont {R.}~\bibnamefont
  {Franceschini}}, \bibinfo {author} {\bibfnamefont {T.}~\bibnamefont
  {Hambye}}, \ and\ \bibinfo {author} {\bibfnamefont {A.}~\bibnamefont
  {Strumia}},\ }\href@noop {} {\bibfield  {journal} {\bibinfo  {journal} {Phys.
  Rev.}\ }\textbf {\bibinfo {volume} {D78}},\ \bibinfo {pages} {033002}
  (\bibinfo {year} {2008})},\ \Eprint {http://arxiv.org/abs/0805.1613}
  {arXiv:0805.1613 [hep-ph]} \BibitemShut {NoStop}%
\bibitem [{\citenamefont {Borzumati}\ and\ \citenamefont
  {Masiero}(1986)}]{borzumati:1986qx}%
  \BibitemOpen
  \bibfield  {author} {\bibinfo {author} {\bibfnamefont {F.}~\bibnamefont
  {Borzumati}}\ and\ \bibinfo {author} {\bibfnamefont {A.}~\bibnamefont
  {Masiero}},\ }\href@noop {} {\bibfield  {journal} {\bibinfo  {journal} {Phys.
  Rev. Lett.}\ }\textbf {\bibinfo {volume} {57}},\ \bibinfo {pages} {961}
  (\bibinfo {year} {1986})}\BibitemShut {NoStop}%
\bibitem [{\citenamefont {Hirsch}\ \emph {et~al.}(2008)\citenamefont {Hirsch}
  \emph {et~al.}}]{hirsch:2008dy}%
  \BibitemOpen
  \bibfield  {author} {\bibinfo {author} {\bibfnamefont {M.}~\bibnamefont
  {Hirsch}} \emph {et~al.},\ }\href@noop {} {\bibfield  {journal} {\bibinfo
  {journal} {Phys. Rev.}\ }\textbf {\bibinfo {volume} {D78}},\ \bibinfo {pages}
  {013006} (\bibinfo {year} {2008})},\ \Eprint {http://arxiv.org/abs/0804.4072}
  {arXiv:0804.4072 [hep-ph]} \BibitemShut {NoStop}%
\bibitem [{\citenamefont {Esteves}\ \emph {et~al.}(2010)\citenamefont
  {Esteves}, \citenamefont {Romao}, \citenamefont {Hirsch}, \citenamefont
  {Vicente}, \citenamefont {Porod} \emph {et~al.}}]{Esteves:2010si}%
  \BibitemOpen
  \bibfield  {author} {\bibinfo {author} {\bibfnamefont {J.}~\bibnamefont
  {Esteves}}, \bibinfo {author} {\bibfnamefont {J.}~\bibnamefont {Romao}},
  \bibinfo {author} {\bibfnamefont {M.}~\bibnamefont {Hirsch}}, \bibinfo
  {author} {\bibfnamefont {A.}~\bibnamefont {Vicente}}, \bibinfo {author}
  {\bibfnamefont {W.}~\bibnamefont {Porod}},  \emph {et~al.},\ }\href@noop {}
  {\bibfield  {journal} {\bibinfo  {journal} {JHEP}\ }\textbf {\bibinfo
  {volume} {1012}},\ \bibinfo {pages} {077} (\bibinfo {year} {2010})},\ \Eprint
  {http://arxiv.org/abs/1011.0348} {arXiv:1011.0348 [hep-ph]} \BibitemShut
  {NoStop}%
\bibitem [{\citenamefont {Fileviez~Perez}\ \emph {et~al.}(2009)\citenamefont
  {Fileviez~Perez}, \citenamefont {Han},\ and\ \citenamefont
  {Li}}]{perez:2009mu}%
  \BibitemOpen
  \bibfield  {author} {\bibinfo {author} {\bibfnamefont {P.}~\bibnamefont
  {Fileviez~Perez}}, \bibinfo {author} {\bibfnamefont {T.}~\bibnamefont {Han}},
  \ and\ \bibinfo {author} {\bibfnamefont {T.}~\bibnamefont {Li}},\ }\href@noop
  {} {\  (\bibinfo {year} {2009})},\ \Eprint {http://arxiv.org/abs/0907.4186}
  {arXiv:0907.4186 [hep-ph]} \BibitemShut {NoStop}%
\bibitem [{\citenamefont {Aristizabal~Sierra}\ \emph
  {et~al.}(2003)\citenamefont {Aristizabal~Sierra} \emph
  {et~al.}}]{AristizabalSierra:2003ix}%
  \BibitemOpen
  \bibfield  {author} {\bibinfo {author} {\bibfnamefont {D.}~\bibnamefont
  {Aristizabal~Sierra}} \emph {et~al.},\ }\href@noop {} {\bibfield  {journal}
  {\bibinfo  {journal} {Phys. Rev.}\ }\textbf {\bibinfo {volume} {D68}},\
  \bibinfo {pages} {033006} (\bibinfo {year} {2003})},\ \Eprint
  {http://arxiv.org/abs/hep-ph/0304141} {hep-ph/0304141} \BibitemShut {NoStop}%
\bibitem [{\citenamefont {Esteves}\ \emph {et~al.}(2009)\citenamefont {Esteves}
  \emph {et~al.}}]{esteves:2009vg}%
  \BibitemOpen
  \bibfield  {author} {\bibinfo {author} {\bibfnamefont {J.~N.}\ \bibnamefont
  {Esteves}} \emph {et~al.},\ }\href@noop {} {\bibfield  {journal} {\bibinfo
  {journal} {JHEP}\ }\textbf {\bibinfo {volume} {05}},\ \bibinfo {pages} {003}
  (\bibinfo {year} {2009})}\BibitemShut {NoStop}%
\bibitem [{\citenamefont {Dev}\ and\ \citenamefont
  {Mohapatra}(2010)}]{Dev:2009aw}%
  \BibitemOpen
  \bibfield  {author} {\bibinfo {author} {\bibfnamefont {P.}~\bibnamefont
  {Dev}}\ and\ \bibinfo {author} {\bibfnamefont {R.}~\bibnamefont
  {Mohapatra}},\ }\href@noop {} {\bibfield  {journal} {\bibinfo  {journal}
  {Phys.Rev.}\ }\textbf {\bibinfo {volume} {D81}},\ \bibinfo {pages} {013001}
  (\bibinfo {year} {2010})},\ \Eprint {http://arxiv.org/abs/0910.3924}
  {arXiv:0910.3924 [hep-ph]} \BibitemShut {NoStop}%
\bibitem [{\citenamefont {Lee}\ \emph {et~al.}(2013)\citenamefont {Lee},
  \citenamefont {Bhupal~Dev},\ and\ \citenamefont {Mohapatra}}]{Dev:2013oxa}%
  \BibitemOpen
  \bibfield  {author} {\bibinfo {author} {\bibfnamefont {C.-H.}\ \bibnamefont
  {Lee}}, \bibinfo {author} {\bibfnamefont {P.}~\bibnamefont {Bhupal~Dev}}, \
  and\ \bibinfo {author} {\bibfnamefont {R.}~\bibnamefont {Mohapatra}},\ }\href
  {\doibase 10.1103/PhysRevD.88.093010} {\bibfield  {journal} {\bibinfo
  {journal} {Phys.Rev.}\ }\textbf {\bibinfo {volume} {D88}},\ \bibinfo {pages}
  {093010} (\bibinfo {year} {2013})},\ \Eprint {http://arxiv.org/abs/1309.0774}
  {arXiv:1309.0774 [hep-ph]} \BibitemShut {NoStop}%
\bibitem [{\citenamefont {Deppisch}\ \emph {et~al.}(2006)\citenamefont
  {Deppisch}, \citenamefont {Kosmas},\ and\ \citenamefont
  {Valle}}]{deppisch:2005zm}%
  \BibitemOpen
  \bibfield  {author} {\bibinfo {author} {\bibfnamefont {F.}~\bibnamefont
  {Deppisch}}, \bibinfo {author} {\bibfnamefont {T.~S.}\ \bibnamefont
  {Kosmas}}, \ and\ \bibinfo {author} {\bibfnamefont {J.~W.~F.}\ \bibnamefont
  {Valle}},\ }\href@noop {} {\bibfield  {journal} {\bibinfo  {journal} {Nucl.
  Phys.}\ }\textbf {\bibinfo {volume} {B752}},\ \bibinfo {pages} {80} (\bibinfo
  {year} {2006})},\ \Eprint {http://arxiv.org/abs/hep-ph/0512360}
  {hep-ph/0512360} \BibitemShut {NoStop}%
\bibitem [{\citenamefont {Deppisch}\ and\ \citenamefont
  {Valle}(2005)}]{Deppisch:2004fa}%
  \BibitemOpen
  \bibfield  {author} {\bibinfo {author} {\bibfnamefont {F.}~\bibnamefont
  {Deppisch}}\ and\ \bibinfo {author} {\bibfnamefont {J.~W.~F.}\ \bibnamefont
  {Valle}},\ }\href@noop {} {\bibfield  {journal} {\bibinfo  {journal} {Phys.
  Rev.}\ }\textbf {\bibinfo {volume} {D72}},\ \bibinfo {pages} {036001}
  (\bibinfo {year} {2005})},\ \Eprint {http://arxiv.org/abs/hep-ph/0406040}
  {hep-ph/0406040} \BibitemShut {NoStop}%
\bibitem [{\citenamefont {Forero}\ \emph {et~al.}(2011)\citenamefont {Forero},
  \citenamefont {Morisi}, \citenamefont {Tortola},\ and\ \citenamefont
  {Valle}}]{Forero:2011pc}%
  \BibitemOpen
  \bibfield  {author} {\bibinfo {author} {\bibfnamefont {D.}~\bibnamefont
  {Forero}}, \bibinfo {author} {\bibfnamefont {S.}~\bibnamefont {Morisi}},
  \bibinfo {author} {\bibfnamefont {M.}~\bibnamefont {Tortola}}, \ and\
  \bibinfo {author} {\bibfnamefont {J.~W.~F.}\ \bibnamefont {Valle}},\
  }\href@noop {} {\bibfield  {journal} {\bibinfo  {journal} {JHEP}\ }\textbf
  {\bibinfo {volume} {1109}},\ \bibinfo {pages} {142} (\bibinfo {year}
  {2011})},\ \Eprint {http://arxiv.org/abs/1107.6009} {arXiv:1107.6009
  [hep-ph]} \BibitemShut {NoStop}%
\bibitem [{\citenamefont {Bernabeu}\ \emph {et~al.}(1987)\citenamefont
  {Bernabeu} \emph {et~al.}}]{bernabeu:1987gr}%
  \BibitemOpen
  \bibfield  {author} {\bibinfo {author} {\bibfnamefont {J.}~\bibnamefont
  {Bernabeu}} \emph {et~al.},\ }\href@noop {} {\bibfield  {journal} {\bibinfo
  {journal} {Phys. Lett.}\ }\textbf {\bibinfo {volume} {B187}},\ \bibinfo
  {pages} {303} (\bibinfo {year} {1987})}\BibitemShut {NoStop}%
\bibitem [{\citenamefont {Gonzalez-Garcia}\ and\ \citenamefont
  {Valle}(1992)}]{gonzalez-garcia:1992be}%
  \BibitemOpen
  \bibfield  {author} {\bibinfo {author} {\bibfnamefont {M.~C.}\ \bibnamefont
  {Gonzalez-Garcia}}\ and\ \bibinfo {author} {\bibfnamefont {J.~W.~F.}\
  \bibnamefont {Valle}},\ }\href@noop {} {\bibfield  {journal} {\bibinfo
  {journal} {Mod. Phys. Lett.}\ }\textbf {\bibinfo {volume} {A7}},\ \bibinfo
  {pages} {477} (\bibinfo {year} {1992})}\BibitemShut {NoStop}%
\bibitem [{\citenamefont {Gonzalez-Garcia}\ and\ \citenamefont
  {Valle}(1989)}]{gonzalezgarcia:1988rw}%
  \BibitemOpen
  \bibfield  {author} {\bibinfo {author} {\bibfnamefont {M.}~\bibnamefont
  {Gonzalez-Garcia}}\ and\ \bibinfo {author} {\bibfnamefont {J.}~\bibnamefont
  {Valle}},\ }\href {\doibase 10.1016/0370-2693(89)91131-3} {\bibfield
  {journal} {\bibinfo  {journal} {Phys.Lett.}\ }\textbf {\bibinfo {volume}
  {B216}},\ \bibinfo {pages} {360} (\bibinfo {year} {1989})}\BibitemShut
  {NoStop}%
\bibitem [{\citenamefont {Branco}\ \emph {et~al.}(1989)\citenamefont {Branco},
  \citenamefont {Rebelo},\ and\ \citenamefont {Valle}}]{branco:1989bn}%
  \BibitemOpen
  \bibfield  {author} {\bibinfo {author} {\bibfnamefont {G.~C.}\ \bibnamefont
  {Branco}}, \bibinfo {author} {\bibfnamefont {M.~N.}\ \bibnamefont {Rebelo}},
  \ and\ \bibinfo {author} {\bibfnamefont {J.~W.~F.}\ \bibnamefont {Valle}},\
  }\href@noop {} {\bibfield  {journal} {\bibinfo  {journal} {Phys. Lett.}\
  }\textbf {\bibinfo {volume} {B225}},\ \bibinfo {pages} {385} (\bibinfo {year}
  {1989})}\BibitemShut {NoStop}%
\bibitem [{\citenamefont {Rius}\ and\ \citenamefont
  {Valle}(1990)}]{rius:1989gk}%
  \BibitemOpen
  \bibfield  {author} {\bibinfo {author} {\bibfnamefont {N.}~\bibnamefont
  {Rius}}\ and\ \bibinfo {author} {\bibfnamefont {J.~W.~F.}\ \bibnamefont
  {Valle}},\ }\href@noop {} {\bibfield  {journal} {\bibinfo  {journal} {Phys.
  Lett.}\ }\textbf {\bibinfo {volume} {B246}},\ \bibinfo {pages} {249}
  (\bibinfo {year} {1990})}\BibitemShut {NoStop}%
\bibitem [{\citenamefont {Bazzocchi}\ \emph {et~al.}(2010)\citenamefont
  {Bazzocchi}, \citenamefont {Cerdeno}, \citenamefont {Munoz},\ and\
  \citenamefont {Valle}}]{Bazzocchi:2009kc}%
  \BibitemOpen
  \bibfield  {author} {\bibinfo {author} {\bibfnamefont {F.}~\bibnamefont
  {Bazzocchi}}, \bibinfo {author} {\bibfnamefont {D.}~\bibnamefont {Cerdeno}},
  \bibinfo {author} {\bibfnamefont {C.}~\bibnamefont {Munoz}}, \ and\ \bibinfo
  {author} {\bibfnamefont {J.~W.~F.}\ \bibnamefont {Valle}},\ }\href@noop {}
  {\bibfield  {journal} {\bibinfo  {journal} {Phys.Rev.}\ }\textbf {\bibinfo
  {volume} {D81}},\ \bibinfo {pages} {051701} (\bibinfo {year} {2010})},\
  \Eprint {http://arxiv.org/abs/0907.1262} {arXiv:0907.1262 [hep-ph]}
  \BibitemShut {NoStop}%
\bibitem [{\citenamefont {Bazzocchi}(2011)}]{Bazzocchi:2010dt}%
  \BibitemOpen
  \bibfield  {author} {\bibinfo {author} {\bibfnamefont {F.}~\bibnamefont
  {Bazzocchi}},\ }\href@noop {} {\bibfield  {journal} {\bibinfo  {journal}
  {Phys.Rev.}\ }\textbf {\bibinfo {volume} {D83}},\ \bibinfo {pages} {093009}
  (\bibinfo {year} {2011})},\ \Eprint {http://arxiv.org/abs/1011.6299}
  {arXiv:1011.6299 [hep-ph]} \BibitemShut {NoStop}%
\bibitem [{\citenamefont {Hettmansperger}\ \emph {et~al.}(2011)\citenamefont
  {Hettmansperger}, \citenamefont {Lindner},\ and\ \citenamefont
  {Rodejohann}}]{Hettmansperger:2011bt}%
  \BibitemOpen
  \bibfield  {author} {\bibinfo {author} {\bibfnamefont {H.}~\bibnamefont
  {Hettmansperger}}, \bibinfo {author} {\bibfnamefont {M.}~\bibnamefont
  {Lindner}}, \ and\ \bibinfo {author} {\bibfnamefont {W.}~\bibnamefont
  {Rodejohann}},\ }\href@noop {} {\bibfield  {journal} {\bibinfo  {journal}
  {JHEP}\ }\textbf {\bibinfo {volume} {1104}},\ \bibinfo {pages} {123}
  (\bibinfo {year} {2011})},\ \Eprint {http://arxiv.org/abs/1102.3432}
  {arXiv:1102.3432 [hep-ph]} \BibitemShut {NoStop}%
\bibitem [{\citenamefont {Das}\ and\ \citenamefont {Okada}(2013)}]{Das:2012ze}%
  \BibitemOpen
  \bibfield  {author} {\bibinfo {author} {\bibfnamefont {A.}~\bibnamefont
  {Das}}\ and\ \bibinfo {author} {\bibfnamefont {N.}~\bibnamefont {Okada}},\
  }\href {\doibase 10.1103/PhysRevD.88.113001} {\bibfield  {journal} {\bibinfo
  {journal} {Phys.Rev.}\ }\textbf {\bibinfo {volume} {D88}},\ \bibinfo {pages}
  {113001} (\bibinfo {year} {2013})},\ \Eprint {http://arxiv.org/abs/1207.3734}
  {arXiv:1207.3734 [hep-ph]} \BibitemShut {NoStop}%
\bibitem [{\citenamefont {De~Romeri}\ \emph {et~al.}(2011)\citenamefont
  {De~Romeri}, \citenamefont {Hirsch},\ and\ \citenamefont
  {Malinsky}}]{DeRomeri:2011ie}%
  \BibitemOpen
  \bibfield  {author} {\bibinfo {author} {\bibfnamefont {V.}~\bibnamefont
  {De~Romeri}}, \bibinfo {author} {\bibfnamefont {M.}~\bibnamefont {Hirsch}}, \
  and\ \bibinfo {author} {\bibfnamefont {M.}~\bibnamefont {Malinsky}},\ }\href
  {\doibase 10.1103/PhysRevD.84.053012} {\bibfield  {journal} {\bibinfo
  {journal} {Phys.Rev.}\ }\textbf {\bibinfo {volume} {D84}},\ \bibinfo {pages}
  {053012} (\bibinfo {year} {2011})},\ \Eprint {http://arxiv.org/abs/1107.3412}
  {arXiv:1107.3412 [hep-ph]} \BibitemShut {NoStop}%
\bibitem [{\citenamefont {Nath}\ \emph {et~al.}(2010)\citenamefont {Nath},
  \citenamefont {Nelson}, \citenamefont {Davoudiasl}, \citenamefont {Dutta},
  \citenamefont {Feldman} \emph {et~al.}}]{Nath:2010zj}%
  \BibitemOpen
  \bibfield  {author} {\bibinfo {author} {\bibfnamefont {P.}~\bibnamefont
  {Nath}}, \bibinfo {author} {\bibfnamefont {B.~D.}\ \bibnamefont {Nelson}},
  \bibinfo {author} {\bibfnamefont {H.}~\bibnamefont {Davoudiasl}}, \bibinfo
  {author} {\bibfnamefont {B.}~\bibnamefont {Dutta}}, \bibinfo {author}
  {\bibfnamefont {D.}~\bibnamefont {Feldman}},  \emph {et~al.},\ }\href@noop {}
  {\bibfield  {journal} {\bibinfo  {journal} {Nucl.Phys.Proc.Suppl.}\ }\textbf
  {\bibinfo {volume} {200-202}},\ \bibinfo {pages} {185} (\bibinfo {year}
  {2010})},\ \Eprint {http://arxiv.org/abs/1001.2693} {arXiv:1001.2693
  [hep-ph]} \BibitemShut {NoStop}%
\bibitem [{\citenamefont {Das}\ \emph {et~al.}(2012)\citenamefont {Das},
  \citenamefont {Deppisch}, \citenamefont {Kittel},\ and\ \citenamefont
  {Valle}}]{Das:2012ii}%
  \BibitemOpen
  \bibfield  {author} {\bibinfo {author} {\bibfnamefont {S.}~\bibnamefont
  {Das}}, \bibinfo {author} {\bibfnamefont {F.}~\bibnamefont {Deppisch}},
  \bibinfo {author} {\bibfnamefont {O.}~\bibnamefont {Kittel}}, \ and\ \bibinfo
  {author} {\bibfnamefont {J.}~\bibnamefont {Valle}},\ }\href@noop {}
  {\bibfield  {journal} {\bibinfo  {journal} {Phys.Rev.}\ }\textbf {\bibinfo
  {volume} {D86}},\ \bibinfo {pages} {055006} (\bibinfo {year} {2012})},\
  \Eprint {http://arxiv.org/abs/1206.0256} {arXiv:1206.0256 [hep-ph]}
  \BibitemShut {NoStop}%
\bibitem [{\citenamefont {Aguilar-Saavedra}\ \emph {et~al.}(2012)\citenamefont
  {Aguilar-Saavedra}, \citenamefont {Deppisch}, \citenamefont {Kittel},\ and\
  \citenamefont {Valle}}]{AguilarSaavedra:2012fu}%
  \BibitemOpen
  \bibfield  {author} {\bibinfo {author} {\bibfnamefont {J.}~\bibnamefont
  {Aguilar-Saavedra}}, \bibinfo {author} {\bibfnamefont {F.}~\bibnamefont
  {Deppisch}}, \bibinfo {author} {\bibfnamefont {O.}~\bibnamefont {Kittel}}, \
  and\ \bibinfo {author} {\bibfnamefont {J.}~\bibnamefont {Valle}},\
  }\href@noop {} {\bibfield  {journal} {\bibinfo  {journal} {Phys.Rev.}\
  }\textbf {\bibinfo {volume} {D85}},\ \bibinfo {pages} {091301} (\bibinfo
  {year} {2012})},\ \Eprint {http://arxiv.org/abs/1203.5998} {arXiv:1203.5998
  [hep-ph]} \BibitemShut {NoStop}%
\bibitem [{\citenamefont {Deppisch}\ \emph {et~al.}(2013)\citenamefont
  {Deppisch}, \citenamefont {Desai},\ and\ \citenamefont
  {Valle}}]{Deppisch:2013cya}%
  \BibitemOpen
  \bibfield  {author} {\bibinfo {author} {\bibfnamefont {F.~F.}\ \bibnamefont
  {Deppisch}}, \bibinfo {author} {\bibfnamefont {N.}~\bibnamefont {Desai}}, \
  and\ \bibinfo {author} {\bibfnamefont {J.~W.~F.}\ \bibnamefont {Valle}},\
  }\href@noop {} {\  (\bibinfo {year} {2013})},\ \Eprint
  {http://arxiv.org/abs/1308.6789} {arXiv:1308.6789 [hep-ph]} \BibitemShut
  {NoStop}%
\bibitem [{\citenamefont {Ma}(2009)}]{Ma:2009kh}%
  \BibitemOpen
  \bibfield  {author} {\bibinfo {author} {\bibfnamefont {E.}~\bibnamefont
  {Ma}},\ }\href@noop {} {\  (\bibinfo {year} {2009})},\ \Eprint
  {http://arxiv.org/abs/0905.2972} {arXiv:0905.2972 [hep-ph]} \BibitemShut
  {NoStop}%
\bibitem [{\citenamefont {Ibanez}\ \emph {et~al.}(2009)\citenamefont {Ibanez},
  \citenamefont {Morisi},\ and\ \citenamefont {Valle}}]{Ibanez:2009du}%
  \BibitemOpen
  \bibfield  {author} {\bibinfo {author} {\bibfnamefont {D.}~\bibnamefont
  {Ibanez}}, \bibinfo {author} {\bibfnamefont {S.}~\bibnamefont {Morisi}}, \
  and\ \bibinfo {author} {\bibfnamefont {J.~W.~F.}\ \bibnamefont {Valle}},\
  }\href@noop {} {\bibfield  {journal} {\bibinfo  {journal} {Phys. Rev.}\
  }\textbf {\bibinfo {volume} {D80}},\ \bibinfo {pages} {053015} (\bibinfo
  {year} {2009})},\ \Eprint {http://arxiv.org/abs/0907.3109} {arXiv:0907.3109
  [hep-ph]} \BibitemShut {NoStop}%
\bibitem [{\citenamefont {Glashow}\ \emph {et~al.}(1970)\citenamefont
  {Glashow}, \citenamefont {Iliopoulos},\ and\ \citenamefont
  {Maiani}}]{glashow:1970gm}%
  \BibitemOpen
  \bibfield  {author} {\bibinfo {author} {\bibfnamefont {S.~L.}\ \bibnamefont
  {Glashow}}, \bibinfo {author} {\bibfnamefont {J.}~\bibnamefont {Iliopoulos}},
  \ and\ \bibinfo {author} {\bibfnamefont {L.}~\bibnamefont {Maiani}},\
  }\href@noop {} {\bibfield  {journal} {\bibinfo  {journal} {Phys. Rev.}\
  }\textbf {\bibinfo {volume} {D2}},\ \bibinfo {pages} {1285} (\bibinfo {year}
  {1970})}\BibitemShut {NoStop}%
\bibitem [{\citenamefont {Fileviez~Perez}\ \emph {et~al.}(2008)\citenamefont
  {Fileviez~Perez}, \citenamefont {Han}, \citenamefont {Huang}, \citenamefont
  {Li},\ and\ \citenamefont {Wang}}]{Perez:2008ha}%
  \BibitemOpen
  \bibfield  {author} {\bibinfo {author} {\bibfnamefont {P.}~\bibnamefont
  {Fileviez~Perez}}, \bibinfo {author} {\bibfnamefont {T.}~\bibnamefont {Han}},
  \bibinfo {author} {\bibfnamefont {G.-y.}\ \bibnamefont {Huang}}, \bibinfo
  {author} {\bibfnamefont {T.}~\bibnamefont {Li}}, \ and\ \bibinfo {author}
  {\bibfnamefont {K.}~\bibnamefont {Wang}},\ }\href@noop {} {\bibfield
  {journal} {\bibinfo  {journal} {Phys. Rev.}\ }\textbf {\bibinfo {volume}
  {D78}},\ \bibinfo {pages} {015018} (\bibinfo {year} {2008})},\ \Eprint
  {http://arxiv.org/abs/0805.3536} {arXiv:0805.3536 [hep-ph]} \BibitemShut
  {NoStop}%
\bibitem [{\citenamefont {del Aguila}\ and\ \citenamefont
  {Aguilar-Saavedra}(2009)}]{delAguila:2008cj}%
  \BibitemOpen
  \bibfield  {author} {\bibinfo {author} {\bibfnamefont {F.}~\bibnamefont {del
  Aguila}}\ and\ \bibinfo {author} {\bibfnamefont {J.~A.}\ \bibnamefont
  {Aguilar-Saavedra}},\ }\href@noop {} {\bibfield  {journal} {\bibinfo
  {journal} {Nucl. Phys.}\ }\textbf {\bibinfo {volume} {B813}},\ \bibinfo
  {pages} {22} (\bibinfo {year} {2009})},\ \Eprint
  {http://arxiv.org/abs/0808.2468} {arXiv:0808.2468 [hep-ph]} \BibitemShut
  {NoStop}%
\bibitem [{\citenamefont {Babu}\ and\ \citenamefont
  {Leung}(2001)}]{Babu:2001ex}%
  \BibitemOpen
  \bibfield  {author} {\bibinfo {author} {\bibfnamefont {K.}~\bibnamefont
  {Babu}}\ and\ \bibinfo {author} {\bibfnamefont {C.~N.}\ \bibnamefont
  {Leung}},\ }\href {\doibase 10.1016/S0550-3213(01)00504-1} {\bibfield
  {journal} {\bibinfo  {journal} {Nucl.Phys.}\ }\textbf {\bibinfo {volume}
  {B619}},\ \bibinfo {pages} {667} (\bibinfo {year} {2001})},\ \Eprint
  {http://arxiv.org/abs/hep-ph/0106054} {arXiv:hep-ph/0106054 [hep-ph]}
  \BibitemShut {NoStop}%
\bibitem [{\citenamefont {Angel}\ \emph {et~al.}(2013)\citenamefont {Angel},
  \citenamefont {Rodd},\ and\ \citenamefont {Volkas}}]{Angel:2012ug}%
  \BibitemOpen
  \bibfield  {author} {\bibinfo {author} {\bibfnamefont {P.~W.}\ \bibnamefont
  {Angel}}, \bibinfo {author} {\bibfnamefont {N.~L.}\ \bibnamefont {Rodd}}, \
  and\ \bibinfo {author} {\bibfnamefont {R.~R.}\ \bibnamefont {Volkas}},\
  }\href {\doibase 10.1103/PhysRevD.87.073007} {\bibfield  {journal} {\bibinfo
  {journal} {Phys.Rev.}\ }\textbf {\bibinfo {volume} {D87}},\ \bibinfo {pages}
  {073007} (\bibinfo {year} {2013})},\ \Eprint {http://arxiv.org/abs/1212.6111}
  {arXiv:1212.6111 [hep-ph]} \BibitemShut {NoStop}%
\bibitem [{\citenamefont {Farzan}\ \emph {et~al.}(2013)\citenamefont {Farzan},
  \citenamefont {Pascoli},\ and\ \citenamefont {Schmidt}}]{Farzan:2012ev}%
  \BibitemOpen
  \bibfield  {author} {\bibinfo {author} {\bibfnamefont {Y.}~\bibnamefont
  {Farzan}}, \bibinfo {author} {\bibfnamefont {S.}~\bibnamefont {Pascoli}}, \
  and\ \bibinfo {author} {\bibfnamefont {M.~A.}\ \bibnamefont {Schmidt}},\
  }\href {\doibase 10.1007/JHEP03(2013)107} {\bibfield  {journal} {\bibinfo
  {journal} {JHEP}\ }\textbf {\bibinfo {volume} {1303}},\ \bibinfo {pages}
  {107} (\bibinfo {year} {2013})},\ \Eprint {http://arxiv.org/abs/1208.2732}
  {arXiv:1208.2732 [hep-ph]} \BibitemShut {NoStop}%
\bibitem [{\citenamefont {Law}\ and\ \citenamefont
  {McDonald}(2014)}]{Law:2013dya}%
  \BibitemOpen
  \bibfield  {author} {\bibinfo {author} {\bibfnamefont {S.~S.}\ \bibnamefont
  {Law}}\ and\ \bibinfo {author} {\bibfnamefont {K.~L.}\ \bibnamefont
  {McDonald}},\ }\href {\doibase 10.1142/S0217751X1450064X} {\bibfield
  {journal} {\bibinfo  {journal} {Int.J.Mod.Phys.}\ }\textbf {\bibinfo {volume}
  {A29}},\ \bibinfo {pages} {1450064} (\bibinfo {year} {2014})},\ \Eprint
  {http://arxiv.org/abs/1303.6384} {arXiv:1303.6384 [hep-ph]} \BibitemShut
  {NoStop}%
\bibitem [{\citenamefont {Pilaftsis}(1992)}]{Pilaftsis:1991ug}%
  \BibitemOpen
  \bibfield  {author} {\bibinfo {author} {\bibfnamefont {A.}~\bibnamefont
  {Pilaftsis}},\ }\href {\doibase 10.1007/BF01482590} {\bibfield  {journal}
  {\bibinfo  {journal} {Z.Phys.}\ }\textbf {\bibinfo {volume} {C55}},\ \bibinfo
  {pages} {275} (\bibinfo {year} {1992})},\ \Eprint
  {http://arxiv.org/abs/hep-ph/9901206} {arXiv:hep-ph/9901206 [hep-ph]}
  \BibitemShut {NoStop}%
\bibitem [{\citenamefont {Dev}\ and\ \citenamefont
  {Pilaftsis}(2012)}]{Dev:2012sg}%
  \BibitemOpen
  \bibfield  {author} {\bibinfo {author} {\bibfnamefont {P.~B.}\ \bibnamefont
  {Dev}}\ and\ \bibinfo {author} {\bibfnamefont {A.}~\bibnamefont
  {Pilaftsis}},\ }\href {\doibase 10.1103/PhysRevD.86.113001} {\bibfield
  {journal} {\bibinfo  {journal} {Phys.Rev.}\ }\textbf {\bibinfo {volume}
  {D86}},\ \bibinfo {pages} {113001} (\bibinfo {year} {2012})},\ \Eprint
  {http://arxiv.org/abs/1209.4051} {arXiv:1209.4051 [hep-ph]} \BibitemShut
  {NoStop}%
\bibitem [{\citenamefont {Fileviez~Perez}\ and\ \citenamefont
  {Wise}(2009)}]{FileviezPerez:2009ud}%
  \BibitemOpen
  \bibfield  {author} {\bibinfo {author} {\bibfnamefont {P.}~\bibnamefont
  {Fileviez~Perez}}\ and\ \bibinfo {author} {\bibfnamefont {M.~B.}\
  \bibnamefont {Wise}},\ }\href@noop {} {\bibfield  {journal} {\bibinfo
  {journal} {Phys. Rev.}\ }\textbf {\bibinfo {volume} {D80}},\ \bibinfo {pages}
  {053006} (\bibinfo {year} {2009})},\ \Eprint {http://arxiv.org/abs/0906.2950}
  {arXiv:0906.2950 [hep-ph]} \BibitemShut {NoStop}%
\bibitem [{\citenamefont {Zee}(1980)}]{Zee:1980ai}%
  \BibitemOpen
  \bibfield  {author} {\bibinfo {author} {\bibfnamefont {A.}~\bibnamefont
  {Zee}},\ }\href@noop {} {\bibfield  {journal} {\bibinfo  {journal} {Phys.
  Lett.}\ }\textbf {\bibinfo {volume} {B93}},\ \bibinfo {pages} {389} (\bibinfo
  {year} {1980})}\BibitemShut {NoStop}%
\bibitem [{\citenamefont {Aristizabal~Sierra}\ and\ \citenamefont
  {Restrepo}(2006)}]{AristizabalSierra:2006ri}%
  \BibitemOpen
  \bibfield  {author} {\bibinfo {author} {\bibfnamefont {D.}~\bibnamefont
  {Aristizabal~Sierra}}\ and\ \bibinfo {author} {\bibfnamefont
  {D.}~\bibnamefont {Restrepo}},\ }\href@noop {} {\bibfield  {journal}
  {\bibinfo  {journal} {JHEP}\ }\textbf {\bibinfo {volume} {08}},\ \bibinfo
  {pages} {036} (\bibinfo {year} {2006})},\ \Eprint
  {http://arxiv.org/abs/hep-ph/0604012} {arXiv:hep-ph/0604012} \BibitemShut
  {NoStop}%
\bibitem [{\citenamefont {Smirnov}\ and\ \citenamefont
  {Tanimoto}(1997)}]{Smirnov:1996bv}%
  \BibitemOpen
  \bibfield  {author} {\bibinfo {author} {\bibfnamefont {A.~Y.}\ \bibnamefont
  {Smirnov}}\ and\ \bibinfo {author} {\bibfnamefont {M.}~\bibnamefont
  {Tanimoto}},\ }\href {\doibase 10.1103/PhysRevD.55.1665} {\bibfield
  {journal} {\bibinfo  {journal} {Phys.Rev.}\ }\textbf {\bibinfo {volume}
  {D55}},\ \bibinfo {pages} {1665} (\bibinfo {year} {1997})},\ \Eprint
  {http://arxiv.org/abs/hep-ph/9604370} {arXiv:hep-ph/9604370 [hep-ph]}
  \BibitemShut {NoStop}%
\bibitem [{\citenamefont {Jarlskog}\ \emph {et~al.}(1999)\citenamefont
  {Jarlskog}, \citenamefont {Matsuda}, \citenamefont {Skadhauge},\ and\
  \citenamefont {Tanimoto}}]{Jarlskog:1998uf}%
  \BibitemOpen
  \bibfield  {author} {\bibinfo {author} {\bibfnamefont {C.}~\bibnamefont
  {Jarlskog}}, \bibinfo {author} {\bibfnamefont {M.}~\bibnamefont {Matsuda}},
  \bibinfo {author} {\bibfnamefont {S.}~\bibnamefont {Skadhauge}}, \ and\
  \bibinfo {author} {\bibfnamefont {M.}~\bibnamefont {Tanimoto}},\ }\href@noop
  {} {\bibfield  {journal} {\bibinfo  {journal} {Phys.Lett.}\ }\textbf
  {\bibinfo {volume} {B449}},\ \bibinfo {pages} {240} (\bibinfo {year}
  {1999})},\ \Eprint {http://arxiv.org/abs/hep-ph/9812282}
  {arXiv:hep-ph/9812282 [hep-ph]} \BibitemShut {NoStop}%
\bibitem [{\citenamefont {Frampton}\ and\ \citenamefont
  {Glashow}(1999)}]{Frampton:1999yn}%
  \BibitemOpen
  \bibfield  {author} {\bibinfo {author} {\bibfnamefont {P.~H.}\ \bibnamefont
  {Frampton}}\ and\ \bibinfo {author} {\bibfnamefont {S.~L.}\ \bibnamefont
  {Glashow}},\ }\href {\doibase 10.1016/S0370-2693(99)00824-2} {\bibfield
  {journal} {\bibinfo  {journal} {Phys.Lett.}\ }\textbf {\bibinfo {volume}
  {B461}},\ \bibinfo {pages} {95} (\bibinfo {year} {1999})},\ \Eprint
  {http://arxiv.org/abs/hep-ph/9906375} {arXiv:hep-ph/9906375 [hep-ph]}
  \BibitemShut {NoStop}%
\bibitem [{\citenamefont {Joshipura}\ and\ \citenamefont
  {Rindani}(1999)}]{Joshipura:1999xe}%
  \BibitemOpen
  \bibfield  {author} {\bibinfo {author} {\bibfnamefont {A.~S.}\ \bibnamefont
  {Joshipura}}\ and\ \bibinfo {author} {\bibfnamefont {S.~D.}\ \bibnamefont
  {Rindani}},\ }\href {\doibase 10.1016/S0370-2693(99)00995-8} {\bibfield
  {journal} {\bibinfo  {journal} {Phys.Lett.}\ }\textbf {\bibinfo {volume}
  {B464}},\ \bibinfo {pages} {239} (\bibinfo {year} {1999})},\ \Eprint
  {http://arxiv.org/abs/hep-ph/9907390} {arXiv:hep-ph/9907390 [hep-ph]}
  \BibitemShut {NoStop}%
\bibitem [{\citenamefont {McLaughlin}\ and\ \citenamefont
  {Ng}(1999)}]{McLaughlin:1999un}%
  \BibitemOpen
  \bibfield  {author} {\bibinfo {author} {\bibfnamefont {G.}~\bibnamefont
  {McLaughlin}}\ and\ \bibinfo {author} {\bibfnamefont {J.}~\bibnamefont
  {Ng}},\ }\href {\doibase 10.1016/S0370-2693(99)00917-X} {\bibfield  {journal}
  {\bibinfo  {journal} {Phys.Lett.}\ }\textbf {\bibinfo {volume} {B464}},\
  \bibinfo {pages} {232} (\bibinfo {year} {1999})},\ \Eprint
  {http://arxiv.org/abs/hep-ph/9907449} {arXiv:hep-ph/9907449 [hep-ph]}
  \BibitemShut {NoStop}%
\bibitem [{\citenamefont {Cheung}\ and\ \citenamefont
  {Kong}(2000)}]{Cheung:1999az}%
  \BibitemOpen
  \bibfield  {author} {\bibinfo {author} {\bibfnamefont {K.-m.}\ \bibnamefont
  {Cheung}}\ and\ \bibinfo {author} {\bibfnamefont {O.~C.}\ \bibnamefont
  {Kong}},\ }\href {\doibase 10.1103/PhysRevD.61.113012} {\bibfield  {journal}
  {\bibinfo  {journal} {Phys.Rev.}\ }\textbf {\bibinfo {volume} {D61}},\
  \bibinfo {pages} {113012} (\bibinfo {year} {2000})},\ \Eprint
  {http://arxiv.org/abs/hep-ph/9912238} {arXiv:hep-ph/9912238 [hep-ph]}
  \BibitemShut {NoStop}%
\bibitem [{\citenamefont {Chang}\ and\ \citenamefont
  {Zee}(2000)}]{Chang:1999hga}%
  \BibitemOpen
  \bibfield  {author} {\bibinfo {author} {\bibfnamefont {D.}~\bibnamefont
  {Chang}}\ and\ \bibinfo {author} {\bibfnamefont {A.}~\bibnamefont {Zee}},\
  }\href {\doibase 10.1103/PhysRevD.61.071303} {\bibfield  {journal} {\bibinfo
  {journal} {Phys.Rev.}\ }\textbf {\bibinfo {volume} {D61}},\ \bibinfo {pages}
  {071303} (\bibinfo {year} {2000})},\ \Eprint
  {http://arxiv.org/abs/hep-ph/9912380} {arXiv:hep-ph/9912380 [hep-ph]}
  \BibitemShut {NoStop}%
\bibitem [{\citenamefont {Dicus}\ \emph {et~al.}(2001)\citenamefont {Dicus},
  \citenamefont {He},\ and\ \citenamefont {Ng}}]{Dicus:2001ph}%
  \BibitemOpen
  \bibfield  {author} {\bibinfo {author} {\bibfnamefont {D.~A.}\ \bibnamefont
  {Dicus}}, \bibinfo {author} {\bibfnamefont {H.-J.}\ \bibnamefont {He}}, \
  and\ \bibinfo {author} {\bibfnamefont {J.~N.}\ \bibnamefont {Ng}},\ }\href
  {\doibase 10.1103/PhysRevLett.87.111803} {\bibfield  {journal} {\bibinfo
  {journal} {Phys.Rev.Lett.}\ }\textbf {\bibinfo {volume} {87}},\ \bibinfo
  {pages} {111803} (\bibinfo {year} {2001})},\ \Eprint
  {http://arxiv.org/abs/hep-ph/0103126} {arXiv:hep-ph/0103126 [hep-ph]}
  \BibitemShut {NoStop}%
\bibitem [{\citenamefont {Balaji}\ \emph {et~al.}(2001)\citenamefont {Balaji},
  \citenamefont {Grimus},\ and\ \citenamefont {Schwetz}}]{Balaji:2001ex}%
  \BibitemOpen
  \bibfield  {author} {\bibinfo {author} {\bibfnamefont {K.}~\bibnamefont
  {Balaji}}, \bibinfo {author} {\bibfnamefont {W.}~\bibnamefont {Grimus}}, \
  and\ \bibinfo {author} {\bibfnamefont {T.}~\bibnamefont {Schwetz}},\ }\href
  {\doibase 10.1016/S0370-2693(01)00532-9} {\bibfield  {journal} {\bibinfo
  {journal} {Phys.Lett.}\ }\textbf {\bibinfo {volume} {B508}},\ \bibinfo
  {pages} {301} (\bibinfo {year} {2001})},\ \Eprint
  {http://arxiv.org/abs/hep-ph/0104035} {arXiv:hep-ph/0104035 [hep-ph]}
  \BibitemShut {NoStop}%
\bibitem [{\citenamefont {Mitsuda}\ and\ \citenamefont
  {Sasaki}(2001)}]{Mitsuda:2001vh}%
  \BibitemOpen
  \bibfield  {author} {\bibinfo {author} {\bibfnamefont {E.}~\bibnamefont
  {Mitsuda}}\ and\ \bibinfo {author} {\bibfnamefont {K.}~\bibnamefont
  {Sasaki}},\ }\href {\doibase 10.1016/S0370-2693(01)00842-5} {\bibfield
  {journal} {\bibinfo  {journal} {Phys.Lett.}\ }\textbf {\bibinfo {volume}
  {B516}},\ \bibinfo {pages} {47} (\bibinfo {year} {2001})}\BibitemShut
  {NoStop}%
\bibitem [{\citenamefont {Ghosal}\ \emph {et~al.}(2001)\citenamefont {Ghosal},
  \citenamefont {Koide},\ and\ \citenamefont {Fusaoka}}]{Ghosal:2001ep}%
  \BibitemOpen
  \bibfield  {author} {\bibinfo {author} {\bibfnamefont {A.}~\bibnamefont
  {Ghosal}}, \bibinfo {author} {\bibfnamefont {Y.}~\bibnamefont {Koide}}, \
  and\ \bibinfo {author} {\bibfnamefont {H.}~\bibnamefont {Fusaoka}},\ }\href
  {\doibase 10.1103/PhysRevD.64.053012} {\bibfield  {journal} {\bibinfo
  {journal} {Phys.Rev.}\ }\textbf {\bibinfo {volume} {D64}},\ \bibinfo {pages}
  {053012} (\bibinfo {year} {2001})},\ \Eprint
  {http://arxiv.org/abs/hep-ph/0104104} {arXiv:hep-ph/0104104 [hep-ph]}
  \BibitemShut {NoStop}%
\bibitem [{\citenamefont {Koide}(2001)}]{Koide:2001xy}%
  \BibitemOpen
  \bibfield  {author} {\bibinfo {author} {\bibfnamefont {Y.}~\bibnamefont
  {Koide}},\ }\href {\doibase 10.1103/PhysRevD.64.077301} {\bibfield  {journal}
  {\bibinfo  {journal} {Phys.Rev.}\ }\textbf {\bibinfo {volume} {D64}},\
  \bibinfo {pages} {077301} (\bibinfo {year} {2001})},\ \Eprint
  {http://arxiv.org/abs/hep-ph/0104226} {arXiv:hep-ph/0104226 [hep-ph]}
  \BibitemShut {NoStop}%
\bibitem [{\citenamefont {Brahmachari}\ and\ \citenamefont
  {Choubey}(2002)}]{Brahmachari:2001rn}%
  \BibitemOpen
  \bibfield  {author} {\bibinfo {author} {\bibfnamefont {B.}~\bibnamefont
  {Brahmachari}}\ and\ \bibinfo {author} {\bibfnamefont {S.}~\bibnamefont
  {Choubey}},\ }\href {\doibase 10.1016/S0370-2693(02)01366-7} {\bibfield
  {journal} {\bibinfo  {journal} {Phys.Lett.}\ }\textbf {\bibinfo {volume}
  {B531}},\ \bibinfo {pages} {99} (\bibinfo {year} {2002})},\ \Eprint
  {http://arxiv.org/abs/hep-ph/0111133} {arXiv:hep-ph/0111133 [hep-ph]}
  \BibitemShut {NoStop}%
\bibitem [{\citenamefont {Kitabayashi}\ and\ \citenamefont
  {Yasue}(2002)}]{Kitabayashi:2001it}%
  \BibitemOpen
  \bibfield  {author} {\bibinfo {author} {\bibfnamefont {T.}~\bibnamefont
  {Kitabayashi}}\ and\ \bibinfo {author} {\bibfnamefont {M.}~\bibnamefont
  {Yasue}},\ }\href {\doibase 10.1142/S0217751X02010959} {\bibfield  {journal}
  {\bibinfo  {journal} {Int.J.Mod.Phys.}\ }\textbf {\bibinfo {volume} {A17}},\
  \bibinfo {pages} {2519} (\bibinfo {year} {2002})},\ \Eprint
  {http://arxiv.org/abs/hep-ph/0112287} {arXiv:hep-ph/0112287 [hep-ph]}
  \BibitemShut {NoStop}%
\bibitem [{\citenamefont {Koide}(2002)}]{Koide:2002uk}%
  \BibitemOpen
  \bibfield  {author} {\bibinfo {author} {\bibfnamefont {Y.}~\bibnamefont
  {Koide}},\ }\href {\doibase 10.1016/S0920-5632(02)01726-7} {\bibfield
  {journal} {\bibinfo  {journal} {Nucl.Phys.Proc.Suppl.}\ }\textbf {\bibinfo
  {volume} {111}},\ \bibinfo {pages} {294} (\bibinfo {year} {2002})},\ \Eprint
  {http://arxiv.org/abs/hep-ph/0201250} {arXiv:hep-ph/0201250 [hep-ph]}
  \BibitemShut {NoStop}%
\bibitem [{\citenamefont {Cheng}\ and\ \citenamefont
  {Cheung}(2002)}]{Cheng:2002hr}%
  \BibitemOpen
  \bibfield  {author} {\bibinfo {author} {\bibfnamefont {M.-Y.}\ \bibnamefont
  {Cheng}}\ and\ \bibinfo {author} {\bibfnamefont {K.-m.}\ \bibnamefont
  {Cheung}},\ }\href@noop {} {\  (\bibinfo {year} {2002})},\ \Eprint
  {http://arxiv.org/abs/hep-ph/0203051} {arXiv:hep-ph/0203051 [hep-ph]}
  \BibitemShut {NoStop}%
\bibitem [{\citenamefont {He}\ and\ \citenamefont {Zee}(2003)}]{He:2003rm}%
  \BibitemOpen
  \bibfield  {author} {\bibinfo {author} {\bibfnamefont {X.~G.}\ \bibnamefont
  {He}}\ and\ \bibinfo {author} {\bibfnamefont {A.}~\bibnamefont {Zee}},\
  }\href {\doibase 10.1016/S0370-2693(03)00390-3} {\bibfield  {journal}
  {\bibinfo  {journal} {Phys.Lett.}\ }\textbf {\bibinfo {volume} {B560}},\
  \bibinfo {pages} {87} (\bibinfo {year} {2003})},\ \Eprint
  {http://arxiv.org/abs/hep-ph/0301092} {arXiv:hep-ph/0301092 [hep-ph]}
  \BibitemShut {NoStop}%
\bibitem [{\citenamefont {Hasegawa}\ \emph {et~al.}(2003)\citenamefont
  {Hasegawa}, \citenamefont {Lim},\ and\ \citenamefont
  {Ogure}}]{Hasegawa:2003by}%
  \BibitemOpen
  \bibfield  {author} {\bibinfo {author} {\bibfnamefont {K.}~\bibnamefont
  {Hasegawa}}, \bibinfo {author} {\bibfnamefont {C.}~\bibnamefont {Lim}}, \
  and\ \bibinfo {author} {\bibfnamefont {K.}~\bibnamefont {Ogure}},\ }\href
  {\doibase 10.1103/PhysRevD.68.053006} {\bibfield  {journal} {\bibinfo
  {journal} {Phys.Rev.}\ }\textbf {\bibinfo {volume} {D68}},\ \bibinfo {pages}
  {053006} (\bibinfo {year} {2003})},\ \Eprint
  {http://arxiv.org/abs/hep-ph/0303252} {arXiv:hep-ph/0303252 [hep-ph]}
  \BibitemShut {NoStop}%
\bibitem [{\citenamefont {Kanemura}\ \emph {et~al.}(2006)\citenamefont
  {Kanemura}, \citenamefont {Ota},\ and\ \citenamefont
  {Tsumura}}]{Kanemura:2005hr}%
  \BibitemOpen
  \bibfield  {author} {\bibinfo {author} {\bibfnamefont {S.}~\bibnamefont
  {Kanemura}}, \bibinfo {author} {\bibfnamefont {T.}~\bibnamefont {Ota}}, \
  and\ \bibinfo {author} {\bibfnamefont {K.}~\bibnamefont {Tsumura}},\ }\href
  {\doibase 10.1103/PhysRevD.73.016006} {\bibfield  {journal} {\bibinfo
  {journal} {Phys.Rev.}\ }\textbf {\bibinfo {volume} {D73}},\ \bibinfo {pages}
  {016006} (\bibinfo {year} {2006})},\ \Eprint
  {http://arxiv.org/abs/hep-ph/0505191} {arXiv:hep-ph/0505191 [hep-ph]}
  \BibitemShut {NoStop}%
\bibitem [{\citenamefont {Brahmachari}\ and\ \citenamefont
  {Choubey}(2006)}]{Brahmachari:2006qm}%
  \BibitemOpen
  \bibfield  {author} {\bibinfo {author} {\bibfnamefont {B.}~\bibnamefont
  {Brahmachari}}\ and\ \bibinfo {author} {\bibfnamefont {S.}~\bibnamefont
  {Choubey}},\ }\href {\doibase 10.1016/j.physletb.2006.10.007} {\bibfield
  {journal} {\bibinfo  {journal} {Phys.Lett.}\ }\textbf {\bibinfo {volume}
  {B642}},\ \bibinfo {pages} {495} (\bibinfo {year} {2006})},\ \Eprint
  {http://arxiv.org/abs/hep-ph/0608089} {arXiv:hep-ph/0608089 [hep-ph]}
  \BibitemShut {NoStop}%
\bibitem [{\citenamefont {Sahu}\ and\ \citenamefont
  {Sarkar}(2008)}]{Sahu:2008aw}%
  \BibitemOpen
  \bibfield  {author} {\bibinfo {author} {\bibfnamefont {N.}~\bibnamefont
  {Sahu}}\ and\ \bibinfo {author} {\bibfnamefont {U.}~\bibnamefont {Sarkar}},\
  }\href {\doibase 10.1103/PhysRevD.78.115013} {\bibfield  {journal} {\bibinfo
  {journal} {Phys.Rev.}\ }\textbf {\bibinfo {volume} {D78}},\ \bibinfo {pages}
  {115013} (\bibinfo {year} {2008})},\ \Eprint {http://arxiv.org/abs/0804.2072}
  {arXiv:0804.2072 [hep-ph]} \BibitemShut {NoStop}%
\bibitem [{\citenamefont {Fukuyama}\ \emph {et~al.}(2011)\citenamefont
  {Fukuyama}, \citenamefont {Sugiyama},\ and\ \citenamefont
  {Tsumura}}]{Fukuyama:2010ff}%
  \BibitemOpen
  \bibfield  {author} {\bibinfo {author} {\bibfnamefont {T.}~\bibnamefont
  {Fukuyama}}, \bibinfo {author} {\bibfnamefont {H.}~\bibnamefont {Sugiyama}},
  \ and\ \bibinfo {author} {\bibfnamefont {K.}~\bibnamefont {Tsumura}},\ }\href
  {\doibase 10.1103/PhysRevD.83.056016} {\bibfield  {journal} {\bibinfo
  {journal} {Phys.Rev.}\ }\textbf {\bibinfo {volume} {D83}},\ \bibinfo {pages}
  {056016} (\bibinfo {year} {2011})},\ \Eprint {http://arxiv.org/abs/1012.4886}
  {arXiv:1012.4886 [hep-ph]} \BibitemShut {NoStop}%
\bibitem [{\citenamefont {del Aguila}\ \emph {et~al.}(2012)\citenamefont {del
  Aguila}, \citenamefont {Aparici}, \citenamefont {Bhattacharya}, \citenamefont
  {Santamaria},\ and\ \citenamefont {Wudka}}]{delAguila:2012nu}%
  \BibitemOpen
  \bibfield  {author} {\bibinfo {author} {\bibfnamefont {F.}~\bibnamefont {del
  Aguila}}, \bibinfo {author} {\bibfnamefont {A.}~\bibnamefont {Aparici}},
  \bibinfo {author} {\bibfnamefont {S.}~\bibnamefont {Bhattacharya}}, \bibinfo
  {author} {\bibfnamefont {A.}~\bibnamefont {Santamaria}}, \ and\ \bibinfo
  {author} {\bibfnamefont {J.}~\bibnamefont {Wudka}},\ }\href {\doibase
  10.1007/JHEP06(2012)146} {\bibfield  {journal} {\bibinfo  {journal} {JHEP}\
  }\textbf {\bibinfo {volume} {1206}},\ \bibinfo {pages} {146} (\bibinfo {year}
  {2012})},\ \Eprint {http://arxiv.org/abs/1204.5986} {arXiv:1204.5986
  [hep-ph]} \BibitemShut {NoStop}%
\bibitem [{\citenamefont {Wolfenstein}(1980)}]{Wolfenstein:1980sy}%
  \BibitemOpen
  \bibfield  {author} {\bibinfo {author} {\bibfnamefont {L.}~\bibnamefont
  {Wolfenstein}},\ }\href {\doibase 10.1016/0550-3213(80)90004-8} {\bibfield
  {journal} {\bibinfo  {journal} {Nucl.Phys.}\ }\textbf {\bibinfo {volume}
  {B175}},\ \bibinfo {pages} {93} (\bibinfo {year} {1980})}\BibitemShut
  {NoStop}%
\bibitem [{\citenamefont {Frampton}\ \emph {et~al.}(2002)\citenamefont
  {Frampton}, \citenamefont {Oh},\ and\ \citenamefont
  {Yoshikawa}}]{Frampton:2001eu}%
  \BibitemOpen
  \bibfield  {author} {\bibinfo {author} {\bibfnamefont {P.~H.}\ \bibnamefont
  {Frampton}}, \bibinfo {author} {\bibfnamefont {M.~C.}\ \bibnamefont {Oh}}, \
  and\ \bibinfo {author} {\bibfnamefont {T.}~\bibnamefont {Yoshikawa}},\ }\href
  {\doibase 10.1103/PhysRevD.65.073014} {\bibfield  {journal} {\bibinfo
  {journal} {Phys.Rev.}\ }\textbf {\bibinfo {volume} {D65}},\ \bibinfo {pages}
  {073014} (\bibinfo {year} {2002})},\ \Eprint
  {http://arxiv.org/abs/hep-ph/0110300} {arXiv:hep-ph/0110300 [hep-ph]}
  \BibitemShut {NoStop}%
\bibitem [{\citenamefont {He}(2004)}]{He:2003ih}%
  \BibitemOpen
  \bibfield  {author} {\bibinfo {author} {\bibfnamefont {X.-G.}\ \bibnamefont
  {He}},\ }\href {\doibase 10.1140/epjc/s2004-01669-8} {\bibfield  {journal}
  {\bibinfo  {journal} {Eur.Phys.J.}\ }\textbf {\bibinfo {volume} {C34}},\
  \bibinfo {pages} {371} (\bibinfo {year} {2004})},\ \Eprint
  {http://arxiv.org/abs/hep-ph/0307172} {arXiv:hep-ph/0307172 [hep-ph]}
  \BibitemShut {NoStop}%
\bibitem [{\citenamefont {Kanemura}\ \emph {et~al.}(2001)\citenamefont
  {Kanemura}, \citenamefont {Kasai}, \citenamefont {Lin}, \citenamefont
  {Okada}, \citenamefont {Tseng} \emph {et~al.}}]{Kanemura:2000bq}%
  \BibitemOpen
  \bibfield  {author} {\bibinfo {author} {\bibfnamefont {S.}~\bibnamefont
  {Kanemura}}, \bibinfo {author} {\bibfnamefont {T.}~\bibnamefont {Kasai}},
  \bibinfo {author} {\bibfnamefont {G.-L.}\ \bibnamefont {Lin}}, \bibinfo
  {author} {\bibfnamefont {Y.}~\bibnamefont {Okada}}, \bibinfo {author}
  {\bibfnamefont {J.-J.}\ \bibnamefont {Tseng}},  \emph {et~al.},\ }\href
  {\doibase 10.1103/PhysRevD.64.053007} {\bibfield  {journal} {\bibinfo
  {journal} {Phys.Rev.}\ }\textbf {\bibinfo {volume} {D64}},\ \bibinfo {pages}
  {053007} (\bibinfo {year} {2001})},\ \Eprint
  {http://arxiv.org/abs/hep-ph/0011357} {arXiv:hep-ph/0011357 [hep-ph]}
  \BibitemShut {NoStop}%
\bibitem [{\citenamefont {Assamagan}\ \emph {et~al.}(2003)\citenamefont
  {Assamagan}, \citenamefont {Deandrea},\ and\ \citenamefont
  {Delsart}}]{Assamagan:2002kf}%
  \BibitemOpen
  \bibfield  {author} {\bibinfo {author} {\bibfnamefont {K.~A.}\ \bibnamefont
  {Assamagan}}, \bibinfo {author} {\bibfnamefont {A.}~\bibnamefont {Deandrea}},
  \ and\ \bibinfo {author} {\bibfnamefont {P.-A.}\ \bibnamefont {Delsart}},\
  }\href {\doibase 10.1103/PhysRevD.67.035001} {\bibfield  {journal} {\bibinfo
  {journal} {Phys.Rev.}\ }\textbf {\bibinfo {volume} {D67}},\ \bibinfo {pages}
  {035001} (\bibinfo {year} {2003})},\ \Eprint
  {http://arxiv.org/abs/hep-ph/0207302} {arXiv:hep-ph/0207302 [hep-ph]}
  \BibitemShut {NoStop}%
\bibitem [{\citenamefont {Babu}\ and\ \citenamefont
  {Julio}(2013)}]{Babu:2013pma}%
  \BibitemOpen
  \bibfield  {author} {\bibinfo {author} {\bibfnamefont {K.}~\bibnamefont
  {Babu}}\ and\ \bibinfo {author} {\bibfnamefont {J.}~\bibnamefont {Julio}},\
  }\href@noop {} {\  (\bibinfo {year} {2013})},\ \Eprint
  {http://arxiv.org/abs/1310.0303} {arXiv:1310.0303 [hep-ph]} \BibitemShut
  {NoStop}%
\bibitem [{\citenamefont {Aranda}\ \emph {et~al.}(2012)\citenamefont {Aranda},
  \citenamefont {Bonilla},\ and\ \citenamefont {Rojas}}]{Aranda:2011rt}%
  \BibitemOpen
  \bibfield  {author} {\bibinfo {author} {\bibfnamefont {A.}~\bibnamefont
  {Aranda}}, \bibinfo {author} {\bibfnamefont {C.}~\bibnamefont {Bonilla}}, \
  and\ \bibinfo {author} {\bibfnamefont {A.~D.}\ \bibnamefont {Rojas}},\ }\href
  {\doibase 10.1103/PhysRevD.85.036004} {\bibfield  {journal} {\bibinfo
  {journal} {Phys.Rev.}\ }\textbf {\bibinfo {volume} {D85}},\ \bibinfo {pages}
  {036004} (\bibinfo {year} {2012})},\ \Eprint {http://arxiv.org/abs/1110.1182}
  {arXiv:1110.1182 [hep-ph]} \BibitemShut {NoStop}%
\bibitem [{\citenamefont {Ma}(2006)}]{Ma:2006km}%
  \BibitemOpen
  \bibfield  {author} {\bibinfo {author} {\bibfnamefont {E.}~\bibnamefont
  {Ma}},\ }\href@noop {} {\bibfield  {journal} {\bibinfo  {journal}
  {Phys.Rev.}\ }\textbf {\bibinfo {volume} {D73}},\ \bibinfo {pages} {077301}
  (\bibinfo {year} {2006})},\ \Eprint {http://arxiv.org/abs/hep-ph/0601225}
  {arXiv:hep-ph/0601225 [hep-ph]} \BibitemShut {NoStop}%
\bibitem [{\citenamefont {Kubo}\ \emph {et~al.}(2006)\citenamefont {Kubo},
  \citenamefont {Ma},\ and\ \citenamefont {Suematsu}}]{Kubo:2006yx}%
  \BibitemOpen
  \bibfield  {author} {\bibinfo {author} {\bibfnamefont {J.}~\bibnamefont
  {Kubo}}, \bibinfo {author} {\bibfnamefont {E.}~\bibnamefont {Ma}}, \ and\
  \bibinfo {author} {\bibfnamefont {D.}~\bibnamefont {Suematsu}},\ }\href
  {\doibase 10.1016/j.physletb.2006.08.085} {\bibfield  {journal} {\bibinfo
  {journal} {Phys.Lett.}\ }\textbf {\bibinfo {volume} {B642}},\ \bibinfo
  {pages} {18} (\bibinfo {year} {2006})},\ \Eprint
  {http://arxiv.org/abs/hep-ph/0604114} {arXiv:hep-ph/0604114 [hep-ph]}
  \BibitemShut {NoStop}%
\bibitem [{\citenamefont {Aristizabal~Sierra}\ \emph
  {et~al.}(2009)\citenamefont {Aristizabal~Sierra}, \citenamefont {Kubo},
  \citenamefont {Restrepo}, \citenamefont {Suematsu},\ and\ \citenamefont
  {Zapata}}]{Sierra:2008wj}%
  \BibitemOpen
  \bibfield  {author} {\bibinfo {author} {\bibfnamefont {D.}~\bibnamefont
  {Aristizabal~Sierra}}, \bibinfo {author} {\bibfnamefont {J.}~\bibnamefont
  {Kubo}}, \bibinfo {author} {\bibfnamefont {D.}~\bibnamefont {Restrepo}},
  \bibinfo {author} {\bibfnamefont {D.}~\bibnamefont {Suematsu}}, \ and\
  \bibinfo {author} {\bibfnamefont {O.}~\bibnamefont {Zapata}},\ }\href
  {\doibase 10.1103/PhysRevD.79.013011} {\bibfield  {journal} {\bibinfo
  {journal} {Phys.Rev.}\ }\textbf {\bibinfo {volume} {D79}},\ \bibinfo {pages}
  {013011} (\bibinfo {year} {2009})},\ \Eprint {http://arxiv.org/abs/0808.3340}
  {arXiv:0808.3340 [hep-ph]} \BibitemShut {NoStop}%
\bibitem [{\citenamefont {Suematsu}\ \emph {et~al.}(2009)\citenamefont
  {Suematsu}, \citenamefont {Toma},\ and\ \citenamefont
  {Yoshida}}]{Suematsu:2009ww}%
  \BibitemOpen
  \bibfield  {author} {\bibinfo {author} {\bibfnamefont {D.}~\bibnamefont
  {Suematsu}}, \bibinfo {author} {\bibfnamefont {T.}~\bibnamefont {Toma}}, \
  and\ \bibinfo {author} {\bibfnamefont {T.}~\bibnamefont {Yoshida}},\ }\href
  {\doibase 10.1103/PhysRevD.79.093004} {\bibfield  {journal} {\bibinfo
  {journal} {Phys.Rev.}\ }\textbf {\bibinfo {volume} {D79}},\ \bibinfo {pages}
  {093004} (\bibinfo {year} {2009})},\ \Eprint {http://arxiv.org/abs/0903.0287}
  {arXiv:0903.0287 [hep-ph]} \BibitemShut {NoStop}%
\bibitem [{\citenamefont {Adulpravitchai}\ \emph {et~al.}(2009)\citenamefont
  {Adulpravitchai}, \citenamefont {Lindner},\ and\ \citenamefont
  {Merle}}]{Adulpravitchai:2009gi}%
  \BibitemOpen
  \bibfield  {author} {\bibinfo {author} {\bibfnamefont {A.}~\bibnamefont
  {Adulpravitchai}}, \bibinfo {author} {\bibfnamefont {M.}~\bibnamefont
  {Lindner}}, \ and\ \bibinfo {author} {\bibfnamefont {A.}~\bibnamefont
  {Merle}},\ }\href {\doibase 10.1103/PhysRevD.80.055031} {\bibfield  {journal}
  {\bibinfo  {journal} {Phys.Rev.}\ }\textbf {\bibinfo {volume} {D80}},\
  \bibinfo {pages} {055031} (\bibinfo {year} {2009})},\ \Eprint
  {http://arxiv.org/abs/0907.2147} {arXiv:0907.2147 [hep-ph]} \BibitemShut
  {NoStop}%
\bibitem [{\citenamefont {Toma}\ and\ \citenamefont
  {Vicente}(2014)}]{Toma:2013zsa}%
  \BibitemOpen
  \bibfield  {author} {\bibinfo {author} {\bibfnamefont {T.}~\bibnamefont
  {Toma}}\ and\ \bibinfo {author} {\bibfnamefont {A.}~\bibnamefont {Vicente}},\
  }\href {\doibase 10.1007/JHEP01(2014)160} {\bibfield  {journal} {\bibinfo
  {journal} {JHEP}\ }\textbf {\bibinfo {volume} {1401}},\ \bibinfo {pages}
  {160} (\bibinfo {year} {2014})},\ \Eprint {http://arxiv.org/abs/1312.2840}
  {arXiv:1312.2840} \BibitemShut {NoStop}%
\bibitem [{\citenamefont {Schmidt}\ \emph {et~al.}(2012)\citenamefont
  {Schmidt}, \citenamefont {Schwetz},\ and\ \citenamefont
  {Toma}}]{Schmidt:2012yg}%
  \BibitemOpen
  \bibfield  {author} {\bibinfo {author} {\bibfnamefont {D.}~\bibnamefont
  {Schmidt}}, \bibinfo {author} {\bibfnamefont {T.}~\bibnamefont {Schwetz}}, \
  and\ \bibinfo {author} {\bibfnamefont {T.}~\bibnamefont {Toma}},\ }\href@noop
  {} {\bibfield  {journal} {\bibinfo  {journal} {Phys.Rev.}\ }\textbf {\bibinfo
  {volume} {D85}},\ \bibinfo {pages} {073009} (\bibinfo {year} {2012})},\
  \Eprint {http://arxiv.org/abs/1201.0906} {arXiv:1201.0906 [hep-ph]}
  \BibitemShut {NoStop}%
\bibitem [{\citenamefont {Bouchand}\ and\ \citenamefont
  {Merle}(2012)}]{Bouchand:2012dx}%
  \BibitemOpen
  \bibfield  {author} {\bibinfo {author} {\bibfnamefont {R.}~\bibnamefont
  {Bouchand}}\ and\ \bibinfo {author} {\bibfnamefont {A.}~\bibnamefont
  {Merle}},\ }\href {\doibase 10.1007/JHEP07(2012)084} {\bibfield  {journal}
  {\bibinfo  {journal} {JHEP}\ }\textbf {\bibinfo {volume} {1207}},\ \bibinfo
  {pages} {084} (\bibinfo {year} {2012})},\ \Eprint
  {http://arxiv.org/abs/1205.0008} {arXiv:1205.0008 [hep-ph]} \BibitemShut
  {NoStop}%
\bibitem [{\citenamefont {Aoki}\ and\ \citenamefont
  {Kanemura}(2010)}]{Aoki:2010tf}%
  \BibitemOpen
  \bibfield  {author} {\bibinfo {author} {\bibfnamefont {M.}~\bibnamefont
  {Aoki}}\ and\ \bibinfo {author} {\bibfnamefont {S.}~\bibnamefont
  {Kanemura}},\ }\href {\doibase 10.1016/j.physletb.2010.04.024} {\bibfield
  {journal} {\bibinfo  {journal} {Phys.Lett.}\ }\textbf {\bibinfo {volume}
  {B689}},\ \bibinfo {pages} {28} (\bibinfo {year} {2010})},\ \Eprint
  {http://arxiv.org/abs/1001.0092} {arXiv:1001.0092 [hep-ph]} \BibitemShut
  {NoStop}%
\bibitem [{\citenamefont {Aoki}\ \emph {et~al.}(2013)\citenamefont {Aoki},
  \citenamefont {Kanemura},\ and\ \citenamefont {Yokoya}}]{Aoki:2013lhm}%
  \BibitemOpen
  \bibfield  {author} {\bibinfo {author} {\bibfnamefont {M.}~\bibnamefont
  {Aoki}}, \bibinfo {author} {\bibfnamefont {S.}~\bibnamefont {Kanemura}}, \
  and\ \bibinfo {author} {\bibfnamefont {H.}~\bibnamefont {Yokoya}},\ }\href
  {\doibase 10.1016/j.physletb.2013.07.011} {\bibfield  {journal} {\bibinfo
  {journal} {Phys.Lett.}\ }\textbf {\bibinfo {volume} {B725}},\ \bibinfo
  {pages} {302} (\bibinfo {year} {2013})},\ \Eprint
  {http://arxiv.org/abs/1303.6191} {arXiv:1303.6191 [hep-ph]} \BibitemShut
  {NoStop}%
\bibitem [{\citenamefont {Ho}\ and\ \citenamefont
  {Tandean}(2014)}]{Ho:2013spa}%
  \BibitemOpen
  \bibfield  {author} {\bibinfo {author} {\bibfnamefont {S.-Y.}\ \bibnamefont
  {Ho}}\ and\ \bibinfo {author} {\bibfnamefont {J.}~\bibnamefont {Tandean}},\
  }\href {\doibase 10.1103/PhysRevD.89.114025} {\bibfield  {journal} {\bibinfo
  {journal} {Phys.Rev.}\ }\textbf {\bibinfo {volume} {D89}},\ \bibinfo {pages}
  {114025} (\bibinfo {year} {2014})},\ \Eprint {http://arxiv.org/abs/1312.0931}
  {arXiv:1312.0931 [hep-ph]} \BibitemShut {NoStop}%
\bibitem [{\citenamefont {Ho}\ and\ \citenamefont
  {Tandean}(2013)}]{Ho:2013hia}%
  \BibitemOpen
  \bibfield  {author} {\bibinfo {author} {\bibfnamefont {S.-Y.}\ \bibnamefont
  {Ho}}\ and\ \bibinfo {author} {\bibfnamefont {J.}~\bibnamefont {Tandean}},\
  }\href {\doibase 10.1103/PhysRevD.87.095015} {\bibfield  {journal} {\bibinfo
  {journal} {Phys.Rev.}\ }\textbf {\bibinfo {volume} {D87}},\ \bibinfo {pages}
  {095015} (\bibinfo {year} {2013})},\ \Eprint {http://arxiv.org/abs/1303.5700}
  {arXiv:1303.5700 [hep-ph]} \BibitemShut {NoStop}%
\bibitem [{\citenamefont {Kajiyama}\ \emph {et~al.}(2007)\citenamefont
  {Kajiyama}, \citenamefont {Kubo},\ and\ \citenamefont
  {Okada}}]{Kajiyama:2006ww}%
  \BibitemOpen
  \bibfield  {author} {\bibinfo {author} {\bibfnamefont {Y.}~\bibnamefont
  {Kajiyama}}, \bibinfo {author} {\bibfnamefont {J.}~\bibnamefont {Kubo}}, \
  and\ \bibinfo {author} {\bibfnamefont {H.}~\bibnamefont {Okada}},\ }\href
  {\doibase 10.1103/PhysRevD.75.033001} {\bibfield  {journal} {\bibinfo
  {journal} {Phys.Rev.}\ }\textbf {\bibinfo {volume} {D75}},\ \bibinfo {pages}
  {033001} (\bibinfo {year} {2007})},\ \Eprint
  {http://arxiv.org/abs/hep-ph/0610072} {arXiv:hep-ph/0610072 [hep-ph]}
  \BibitemShut {NoStop}%
\bibitem [{\citenamefont {Suematsu}\ \emph {et~al.}(2010)\citenamefont
  {Suematsu}, \citenamefont {Toma},\ and\ \citenamefont
  {Yoshida}}]{Suematsu:2010gv}%
  \BibitemOpen
  \bibfield  {author} {\bibinfo {author} {\bibfnamefont {D.}~\bibnamefont
  {Suematsu}}, \bibinfo {author} {\bibfnamefont {T.}~\bibnamefont {Toma}}, \
  and\ \bibinfo {author} {\bibfnamefont {T.}~\bibnamefont {Yoshida}},\ }\href
  {\doibase 10.1103/PhysRevD.82.013012} {\bibfield  {journal} {\bibinfo
  {journal} {Phys.Rev.}\ }\textbf {\bibinfo {volume} {D82}},\ \bibinfo {pages}
  {013012} (\bibinfo {year} {2010})},\ \Eprint {http://arxiv.org/abs/1002.3225}
  {arXiv:1002.3225 [hep-ph]} \BibitemShut {NoStop}%
\bibitem [{\citenamefont {Kajiyama}\ \emph
  {et~al.}(2011{\natexlab{a}})\citenamefont {Kajiyama}, \citenamefont {Okada},\
  and\ \citenamefont {Toma}}]{Kajiyama:2011fe}%
  \BibitemOpen
  \bibfield  {author} {\bibinfo {author} {\bibfnamefont {Y.}~\bibnamefont
  {Kajiyama}}, \bibinfo {author} {\bibfnamefont {H.}~\bibnamefont {Okada}}, \
  and\ \bibinfo {author} {\bibfnamefont {T.}~\bibnamefont {Toma}},\ }\href
  {\doibase 10.1140/epjc/s10052-011-1688-0} {\bibfield  {journal} {\bibinfo
  {journal} {Eur.Phys.J.}\ }\textbf {\bibinfo {volume} {C71}},\ \bibinfo
  {pages} {1688} (\bibinfo {year} {2011}{\natexlab{a}})},\ \Eprint
  {http://arxiv.org/abs/1104.0367} {arXiv:1104.0367 [hep-ph]} \BibitemShut
  {NoStop}%
\bibitem [{\citenamefont {Kajiyama}\ \emph
  {et~al.}(2011{\natexlab{b}})\citenamefont {Kajiyama}, \citenamefont {Okada},\
  and\ \citenamefont {Toma}}]{Kajiyama:2011fx}%
  \BibitemOpen
  \bibfield  {author} {\bibinfo {author} {\bibfnamefont {Y.}~\bibnamefont
  {Kajiyama}}, \bibinfo {author} {\bibfnamefont {H.}~\bibnamefont {Okada}}, \
  and\ \bibinfo {author} {\bibfnamefont {T.}~\bibnamefont {Toma}},\ }\href@noop
  {} {\  (\bibinfo {year} {2011}{\natexlab{b}})},\ \Eprint
  {http://arxiv.org/abs/1109.2722} {arXiv:1109.2722 [hep-ph]} \BibitemShut
  {NoStop}%
\bibitem [{\citenamefont {Klasen}\ \emph {et~al.}(2013)\citenamefont {Klasen},
  \citenamefont {Yaguna}, \citenamefont {Ruiz-Alvarez}, \citenamefont
  {Restrepo},\ and\ \citenamefont {Zapata}}]{Klasen:2013jpa}%
  \BibitemOpen
  \bibfield  {author} {\bibinfo {author} {\bibfnamefont {M.}~\bibnamefont
  {Klasen}}, \bibinfo {author} {\bibfnamefont {C.~E.}\ \bibnamefont {Yaguna}},
  \bibinfo {author} {\bibfnamefont {J.~D.}\ \bibnamefont {Ruiz-Alvarez}},
  \bibinfo {author} {\bibfnamefont {D.}~\bibnamefont {Restrepo}}, \ and\
  \bibinfo {author} {\bibfnamefont {O.}~\bibnamefont {Zapata}},\ }\href
  {\doibase 10.1088/1475-7516/2013/04/044} {\bibfield  {journal} {\bibinfo
  {journal} {JCAP}\ }\textbf {\bibinfo {volume} {1304}},\ \bibinfo {pages}
  {044} (\bibinfo {year} {2013})},\ \Eprint {http://arxiv.org/abs/1302.5298}
  {arXiv:1302.5298 [hep-ph]} \BibitemShut {NoStop}%
\bibitem [{\citenamefont {Hu}(2012)}]{Hu:2012az}%
  \BibitemOpen
  \bibfield  {author} {\bibinfo {author} {\bibfnamefont {P.-K.}\ \bibnamefont
  {Hu}},\ }\href@noop {} {\  (\bibinfo {year} {2012})},\ \Eprint
  {http://arxiv.org/abs/1208.2613} {arXiv:1208.2613 [hep-ph]} \BibitemShut
  {NoStop}%
\bibitem [{\citenamefont {Hirsch}\ \emph {et~al.}(2013)\citenamefont {Hirsch},
  \citenamefont {Lineros}, \citenamefont {Morisi}, \citenamefont {Palacio},
  \citenamefont {Rojas} \emph {et~al.}}]{Hirsch:2013ola}%
  \BibitemOpen
  \bibfield  {author} {\bibinfo {author} {\bibfnamefont {M.}~\bibnamefont
  {Hirsch}}, \bibinfo {author} {\bibfnamefont {R.}~\bibnamefont {Lineros}},
  \bibinfo {author} {\bibfnamefont {S.}~\bibnamefont {Morisi}}, \bibinfo
  {author} {\bibfnamefont {J.}~\bibnamefont {Palacio}}, \bibinfo {author}
  {\bibfnamefont {N.}~\bibnamefont {Rojas}},  \emph {et~al.},\ }\href@noop {}
  {\bibfield  {journal} {\bibinfo  {journal} {JHEP}\ }\textbf {\bibinfo
  {volume} {1310}},\ \bibinfo {pages} {149} (\bibinfo {year} {2013})},\ \Eprint
  {http://arxiv.org/abs/1307.8134} {arXiv:1307.8134 [hep-ph]} \BibitemShut
  {NoStop}%
\bibitem [{\citenamefont {Brdar}\ \emph {et~al.}(2014)\citenamefont {Brdar},
  \citenamefont {Picek},\ and\ \citenamefont {Radovcic}}]{Brdar:2013iea}%
  \BibitemOpen
  \bibfield  {author} {\bibinfo {author} {\bibfnamefont {V.}~\bibnamefont
  {Brdar}}, \bibinfo {author} {\bibfnamefont {I.}~\bibnamefont {Picek}}, \ and\
  \bibinfo {author} {\bibfnamefont {B.}~\bibnamefont {Radovcic}},\ }\href
  {\doibase 10.1016/j.physletb.2013.11.045} {\bibfield  {journal} {\bibinfo
  {journal} {Phys.Lett.}\ }\textbf {\bibinfo {volume} {B728}},\ \bibinfo
  {pages} {198} (\bibinfo {year} {2014})},\ \Eprint
  {http://arxiv.org/abs/1310.3183} {arXiv:1310.3183 [hep-ph]} \BibitemShut
  {NoStop}%
\bibitem [{\citenamefont {Restrepo}\ \emph {et~al.}(2013)\citenamefont
  {Restrepo}, \citenamefont {Zapata},\ and\ \citenamefont
  {Yaguna}}]{Restrepo:2013aga}%
  \BibitemOpen
  \bibfield  {author} {\bibinfo {author} {\bibfnamefont {D.}~\bibnamefont
  {Restrepo}}, \bibinfo {author} {\bibfnamefont {O.}~\bibnamefont {Zapata}}, \
  and\ \bibinfo {author} {\bibfnamefont {C.~E.}\ \bibnamefont {Yaguna}},\
  }\href {\doibase 10.1007/JHEP11(2013)011} {\bibfield  {journal} {\bibinfo
  {journal} {JHEP}\ }\textbf {\bibinfo {volume} {1311}},\ \bibinfo {pages}
  {011} (\bibinfo {year} {2013})},\ \Eprint {http://arxiv.org/abs/1308.3655}
  {arXiv:1308.3655 [hep-ph]} \BibitemShut {NoStop}%
\bibitem [{\citenamefont {Law}\ and\ \citenamefont
  {McDonald}(2013)}]{Law:2013saa}%
  \BibitemOpen
  \bibfield  {author} {\bibinfo {author} {\bibfnamefont {S.~S.}\ \bibnamefont
  {Law}}\ and\ \bibinfo {author} {\bibfnamefont {K.~L.}\ \bibnamefont
  {McDonald}},\ }\href {\doibase 10.1007/JHEP09(2013)092} {\bibfield  {journal}
  {\bibinfo  {journal} {JHEP}\ }\textbf {\bibinfo {volume} {1309}},\ \bibinfo
  {pages} {092} (\bibinfo {year} {2013})},\ \Eprint
  {http://arxiv.org/abs/1305.6467} {arXiv:1305.6467 [hep-ph]} \BibitemShut
  {NoStop}%
\bibitem [{\citenamefont {Zee}(1986)}]{Zee:1985id}%
  \BibitemOpen
  \bibfield  {author} {\bibinfo {author} {\bibfnamefont {A.}~\bibnamefont
  {Zee}},\ }\href@noop {} {\bibfield  {journal} {\bibinfo  {journal} {Nucl.
  Phys.}\ }\textbf {\bibinfo {volume} {B264}},\ \bibinfo {pages} {99} (\bibinfo
  {year} {1986})}\BibitemShut {NoStop}%
\bibitem [{\citenamefont {Schechter}\ and\ \citenamefont
  {Valle}(1982{\natexlab{b}})}]{Schechter:1982bd}%
  \BibitemOpen
  \bibfield  {author} {\bibinfo {author} {\bibfnamefont {J.}~\bibnamefont
  {Schechter}}\ and\ \bibinfo {author} {\bibfnamefont {J.~W.~F.}\ \bibnamefont
  {Valle}},\ }\href@noop {} {\bibfield  {journal} {\bibinfo  {journal} {Phys.
  Rev.}\ }\textbf {\bibinfo {volume} {D25}},\ \bibinfo {pages} {2951} (\bibinfo
  {year} {1982}{\natexlab{b}})}\BibitemShut {NoStop}%
\bibitem [{\citenamefont {Peltoniemi}\ and\ \citenamefont
  {Valle}(1993)}]{peltoniemi:1993pd}%
  \BibitemOpen
  \bibfield  {author} {\bibinfo {author} {\bibfnamefont {J.~T.}\ \bibnamefont
  {Peltoniemi}}\ and\ \bibinfo {author} {\bibfnamefont {J.~W.~F.}\ \bibnamefont
  {Valle}},\ }\href@noop {} {\bibfield  {journal} {\bibinfo  {journal} {Phys.
  Lett.}\ }\textbf {\bibinfo {volume} {B304}},\ \bibinfo {pages} {147}
  (\bibinfo {year} {1993})},\ \Eprint {http://arxiv.org/abs/hep-ph/9301231}
  {hep-ph/9301231} \BibitemShut {NoStop}%
\bibitem [{\citenamefont {Nebot}\ \emph {et~al.}(2007)\citenamefont {Nebot},
  \citenamefont {Oliver}, \citenamefont {Palao},\ and\ \citenamefont
  {Santamaria}}]{Nebot:2007bc}%
  \BibitemOpen
  \bibfield  {author} {\bibinfo {author} {\bibfnamefont {M.}~\bibnamefont
  {Nebot}}, \bibinfo {author} {\bibfnamefont {J.~F.}\ \bibnamefont {Oliver}},
  \bibinfo {author} {\bibfnamefont {D.}~\bibnamefont {Palao}}, \ and\ \bibinfo
  {author} {\bibfnamefont {A.}~\bibnamefont {Santamaria}},\ }\href@noop {} {\
  (\bibinfo {year} {2007})},\ \Eprint {http://arxiv.org/abs/0711.0483}
  {arXiv:0711.0483 [hep-ph]} \BibitemShut {NoStop}%
\bibitem [{\citenamefont {Schmidt}\ \emph {et~al.}(2014)\citenamefont
  {Schmidt}, \citenamefont {Schwetz},\ and\ \citenamefont
  {Zhang}}]{Schmidt:2014zoa}%
  \BibitemOpen
  \bibfield  {author} {\bibinfo {author} {\bibfnamefont {D.}~\bibnamefont
  {Schmidt}}, \bibinfo {author} {\bibfnamefont {T.}~\bibnamefont {Schwetz}}, \
  and\ \bibinfo {author} {\bibfnamefont {H.}~\bibnamefont {Zhang}},\
  }\href@noop {} {\  (\bibinfo {year} {2014})},\ \Eprint
  {http://arxiv.org/abs/1402.2251} {arXiv:1402.2251 [hep-ph]} \BibitemShut
  {NoStop}%
\bibitem [{\citenamefont {Babu}\ and\ \citenamefont
  {Macesanu}(2003)}]{Babu:2002uu}%
  \BibitemOpen
  \bibfield  {author} {\bibinfo {author} {\bibfnamefont {K.~S.}\ \bibnamefont
  {Babu}}\ and\ \bibinfo {author} {\bibfnamefont {C.}~\bibnamefont
  {Macesanu}},\ }\href@noop {} {\bibfield  {journal} {\bibinfo  {journal}
  {Phys. Rev.}\ }\textbf {\bibinfo {volume} {D67}},\ \bibinfo {pages} {073010}
  (\bibinfo {year} {2003})},\ \Eprint {http://arxiv.org/abs/hep-ph/0212058}
  {hep-ph/0212058} \BibitemShut {NoStop}%
\bibitem [{\citenamefont {Aristizabal~Sierra}\ and\ \citenamefont
  {Hirsch}(2006)}]{AristizabalSierra:2006gb}%
  \BibitemOpen
  \bibfield  {author} {\bibinfo {author} {\bibfnamefont {D.}~\bibnamefont
  {Aristizabal~Sierra}}\ and\ \bibinfo {author} {\bibfnamefont
  {M.}~\bibnamefont {Hirsch}},\ }\href@noop {} {\bibfield  {journal} {\bibinfo
  {journal} {JHEP}\ }\textbf {\bibinfo {volume} {12}},\ \bibinfo {pages} {052}
  (\bibinfo {year} {2006})},\ \Eprint {http://arxiv.org/abs/hep-ph/0609307}
  {hep-ph/0609307} \BibitemShut {NoStop}%
\bibitem [{\citenamefont {Herrero-Garcia}\ \emph {et~al.}(2014)\citenamefont
  {Herrero-Garcia}, \citenamefont {Nebot}, \citenamefont {Rius},\ and\
  \citenamefont {Santamaria}}]{Herrero-Garcia:2014hfa}%
  \BibitemOpen
  \bibfield  {author} {\bibinfo {author} {\bibfnamefont {J.}~\bibnamefont
  {Herrero-Garcia}}, \bibinfo {author} {\bibfnamefont {M.}~\bibnamefont
  {Nebot}}, \bibinfo {author} {\bibfnamefont {N.}~\bibnamefont {Rius}}, \ and\
  \bibinfo {author} {\bibfnamefont {A.}~\bibnamefont {Santamaria}},\
  }\href@noop {} {\  (\bibinfo {year} {2014})},\ \Eprint
  {http://arxiv.org/abs/1402.4491} {arXiv:1402.4491 [hep-ph]} \BibitemShut
  {NoStop}%
\bibitem [{\citenamefont {Krauss}\ \emph {et~al.}(2003)\citenamefont {Krauss},
  \citenamefont {Nasri},\ and\ \citenamefont {Trodden}}]{Krauss:2002px}%
  \BibitemOpen
  \bibfield  {author} {\bibinfo {author} {\bibfnamefont {L.~M.}\ \bibnamefont
  {Krauss}}, \bibinfo {author} {\bibfnamefont {S.}~\bibnamefont {Nasri}}, \
  and\ \bibinfo {author} {\bibfnamefont {M.}~\bibnamefont {Trodden}},\ }\href
  {\doibase 10.1103/PhysRevD.67.085002} {\bibfield  {journal} {\bibinfo
  {journal} {Phys.Rev.}\ }\textbf {\bibinfo {volume} {D67}},\ \bibinfo {pages}
  {085002} (\bibinfo {year} {2003})},\ \Eprint
  {http://arxiv.org/abs/hep-ph/0210389} {arXiv:hep-ph/0210389 [hep-ph]}
  \BibitemShut {NoStop}%
\bibitem [{\citenamefont {Ng}\ and\ \citenamefont {de~la
  Puente}(2013)}]{Ng:2013xja}%
  \BibitemOpen
  \bibfield  {author} {\bibinfo {author} {\bibfnamefont {J.~N.}\ \bibnamefont
  {Ng}}\ and\ \bibinfo {author} {\bibfnamefont {A.}~\bibnamefont {de~la
  Puente}},\ }\href {\doibase 10.1016/j.physletb.2013.10.034} {\bibfield
  {journal} {\bibinfo  {journal} {Phys.Lett.}\ }\textbf {\bibinfo {volume}
  {B727}},\ \bibinfo {pages} {204} (\bibinfo {year} {2013})},\ \Eprint
  {http://arxiv.org/abs/1307.2606} {arXiv:1307.2606} \BibitemShut {NoStop}%
\bibitem [{\citenamefont {Ahriche}\ \emph {et~al.}(2014)\citenamefont
  {Ahriche}, \citenamefont {Chen}, \citenamefont {McDonald},\ and\
  \citenamefont {Nasri}}]{Ahriche:2014cda}%
  \BibitemOpen
  \bibfield  {author} {\bibinfo {author} {\bibfnamefont {A.}~\bibnamefont
  {Ahriche}}, \bibinfo {author} {\bibfnamefont {C.-S.}\ \bibnamefont {Chen}},
  \bibinfo {author} {\bibfnamefont {K.~L.}\ \bibnamefont {McDonald}}, \ and\
  \bibinfo {author} {\bibfnamefont {S.}~\bibnamefont {Nasri}},\ }\href@noop {}
  {\  (\bibinfo {year} {2014})},\ \Eprint {http://arxiv.org/abs/1404.2696}
  {arXiv:1404.2696 [hep-ph]} \BibitemShut {NoStop}%
\bibitem [{\citenamefont {Aoki}\ \emph
  {et~al.}(2009{\natexlab{a}})\citenamefont {Aoki}, \citenamefont {Kanemura},\
  and\ \citenamefont {Seto}}]{Aoki:2008av}%
  \BibitemOpen
  \bibfield  {author} {\bibinfo {author} {\bibfnamefont {M.}~\bibnamefont
  {Aoki}}, \bibinfo {author} {\bibfnamefont {S.}~\bibnamefont {Kanemura}}, \
  and\ \bibinfo {author} {\bibfnamefont {O.}~\bibnamefont {Seto}},\ }\href
  {\doibase 10.1103/PhysRevLett.102.051805} {\bibfield  {journal} {\bibinfo
  {journal} {Phys.Rev.Lett.}\ }\textbf {\bibinfo {volume} {102}},\ \bibinfo
  {pages} {051805} (\bibinfo {year} {2009}{\natexlab{a}})},\ \Eprint
  {http://arxiv.org/abs/0807.0361} {arXiv:0807.0361 [hep-ph]} \BibitemShut
  {NoStop}%
\bibitem [{\citenamefont {Aoki}\ \emph
  {et~al.}(2009{\natexlab{b}})\citenamefont {Aoki}, \citenamefont {Kanemura},\
  and\ \citenamefont {Seto}}]{Aoki:2009vf}%
  \BibitemOpen
  \bibfield  {author} {\bibinfo {author} {\bibfnamefont {M.}~\bibnamefont
  {Aoki}}, \bibinfo {author} {\bibfnamefont {S.}~\bibnamefont {Kanemura}}, \
  and\ \bibinfo {author} {\bibfnamefont {O.}~\bibnamefont {Seto}},\ }\href
  {\doibase 10.1103/PhysRevD.80.033007} {\bibfield  {journal} {\bibinfo
  {journal} {Phys.Rev.}\ }\textbf {\bibinfo {volume} {D80}},\ \bibinfo {pages}
  {033007} (\bibinfo {year} {2009}{\natexlab{b}})},\ \Eprint
  {http://arxiv.org/abs/0904.3829} {arXiv:0904.3829 [hep-ph]} \BibitemShut
  {NoStop}%
\bibitem [{\citenamefont {Gustafsson}\ \emph {et~al.}(2013)\citenamefont
  {Gustafsson}, \citenamefont {No},\ and\ \citenamefont
  {Rivera}}]{Gustafsson:2012vj}%
  \BibitemOpen
  \bibfield  {author} {\bibinfo {author} {\bibfnamefont {M.}~\bibnamefont
  {Gustafsson}}, \bibinfo {author} {\bibfnamefont {J.~M.}\ \bibnamefont {No}},
  \ and\ \bibinfo {author} {\bibfnamefont {M.~A.}\ \bibnamefont {Rivera}},\
  }\href {\doibase 10.1103/PhysRevLett.110.211802} {\bibfield  {journal}
  {\bibinfo  {journal} {Phys.Rev.Lett.}\ }\textbf {\bibinfo {volume} {110}},\
  \bibinfo {pages} {211802} (\bibinfo {year} {2013})},\ \Eprint
  {http://arxiv.org/abs/1212.4806} {arXiv:1212.4806 [hep-ph]} \BibitemShut
  {NoStop}%
\bibitem [{\citenamefont {Chen}\ \emph {et~al.}(2014)\citenamefont {Chen},
  \citenamefont {McDonald},\ and\ \citenamefont {Nasri}}]{Chen:2014ska}%
  \BibitemOpen
  \bibfield  {author} {\bibinfo {author} {\bibfnamefont {C.-S.}\ \bibnamefont
  {Chen}}, \bibinfo {author} {\bibfnamefont {K.~L.}\ \bibnamefont {McDonald}},
  \ and\ \bibinfo {author} {\bibfnamefont {S.}~\bibnamefont {Nasri}},\
  }\href@noop {} {\  (\bibinfo {year} {2014})},\ \Eprint
  {http://arxiv.org/abs/1404.6033} {arXiv:1404.6033 [hep-ph]} \BibitemShut
  {NoStop}%
\bibitem [{\citenamefont {Martin}(1997)}]{martin:1997ns}%
  \BibitemOpen
  \bibfield  {author} {\bibinfo {author} {\bibfnamefont {S.~P.}\ \bibnamefont
  {Martin}},\ }\href@noop {} {\  (\bibinfo {year} {1997})},\ \Eprint
  {http://arxiv.org/abs/hep-ph/9709356} {hep-ph/9709356} \BibitemShut {NoStop}%
\bibitem [{\citenamefont {Barbier}\ \emph {et~al.}(2005)\citenamefont {Barbier}
  \emph {et~al.}}]{barbier:2004ez}%
  \BibitemOpen
  \bibfield  {author} {\bibinfo {author} {\bibfnamefont {R.}~\bibnamefont
  {Barbier}} \emph {et~al.},\ }\href@noop {} {\bibfield  {journal} {\bibinfo
  {journal} {Phys. Rept.}\ }\textbf {\bibinfo {volume} {420}},\ \bibinfo
  {pages} {1} (\bibinfo {year} {2005})},\ \Eprint
  {http://arxiv.org/abs/hep-ph/0406039} {arXiv:hep-ph/0406039} \BibitemShut
  {NoStop}%
\bibitem [{\citenamefont {Barger}\ \emph {et~al.}(1989)\citenamefont {Barger},
  \citenamefont {Giudice},\ and\ \citenamefont {Han}}]{barger:1989rk}%
  \BibitemOpen
  \bibfield  {author} {\bibinfo {author} {\bibfnamefont {V.~D.}\ \bibnamefont
  {Barger}}, \bibinfo {author} {\bibfnamefont {G.~F.}\ \bibnamefont {Giudice}},
  \ and\ \bibinfo {author} {\bibfnamefont {T.}~\bibnamefont {Han}},\
  }\href@noop {} {\bibfield  {journal} {\bibinfo  {journal} {Phys. Rev.}\
  }\textbf {\bibinfo {volume} {D40}},\ \bibinfo {pages} {2987} (\bibinfo {year}
  {1989})}\BibitemShut {NoStop}%
\bibitem [{\citenamefont {Baer}\ and\ \citenamefont
  {Tata}(2006)}]{Baer:2006rs}%
  \BibitemOpen
  \bibfield  {author} {\bibinfo {author} {\bibfnamefont {H.}~\bibnamefont
  {Baer}}\ and\ \bibinfo {author} {\bibfnamefont {X.}~\bibnamefont {Tata}},\
  }\href@noop {} {\emph {\bibinfo {title} {{Weak scale supersymmetry: From
  superfields to scattering events}}}}\ (\bibinfo  {publisher} {Cambridge
  University Press, UK},\ \bibinfo {year} {2006})\BibitemShut {NoStop}%
\bibitem [{\citenamefont {Nogueira}\ \emph {et~al.}(1990)\citenamefont
  {Nogueira}, \citenamefont {Romao},\ and\ \citenamefont
  {Valle}}]{Nogueira:1990wz}%
  \BibitemOpen
  \bibfield  {author} {\bibinfo {author} {\bibfnamefont {P.}~\bibnamefont
  {Nogueira}}, \bibinfo {author} {\bibfnamefont {J.}~\bibnamefont {Romao}}, \
  and\ \bibinfo {author} {\bibfnamefont {J.}~\bibnamefont {Valle}},\ }\href
  {\doibase 10.1016/0370-2693(90)90244-Z} {\bibfield  {journal} {\bibinfo
  {journal} {Phys.Lett.}\ }\textbf {\bibinfo {volume} {B251}},\ \bibinfo
  {pages} {142} (\bibinfo {year} {1990})}\BibitemShut {NoStop}%
\bibitem [{\citenamefont {Hall}\ and\ \citenamefont
  {Suzuki}(1984)}]{Hall:1983id}%
  \BibitemOpen
  \bibfield  {author} {\bibinfo {author} {\bibfnamefont {L.~J.}\ \bibnamefont
  {Hall}}\ and\ \bibinfo {author} {\bibfnamefont {M.}~\bibnamefont {Suzuki}},\
  }\href@noop {} {\bibfield  {journal} {\bibinfo  {journal} {Nucl. Phys.}\
  }\textbf {\bibinfo {volume} {B231}},\ \bibinfo {pages} {419} (\bibinfo {year}
  {1984})}\BibitemShut {NoStop}%
\bibitem [{\citenamefont {Diaz}\ \emph {et~al.}(1998)\citenamefont {Diaz},
  \citenamefont {Romao},\ and\ \citenamefont {Valle}}]{diaz:1997xc}%
  \BibitemOpen
  \bibfield  {author} {\bibinfo {author} {\bibfnamefont {M.~A.}\ \bibnamefont
  {Diaz}}, \bibinfo {author} {\bibfnamefont {J.~C.}\ \bibnamefont {Romao}}, \
  and\ \bibinfo {author} {\bibfnamefont {J.~W.~F.}\ \bibnamefont {Valle}},\
  }\href@noop {} {\bibfield  {journal} {\bibinfo  {journal} {Nucl. Phys.}\
  }\textbf {\bibinfo {volume} {B524}},\ \bibinfo {pages} {23} (\bibinfo {year}
  {1998})}\BibitemShut {NoStop}%
\bibitem [{\citenamefont {Romao}\ \emph {et~al.}(1997)\citenamefont {Romao},
  \citenamefont {Ioannisian},\ and\ \citenamefont {Valle}}]{romao:1997xf}%
  \BibitemOpen
  \bibfield  {author} {\bibinfo {author} {\bibfnamefont {J.~C.}\ \bibnamefont
  {Romao}}, \bibinfo {author} {\bibfnamefont {A.}~\bibnamefont {Ioannisian}}, \
  and\ \bibinfo {author} {\bibfnamefont {J.~W.~F.}\ \bibnamefont {Valle}},\
  }\href@noop {} {\bibfield  {journal} {\bibinfo  {journal} {Phys. Rev.}\
  }\textbf {\bibinfo {volume} {D55}},\ \bibinfo {pages} {427} (\bibinfo {year}
  {1997})},\ \Eprint {http://arxiv.org/abs/hep-ph/9607401} {hep-ph/9607401}
  \BibitemShut {NoStop}%
\bibitem [{\citenamefont {Hirsch}\ \emph {et~al.}(2000)\citenamefont {Hirsch}
  \emph {et~al.}}]{Hirsch:2000ef}%
  \BibitemOpen
  \bibfield  {author} {\bibinfo {author} {\bibfnamefont {M.}~\bibnamefont
  {Hirsch}} \emph {et~al.},\ }\href@noop {} {\bibfield  {journal} {\bibinfo
  {journal} {Phys. Rev.}\ }\textbf {\bibinfo {volume} {D62}},\ \bibinfo {pages}
  {113008} (\bibinfo {year} {2000})},\ \bibinfo {note} {err-ibid. {\bf
  D65}:119901,2002},\ \Eprint {http://arxiv.org/abs/hep-ph/0004115}
  {hep-ph/0004115} \BibitemShut {NoStop}%
\bibitem [{\citenamefont {Diaz}\ \emph {et~al.}(2003)\citenamefont {Diaz} \emph
  {et~al.}}]{Diaz:2003as}%
  \BibitemOpen
  \bibfield  {author} {\bibinfo {author} {\bibfnamefont {M.~A.}\ \bibnamefont
  {Diaz}} \emph {et~al.},\ }\href@noop {} {\bibfield  {journal} {\bibinfo
  {journal} {Phys. Rev.}\ }\textbf {\bibinfo {volume} {D68}},\ \bibinfo {pages}
  {013009} (\bibinfo {year} {2003})},\ \Eprint
  {http://arxiv.org/abs/hep-ph/0302021} {hep-ph/0302021} \BibitemShut {NoStop}%
\bibitem [{\citenamefont {Chun}\ and\ \citenamefont
  {Kang}(2000)}]{Chun:1999bq}%
  \BibitemOpen
  \bibfield  {author} {\bibinfo {author} {\bibfnamefont {E.~J.}\ \bibnamefont
  {Chun}}\ and\ \bibinfo {author} {\bibfnamefont {S.~K.}\ \bibnamefont
  {Kang}},\ }\href@noop {} {\bibfield  {journal} {\bibinfo  {journal} {Phys.
  Rev.}\ }\textbf {\bibinfo {volume} {D61}},\ \bibinfo {pages} {075012}
  (\bibinfo {year} {2000})},\ \Eprint {http://arxiv.org/abs/hep-ph/9909429}
  {hep-ph/9909429} \BibitemShut {NoStop}%
\bibitem [{\citenamefont {Mukhopadhyaya}\ \emph {et~al.}(1998)\citenamefont
  {Mukhopadhyaya}, \citenamefont {Roy},\ and\ \citenamefont
  {Vissani}}]{Mukhopadhyaya:1998xj}%
  \BibitemOpen
  \bibfield  {author} {\bibinfo {author} {\bibfnamefont {B.}~\bibnamefont
  {Mukhopadhyaya}}, \bibinfo {author} {\bibfnamefont {S.}~\bibnamefont {Roy}},
  \ and\ \bibinfo {author} {\bibfnamefont {F.}~\bibnamefont {Vissani}},\
  }\href@noop {} {\bibfield  {journal} {\bibinfo  {journal} {Phys. Lett.}\
  }\textbf {\bibinfo {volume} {B443}},\ \bibinfo {pages} {191} (\bibinfo {year}
  {1998})}\BibitemShut {NoStop}%
\bibitem [{\citenamefont {de~Campos}\ \emph {et~al.}(2012)\citenamefont
  {de~Campos}, \citenamefont {Eboli}, \citenamefont {Magro}, \citenamefont
  {Porod}, \citenamefont {Restrepo} \emph {et~al.}}]{deCampos:2012pf}%
  \BibitemOpen
  \bibfield  {author} {\bibinfo {author} {\bibfnamefont {F.}~\bibnamefont
  {de~Campos}}, \bibinfo {author} {\bibfnamefont {O.}~\bibnamefont {Eboli}},
  \bibinfo {author} {\bibfnamefont {M.}~\bibnamefont {Magro}}, \bibinfo
  {author} {\bibfnamefont {W.}~\bibnamefont {Porod}}, \bibinfo {author}
  {\bibfnamefont {D.}~\bibnamefont {Restrepo}},  \emph {et~al.},\ }\href@noop
  {} {\bibfield  {journal} {\bibinfo  {journal} {Phys.Rev.}\ }\textbf {\bibinfo
  {volume} {D86}},\ \bibinfo {pages} {075001} (\bibinfo {year} {2012})},\
  \Eprint {http://arxiv.org/abs/1206.3605} {arXiv:1206.3605 [hep-ph]}
  \BibitemShut {NoStop}%
\bibitem [{\citenamefont {De~Campos}\ \emph {et~al.}(2010)\citenamefont
  {De~Campos} \emph {et~al.}}]{DeCampos:2010yu}%
  \BibitemOpen
  \bibfield  {author} {\bibinfo {author} {\bibfnamefont {F.}~\bibnamefont
  {De~Campos}} \emph {et~al.},\ }\href@noop {} {\bibfield  {journal} {\bibinfo
  {journal} {Phys. Rev.}\ }\textbf {\bibinfo {volume} {D82}},\ \bibinfo {pages}
  {075002} (\bibinfo {year} {2010})},\ \Eprint {http://arxiv.org/abs/1006.5075}
  {arXiv:1006.5075 [hep-ph]} \BibitemShut {NoStop}%
\end{thebibliography}%
\end{document}